\documentclass[12pt,preprint]{aastex}

\usepackage{mdframed}
\usepackage{natbib}
\usepackage{graphicx}
\usepackage[table,xcdraw]{xcolor}
\usepackage{hyperref}
\usepackage[absolute]{textpos}

\hypersetup{
    colorlinks=true,
    linkcolor=blue,
    filecolor=magenta,      
    urlcolor=cyan,
    linkbordercolor=white,
}
 
\usepackage[english]{babel}
\usepackage{float}

\begin{document}
 
\newpage 
\begin{textblock*}{290mm}(-12mm,-8mm)
	\includegraphics[width=1.1 \paperwidth]{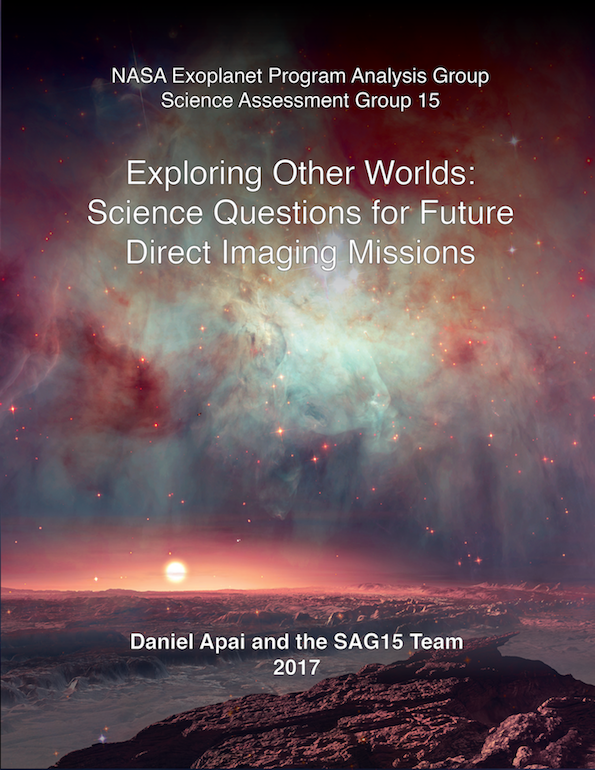}
\end{textblock*}

%\begin{titlepage}

\vfill

\newpage
\thispagestyle{empty}
\mbox{}

\newpage
\thispagestyle{empty}
\mbox{}

\newpage
\thispagestyle{empty}
\mbox{}

\tableofcontents
\newpage

\newpage

\section{The SAG15 Team and Contributors}

{\bf Chair:} D\'aniel Apai, University of Arizona (apai@arizona.edu)

{\bf Authors:}
\begin{minipage}[t]{0.5\textwidth}
{\small 

Travis Barman, University of Arizona

Alan Boss, Carnegie DTM

James Breckinridge, Caltech

Ian Crossfield, UC Santa Cruz

Nicolas Cowan, McGill University

William Danchi, NASA GSFC

Eric Ford, Pennsylvania State University

Anthony del Genio, NASA GISS, NExSS

Shawn Domagal-Goldman, NASA GFSC, NExSS

Yuka Fujii, NASA GISS, NExSS

Renyu Hu, Jet Propulsion Laboratory

Nicolas Iro, University of Hamburg

Stephen Kane, San Francisco State University

Theodora Karalidi, University of Arizona 

Markus Kasper, ESO

James Kasting, Penn State University

Thaddeus Komacek, University of Arizona}
\end{minipage}
\begin{minipage}[t]{0.5\textwidth}
{\small 
Ravikumar Kopparapu, NASA GSFC, NExSS

Patrick Lowrance, Caltech/IPAC

Nikku Madhusudhan, Cambridge University

Eric Mamajek, JPL, NExSS

Avi Mandell, NASA GSFC, NExSS

Mark Marley, NASA Ames, NExSS

Michael McElwain, NASA GSFC

Caroline Morley, Harvard University

William Moore, Hampton University, NExSS

Ilaria Pascucci, University of Arizona, NExSS

Peter Plavchan, Missouri State University

Aki Roberge, NASA GSFC, NExSS

%Leslie Rogers, University of Chicago, NExSS TBC

Glenn Schneider, University of Arizona

Adam Showman, University of Arizona

%Karl Stapelfeldt, JPL

%Mark Swain, JPL

Margaret Turnbull, SETI Institute}
\end{minipage}

\vskip 1cm

{\bf Major Contributors:}

\begin{minipage}[t]{0.5\textwidth}
{\small 

D\'aniel Apai, University of Arizona

Nicolas Cowan, McGill University

Anthony del Genio, NASA GISS, NExSS

Shawn Domagal-Goldman, NASA GFSC, NExSS

Yuka Fujii, NASA GISS, NExSS

Renyu Hu, Jet Propulsion Laboratory

Stephen Kane, San Francisco State University

Theodora Karalidi, University of Arizona}
\end{minipage}
\begin{minipage}[t]{0.5\textwidth}
{\small 
Markus Kasper, ESO

Thaddeus Komacek, University of Arizona

Ravikumar Kopparapu, NASA GSFC, NExSS

Eric Mamajek, JPL, NExSS

Avi Mandell, NASA GSFC, NExSS

Mark Marley, NASA Ames, NExSS

Caroline Morley, Harvard University

Leslie Rogers, University of Chicago, NExSS}

\end{minipage}

%David Ciardi, IPAC/Caltech
%Charley Noecker, JPL
%Philip Stahl, NASA MSFC

\vskip 1cm

{\bf SAG15 Website with up-to-date draft and related documents:}

http://eos-nexus.org/sag15/
\vfill

\newpage

\section{Executive Summary}

	The SAG15 team has solicited, collected, and organized community input on high-level science questions that could be addressed with future direct imaging exoplanet missions and the type and quality of data answering these questions will require (see Appendix A: SAG15 Charter for details).  Input was solicited through a variety of forums and the report draft was shared with the exoplanet community continuously during the period of the report development (Nov 2015 -- May 2017). The report benefitted from the input of over 50 exoplanet scientists and from multiple open-forum discussions at exoplanet and astrobiology meetings.

	The report considered the expected science yield of current and approved major space- and ground-based telescopes and instruments within the timeframe relevant for future direct imaging missions. In particular, Gaia, JWST, WFIRST, Plato, and the 30m-class ground-based telescopes will provide important answers to a variety of exoplanet science questions. The authors agreed that some science questions that are important today will be partly or fully answered by the time new direct imaging exoplanet missions may fly; while some questions that cannot be addressed with existing technology are expected to emerge as central questions in the next decades.
	
\begin{figure}[H]
\begin{center}
\includegraphics[width=1.0 \textwidth]{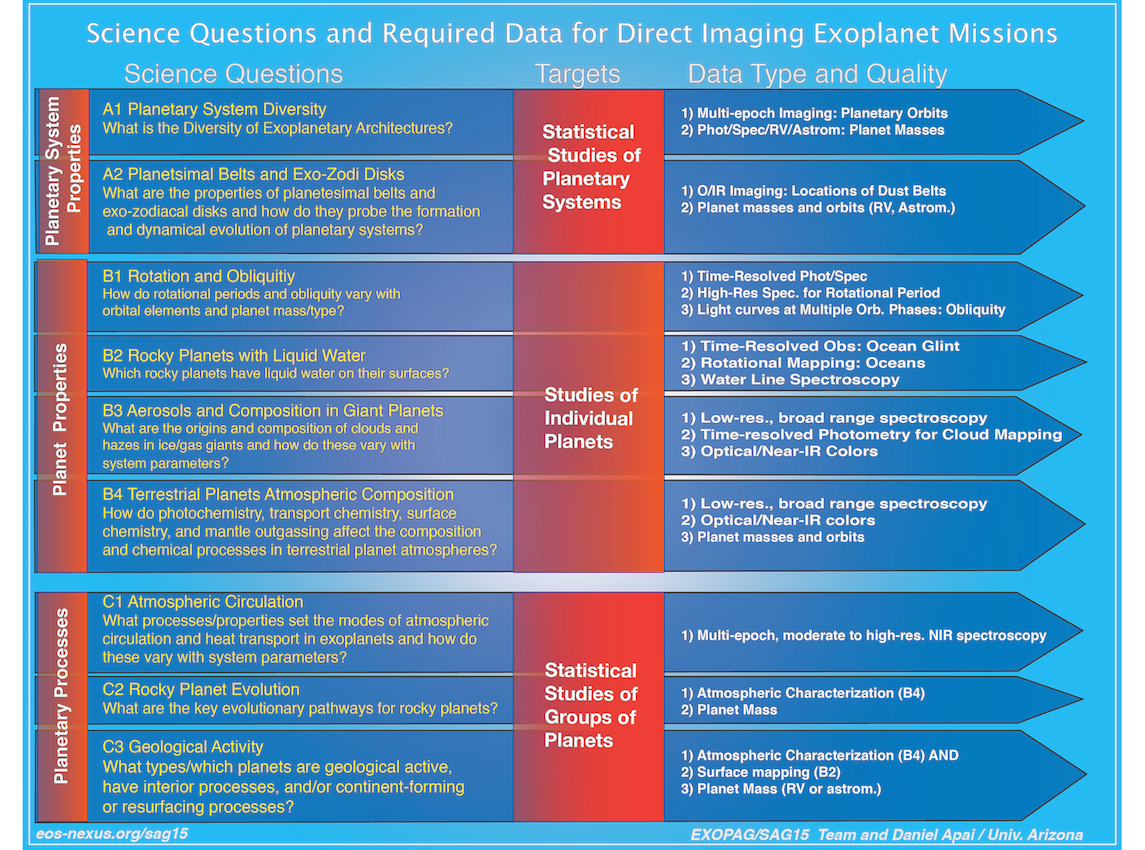}
\caption{Detailed overview of the key high-level science questions and the type of data required to answer them, as identified in the SAG15 Report.
\label{Fig:Overview}}
\end{center}
\end{figure}

	The SAG15 team has identified three group of questions, those that focus on the properties of planetary systems (Questions A1--A2), those that focus on the properties of individual planets (Questions B1--B4), and questions that relate to planetary processes (Questions C1--C4). The questions in categories A, B, and C require different target samples and often different observational approaches. Figure~\ref{Fig:Overview} provides a visual summary of the key questions, the type of targets, and the types of data required for answering them.

The two set of questions on the properties of planetary systems aim to explore the architecture and diversity of exoplanets (massive and low-mass, detected and undetectable) and planetesimals and planetesimal belts. Specifically, Question A1 seek to determine the diversity of planetary architectures, and to identify if there are typical classes of planetary architectures. Answering these questions will also naturally establish the Solar System's relation to the multitude of planetary systems. Question A2 focuses on the distribution and properties of planetesimal belts and exo-zodiacal disks, tracers of the planetesimal population and "fossils" from the planet formation process. 

Questions in the second group focus on the properties of individual planets. Question B1 explores how rotational periods, planetary obliquity, orbital elements, and planet mass/composition are connected. Identifying correlations between these sets of parameters will provide constraints on the formation and evolution of individual planets, and determining these parameters is also important for atmospheric modeling of the climate and atmospheric evolution of habitable planets. Question B2 seeks to understand which rocky planets harbor liquid water on their surface. Question B3 explores the origins and composition of clouds and hazes in ice and gas giant exoplanets and their dependence on the fundamental atmospheric and orbital parameters. Question B4 focuses on rocky planets (habitable and non-habitable) and aims to understand the interplay of photochemistry, transport chemistry, surface chemistry, and mantle outgassing, and their effects on the atmospheric compositions of these planets.

Questions in the third group focus on the evolution of exoplanets and on processes that drive the evolution; questions in this group often require observations that are likely to partly or completely exceed the capabilities of next-generational direct imaging telescopes but may nevertheless be important for missions on the longer-term horizon. Specifically, Question C1 asks what processes and properties set the modes of atmospheric circulation and heat transport in exoplanets and how do these vary with system parameters. Question C2 focuses on rocky planets and seeks to understand the types of evolutionary pathways that are possible for these bodies and what factors determine which pathway will be followed by a given planet. Question C3 seeks understanding of geophysical/geological activity and interior processes in rocky planets, in part to probe the presence of plate tectonics and continent forming/resurfacing processes.

For each questions we summarize the current body of knowledge, the available and future observational approaches that can directly or indirectly contribute to answering the question, and provide examples and general considerations for the target sample required. The questions identified in this report suggest a trend in which questions will increasingly aim to understand complex processes that occur in (or set the properties of) planetary atmospheres and, to some extent, interiors, rather than only exploring the planets' physical properties (mass, density, orbital elements). The community also identified the need for developing a contextual understanding of rocky planets as a prerequisite to correctly interpreting biosignatures (covered in the SAG16 report). 

Our report also provides guidelines and examples for the types and quality of observations required for addressing the individual questions, but a comprehensive and detailed study of the set of required instrumental capabilities is beyond the scope of the current study. Furthermore, we discuss the importance of auxiliary datasets for individual questions and the impact of the independent measurements of planetary mass.

The wide-ranging questions discussed in our report demonstrate how little we know about extrasolar planets, planetary systems, and habitable worlds; and they also highlight the enormous science gains future direct imaging missions will lead to, as well as the fact that comprehensive, multi-mission/multi-instrument studies are often required to provide a thorough understanding of other planets.
	
%	Mass determination
%	Wavelength coverage

\begin{figure}[H]
\begin{center}
\includegraphics[width=1.0 \textwidth]{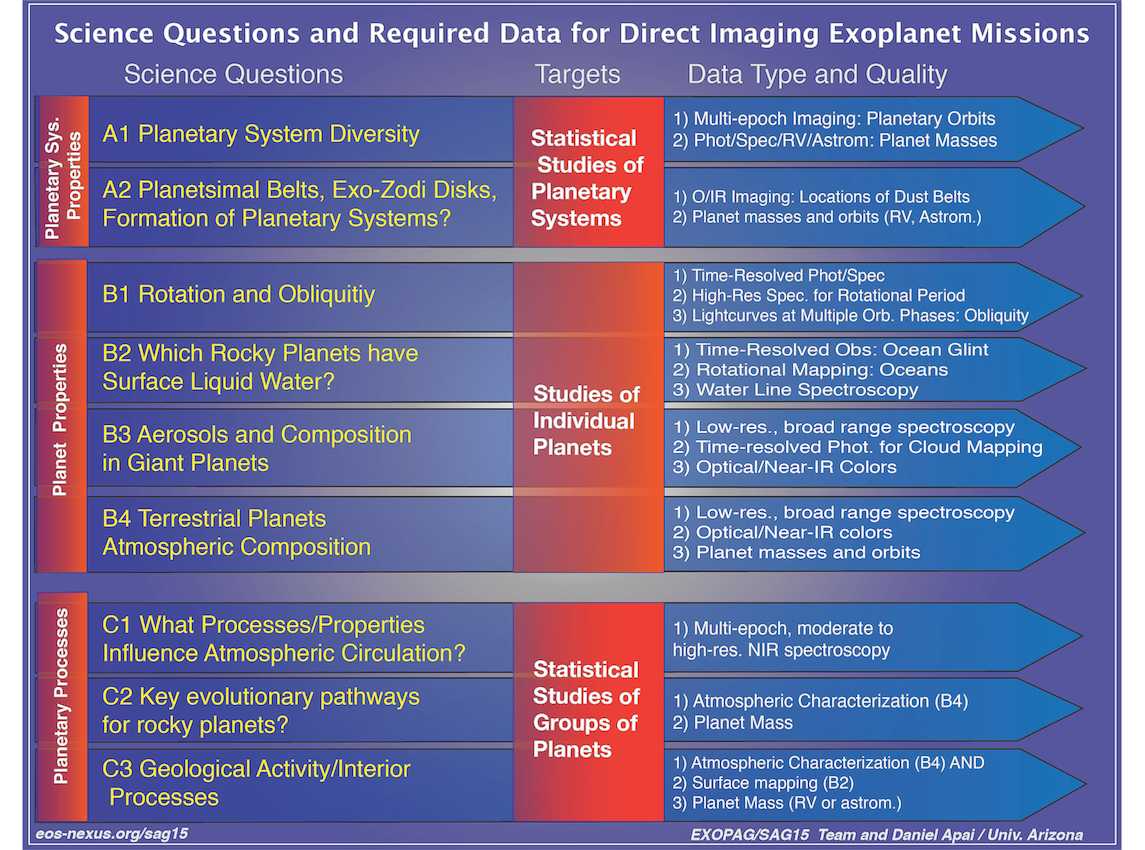}
\caption{Concise overview of the key high-level science questions and the type of data required to answer them, as identified in the SAG15 Report.
\label{Fig:OverviewSimple}}
\end{center}
\end{figure}

\newpage

\section{Introduction}

This report presents organized input from the international exoplanet community on science questions that can be answered by direct imaging missions. 

For each science question we also explore the types and quality of datasets that are either required to answer the question or greatly enhance the quality of the answer. We also highlight questions that require or benefit from complementary (non-direct imaging) observations.

In preparing the report no specific mission architecture or requirements were assumed or advocated for; however, where obvious connections to planned or possible future mission existed these were identified. More detailed evaluations of the capabilities of specific mission architectures are provided in other SAG reports and by ongoing NASA STDTs studies.  The SAG15 report does not include discussion of biosignatures or planets transformed by life, which are discussed in the ongoing SAG16 study, however, the SAG15 reports does include discussion of the characterization of habitable zone earth-sized planets.

{\em Community input:} Input for this report has been collected and comments on the different report drafts have been solicited through a range of channels, including: i) SAG15 website (http://eos-nexus.org/sag15); ii) monthly SAG15 telecons; iii) breakout and discussion sessions during related workshops and meetings; iv) direct requests from topical experts; v) email invitations and solicitations via the EXOPAG and NExSS mailing lists.

{\em Author list and contributor list:} The final report will represent the full endorsement of each author, based on their explicit written statements. In contrast, the SAG15 team list and list of contributors provided in the interim drafts only represents experts who provided input or joined the SAG15 team. The contributor list in the report drafts, therefore, does not represent the endorsement of the draft report and its findings by the contributors.

\vfill

\section{Overview of Science Questions}

The science questions in this report are divided into three categories (see Table~\ref{T:SciQuestions} and Figure~\ref{Fig:Overview}). Questions in Category A aim at the statistical characterization of the formation, evolution, and properties of planetary systems. Questions in Category B aim at the quantitative characterization of individual planets or small groups of planets. Questions in Category C aim at understanding processes that shape planets and planetary atmospheres through comparative studies of planets.

\begin{table}[htp]
\caption{Overview of the science questions discussed in the SAG15 report.\label{T:SciQuestions}}
\begin{center}
%\begin{tabular}{|l|c|}
%\begin{tabular}{p{13cm} l p{4cm}|}
\begin{tabular}{p{13cm}}
{\bf High-Level Science Questions} \\% & {\bf Priority}\\
\hline
\cellcolor{blue!25} {\bf Science questions on exoplanetary system architectures and population } \\ %& \cellcolor{blue!25}\\
\hline
{\bf A1.} What is the diversity of planetary architectures? Are there typical classes/types of planetary architectures? How common are planetary architectures resembling the Solar System? \\
{\bf A2.} What are the distributions and properties of planetesimal belts and exo-zodiacal disks in exoplanetary systems and what can these tell about the formation and dynamical evolution of planetary systems? \\
\hline
\cellcolor{blue!25}{\bf Science questions on exoplanet properties }\\
\hline
{\bf B1.} How do rotational periods and obliquity vary with orbital elements and planet mass/type? \\
{\bf B2.} Which rocky planets have liquid water on their surfaces? \\
{\bf B3.} What are the origins and composition of clouds and hazes in ice/gas giants and how do these vary with system parameters?\\
{\bf B4.} How do photochemistry, transport chemistry, surface chemistry, and mantle outgassing affect the composition and chemical processes in terrestrial planet atmospheres (both habitable and non-habitable)? \\
\hline
\cellcolor{blue!25}{\bf Science questions on exoplanet evolution and processes}\\
\hline
{\bf C1.} What processes/properties set the modes of atmospheric circulation and heat transport in exoplanets and how do these vary with system parameters?  \\
{\bf C2.} What are the key evolutionary pathways for rocky planets?  \\
{\bf C3.} What types/which planets have geological activity, active interior processes, and/or continent-forming/resurfacing processes? \\
\hline
\end{tabular}
\end{center}
\end{table}%

\vfill

\section{Exoplanetary System Characterization}

\subsection{A1. What is the diversity of planetary architectures? Are there typical classes/types of planetary architectures? How common are Solar System-like planetary architectures?}

{\bf Contributors:} Daniel Apai, Nicolas Cowan, Eric Ford, Renyu Hu, Markus Kasper, Timo Prusti

%{\bf Suggested referees:} John Johnson, Tim Morton

The term {\em planetary system architecture} is used here as a descriptor of the high-level structure of a planetary system as given by the stellar mass, the orbits and mass of the planets, as well as the location and mass of its planetesimal belts. 

Understanding the diversity of planetary architectures is important for at least the following two reasons: i) The diversity of planetary system architectures is expected to reflect {\em the range of possible formation and evolution pathways} of planetary systems. ii) To understand how common true {\em Earth analogs} are we must understand {\em how common are planetary systems with architectures similar to that of the Solar System}.

Our current picture of planetary system architectures builds on five sources: 1) Solar System; 2) Data from transiting exoplanets, primarily the Kepler Space Telescope, which probe the inner planetary systems (typically up to periods of 1 year); 3) radial velocity surveys, which provide data on planets with masses typically larger than those accessible to Kepler observations, but over multi-year periods; 4) microlensing surveys, which are also sensitive to small rocky planets at intermediate periods, but provide as yet limited statistics; 5) direct imaging surveys: capable of probing young giant exoplanets at semi-major axes of 8 au or greater.\\

Based on the extrapolation of the close-in exoplanet population detected by the Kepler mission it is very likely that we do not yet have an efficient planet detection method to sample the {\em majority} of exoplanets that exists (at intermediate to large periods and with masses comparable to Earth). ESA's Gaia mission will increase the census of known intermediate- to long-period giant planets by about $\sim$20,000 new discoveries. In addition, the proper motion information for the Solar neighborhood will improve the identification and age-dating of co-moving stellar groups which, in turn, will greatly reduce the uncertainties in the giant planet mass--to--luminosity conversion used by ground-based direct exoplanet imaging surveys, thereby improving the long-period giant planet occurrence rate and mass distribution measurements.

Furthermore, the gradually extending baselines and improving accuracy of radial velocity measurements will also further improve the occurrence rates for short and intermediate-orbit planets (most significantly for neptune-mass and larger planets).
In spite of these significant improvements the occurrence rates of the sub-neptune planets (including rocky and icy planets) at intermediate- to long-period orbits is presently poorly known and may remain unconstrained in the near future.

A powerful direct imaging mission would be powerful in surveying low-mass planets at intermediate and long orbits ($\sim$1 to 5 au), establishing their orbits or constraining their orbital parameters, and measuring or deducing their masses and sizes.

{ Although different techniques will sample different planet populations around different set of stars, a capable direct imaging mission can have the capability of providing a more complete census of planets in the targeted systems than current methods. Direct imaging will survey planets in a range of orbital distances from their host stars, determined by the planet-star separation and photometry contrast. In addition, multiple visits are required to build a a more complete census in the search range of orbital distances, because of the planets' changing orbital phase \citep[][]{2015ApJ...808..172G}.}

\begin{figure}[H]
\begin{center}
\includegraphics[width=0.45 \textwidth]{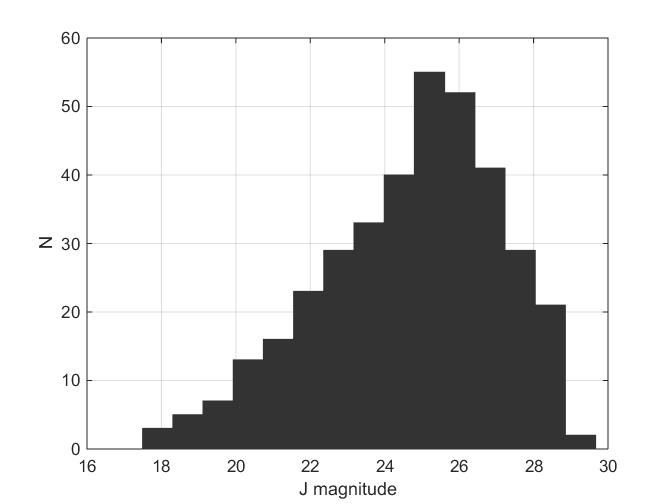}
\includegraphics[width=0.45 \textwidth]{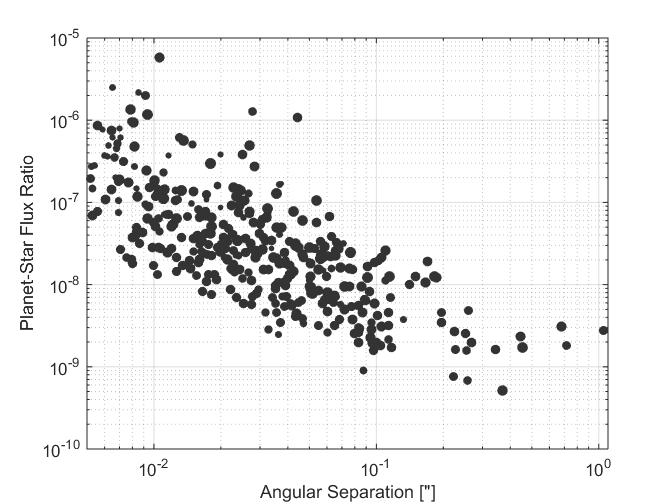}
\caption{Apparent J-band magnitudes and angular separations of known exoplanets. The J-band magnitudes have been estimated assuming a Bond-albedo of 30\%. Courtesy: M. Kasper. \label{Fig:Jband}}
\end{center}
\end{figure}

{\bf Example Questions:}
\begin{itemize}

\item {\em What is the diversity of planetary architectures?} The statistical assessment of the occurrence rate and mass distribution of planets as a function of system parameters (e.g., stellar mass, composition) can constrain and/or verify the predictions of planet formation models. The dispersion in different parameters (from data corrected for selection effects and biases) can be used to quantify the diversity of the architectures.

\item {\em Are there typical classes/types of planetary architectures?} If there are different typical planet formation or evolution pathways, these may lead to the emergence of different classes of planetary architectures (e.g., planetary systems with hot jupiters). The presence of classes of planetary systems may be identified as clustering in the multi-dimensional parameter space that describes planetary architectures, considering the detection and selection biases. Direct imaging will help identify typical planetary architectures but will only provide incomplete and strongly biased information about the existence of planets. 
%primarily provide information about the {\em inner} planets in the planetary systems.

\item {\em How common are Solar System-like planetary architectures?} The {\em local density} of the systems in the multi-dimensional parameter space that describes planetary architectures, determined at the location of the Solar System, provides a measure of the occurrence rate of Solar System-like architectures. Furthermore, in this multi-dimensional parameter space metrics can be defined to quantify the similarity of any two planetary systems. Although non-unique, such metrics may be used to explore the frequency of systems as a function of distance from the Solar System to establish which nearby systems are the most similar to ours.

\end{itemize}

\subsubsection{Complementary Non-Imaging Data}
\begin{itemize}
\item {\em Radial velocity:} Radial velocity precursor measurements can identify sub-stellar companions for future direct imaging (e.g., the TRENDS survey, \citealt[][]{2012ApJ...761...39C} or the California Planet Survey, \citealt[][]{2015ApJ...800...22F}).  Constraints from contemporaneous radial velocity measurements can reduce the number of direct imaging epochs required to establish the orbital elements of the planets. Radial velocity follow-up measurements can also constrain or determine the mass of exoplanets discovered via direct imaging. Furthermore, radial velocity data will be important to discover planets on orbits with semi-major axes smaller than the inner working angle of the direct imaging mission and, thus, important for the goal of assessing the architectures of planetary systems. 

The expected capabilities and impact of near-future radial velocity studies is assessed in the EXOPAG SAG8 report (Latham \& Plavchan and the SAG8 team). Current generation instruments and data analysis techniques are limited in sensitivity to a reflex motion of K$\sim$1-2 m/s by stellar activity and instrument systematics.  Over the next 5--10 years, future generation instruments with reduced instrument systematics (e.g., ESPRESSO, NEID), and improved data analysis techniques for mitigating stellar activity may allow for the detection of reflex motions of less than K$\sim$1 m/s.  For reference, the Earth at 1~au produces a Solar reflex motion of K$\sim$9 cm/s. If this increased radial velocity sensitivity is optimistically realized, the nearest several hundred stars later than F2 in spectral type may be surveyed for Neptune-mass planets amenable to direct imaging (K=28 cm/s for Neptune), and for super-Earths within the inner working angle of a direct imaging mission. In this optimistic scenario, the time baseline of radial velocity observations will be a limiting factor. For example, a decade-long radial velocity survey will only observe linear trends in radial velocity for candidate exoplanets beyond $\sim$10 au.  If there are no further improvements in radial velocity sensitivity from future generation instruments due to stellar activity, then only Jovian analog companions orbiting the nearest stars will be known a priori from radial velocities and amenable to direct imaging.

%Constraints from radial velocity measurements can reduce the number of direct imaging epochs required to establish the orbital elements of the planets. These measurements can also constrain or determine the mass of the target planets. Furthermore, radial velocity data will be important to discover planets on orbits with semi-major axes smaller than the inner working angle of the direct imaging mission and, thus, important for the goal of assessing the architectures of planetary systems. The expected capabilities and impact of near-future radial velocity studies is assessed in the \href{http://adsabs.harvard.edu/abs/2015arXiv150301770P}{SAG8 report} (Latham \& Plavchan and the SAG8 team).

\item {\em Microlensing:} Statistical constraints from the WFIRST-Microlensing (ML) survey on the WFIRST mission (expected to launch in 2026) will provide important context for the frequency of medium-separation low-mass planets. The WFIRST-ML will be sensitive to planets with masses down to $\sim$0.1 M$_{Earth}$ and at separations greater than 0.5~au. The mission will provide complementary information to the Kepler-determined exoplanet population demographics (Figure~\ref{Fig:WFIRST_ML}), albeit for a different population of planet host stars.

\begin{figure}[H]
\begin{center}
\includegraphics[width=0.75 \textwidth]{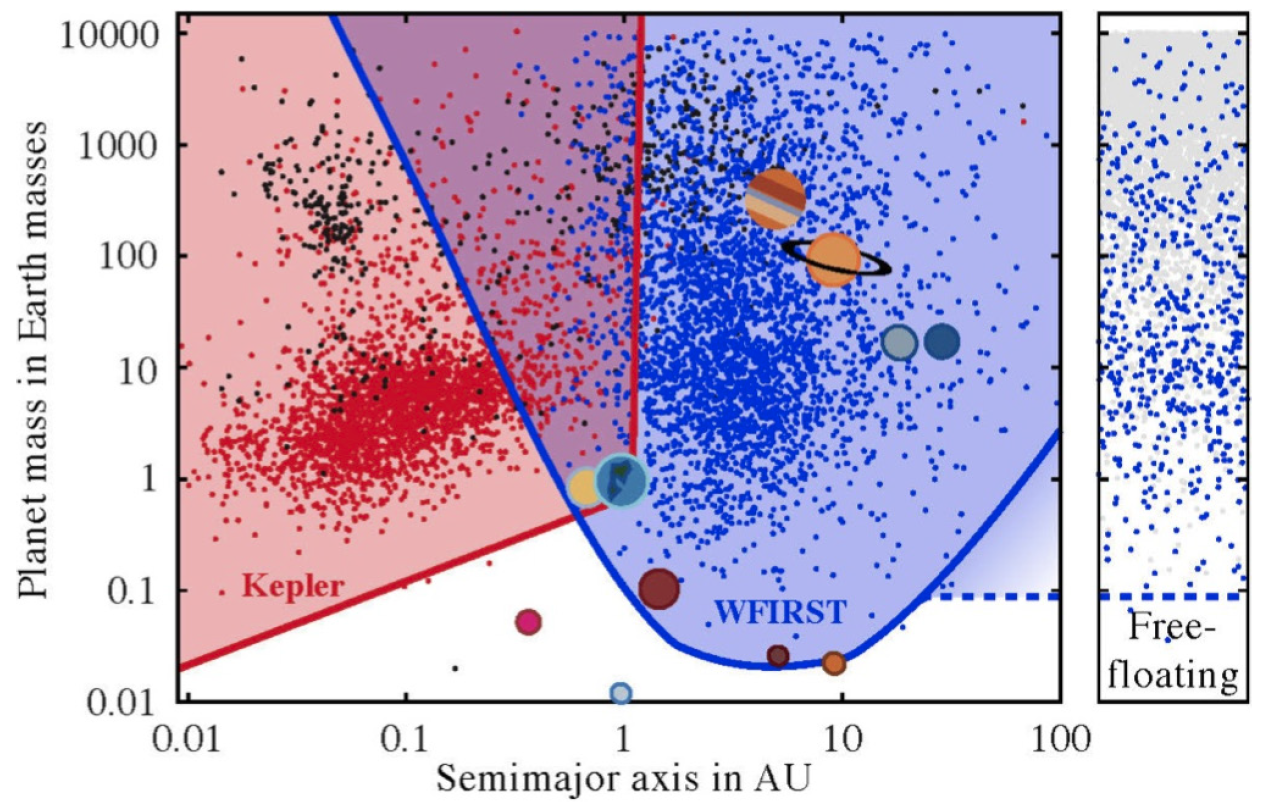}
\caption{The microlensing survey of WFIRST will provide a statistical census for exoplanets with masses down to $\sim$0.1 M$_{Earth}$ and at separations greater than 0.5~au. WFIRST will also be able to probe the population of unbound planets. 
Source: \href{https://wfirst.gsfc.nasa.gov/science/WFIRST_FactSheet_final.pdf}{WFIRST Mission}.
\label{Fig:WFIRST_ML}}
\end{center}
\end{figure}

\begin{figure}[H]
\begin{center}
\includegraphics[width=0.75 \textwidth]{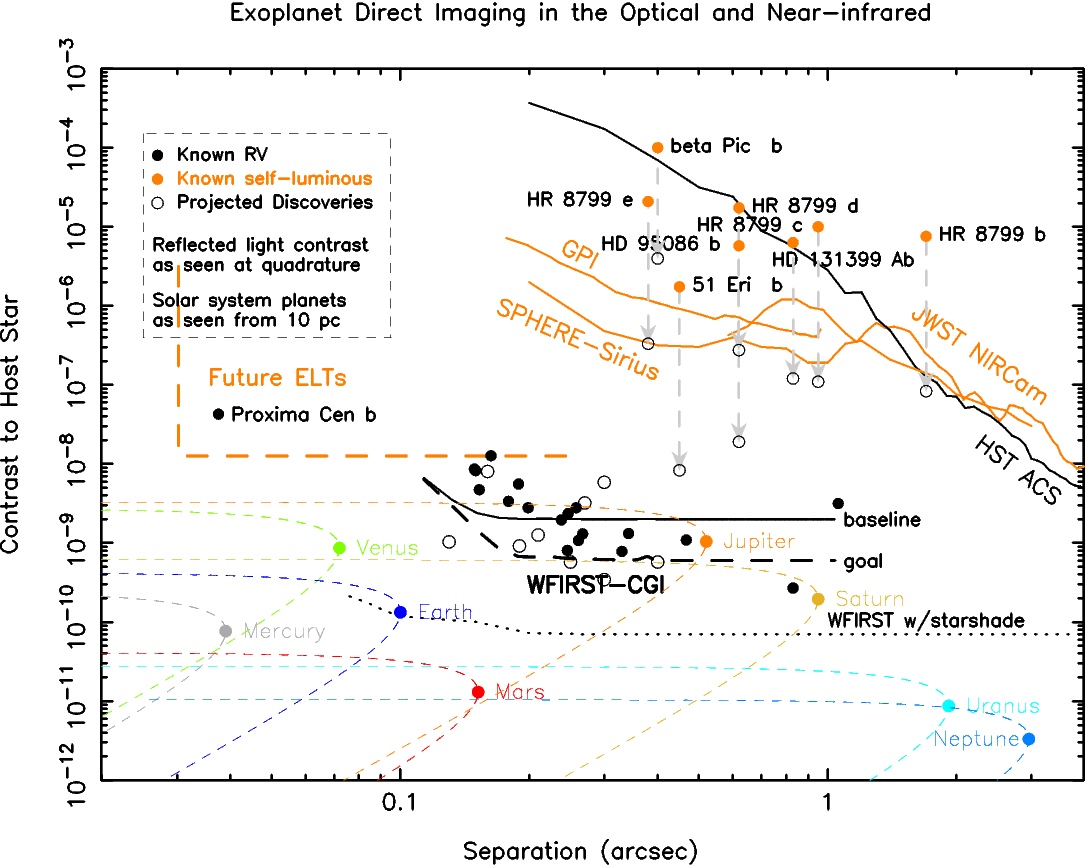}
\caption{The goal of the WFIRST's coronagraphic instrument is to image giant exoplanets and disks at 5 au and greater separations. The sample of planets that WFIRST will be able to image include known radial velocity planets. The Solar System planets (seen at 10 pc) are plotted for reference. Figure by K. Stappelfeldt (JPL). \label{Fig:WFIRST_C}}
\end{center}
\end{figure}

\item {\em WFIRST Coronagraph:} WFIRST will also host coronagraphic capabilities, which primarily aimed at the detection and characterization of giant exoplanets in reflected light. WFIRST will be able to detect giant planets at apparent separations of 0.2" to 1" (Figure~\ref{Fig:WFIRST_C}).

\item {\em Ground-based adaptive optics imaging:} These observations may be capable of discovering earths, super-earths, and neptune-like planets in the inner $\sim$1~au, and gas giant exoplanets at intermediate orbits (up to $\sim$8~au) in nearby systems. By providing positions at additional epochs they will place constraints on the orbits of the planets. In combination with Gaia astrometry the planetary architectures of nearby systems will be relatively well explored in 10-20 years.

{\em Overview of the planned EELT instruments MICADO, HARMONI, METIS, PCS:} The Planetary Camera and Spectrograph (PCS) is expected to have to have first light at the EELT around 2030. It is an visual--NIR high-contrast imager with 15~mas spatial resolution providing an imaging contrast better than $10^{-8}$. Its two main science objectives are: 1) Characterization of Exoplanets from Jupiter-mass to Earth masses and spectroscopy of rocky planets in the habitable zones of nearby late-type stars.

The three first light E-ELT instruments (commissioning around 2025) will provide exoplanet imaging capabilities complementary to PCS. By adopting contrast performance requirements for HARMONI \citep[][]{2014SPIE.9147E..25T,2016SPIE.9908E..1XT} and MICADO \citep[][]{2016SPIE.9908E..1ZD} of $10^{-7}$ at 0.1" and $10^{-8}$ at 0.5" in the near-infrared \citet[][]{2013aoel.confE.121C}, both instruments are well suited for in-depth characterization of self-luminous giant planets, e.g., those discovered by current high-contrast imaging instruments on 8m-class telescopes. HARMONI and MICADO will also be able to exploit the spatial resolution of the E-ELT to resolve the iceline (few au in the nearest star forming regions such as Taurus) and observe forming self-luminous planets at moderately high contrasts. Such data will be key to disentangle the mass-luminosity-age degeneracy and to calibrate evolutionary and atmosphere models of giant planets. MICADO will also provide high precision astrometry at the level of 10 mas, sufficient to detect giant and neptune-mass planets around nearby stars. The sweet spot complementary to the GAIA discovery range are very bright stars in the solar neighborhood and  very late-type stars too faint in the visual wavelengths for GAIA to observe. In this mode MICADO is expected to provide a number of highly interesting targets for follow-up characterization with PCS. 

The instrument METIS \citep[][]{2014SPIE.9147E..21B} will be able to contribute to self-luminous giant planet detection and characterization through mid-infrared imaging and spectroscopy. Older giant planets in the solar neighborhood (up to about 10-20 pc) are accessible if they are sufficiently bright, i.e., heated by the star to T$_{eff}  > 200$~K and seen at a large enough angular separations from their host stars. METIS' exoplanet imaging capabilities are ultimately limited by the large inner working angle ($\sim$50 mas in the L-band, $\sim$150 mas in the N-band) and the sky background-limited sensitivity of a few tens of microJy in N-band (5$\sigma$ in 1 hour). Given that an Earth analogue (a rocky planet in HZ) at 10 pc would provide an N-band flux of about 0.4 microJy, METIS could be able to detect terrestrial planets in the HZ around a handful of the nearest solar-type stars and thereby provide data complementary to PCS for these systems. Through high-resolution spectroscopy, METIS will also be able to measure exoplanet rotation rates and to analyze exoplanet atmospheres.

\item {\em Gaia Astrometry:} The ESA-led precision astrometry mission, launched in 2013, is expected to detect about 21,000 ($\pm$6,000) exoplanets (mostly with masses and orbit similar to Jupiter) in its current 5-year mission, up to a distance of $\sim$500 pc. An extended mission (10-year baseline) would yield about 70,000 planets \citep[][]{2014ApJ...797...14P}. Of these a significant fraction ($\sim$1,000-1,500 planets) will be detected around M-dwarfs, probing such systems up to $\sim$100 pc. Given that the identification of the exoplanets' astrometric signatures is only possible once the parallax and proper motion of the host stars is accurately determined, most of the exoplanet detections are expected to emerge close to the mission's nominal lifetime. 

This dataset will provide orbital elements and masses for a large number of intermediate-period gas giant planets, an important statistical context for the planets to be discovered by the direct imaging mission. However, for the long-period planets ($a>$6 au) Gaia will only be able to measure stellar accelerations, which will only place lower limits on the number, mass, and orbital periods of the planets.

Furthermore, Gaia will not be able to efficiently probe sub-jovian planets, i.e., on its own it will not allow the study of planetary architectures; nevertheless, it will provide a uniquely exhaustive catalogue of jovian exoplanets including measurements of their masses and orbital parameters \citep[][]{2014ApJ...797...14P}.

\end{itemize}

%{\bf From Timo Prusti: Happy to provide the Gaia info to the SAG15 report. The most up to date estimate of exoplanet numbers is that of Perryman et al. 2014 ApJ 797 14P.
%Couple of quick points. Gaia is good with ?Jupiters?. Estimates of the above mentioned paper are about 20,000. It is worth noting that a 10-year Gaia would yield 70,000. We used this also as one of the science cases to motivate extension of Gaia (the estimated life time of Gaia is till end-2023±1 year as far as consumables are concerned). The key message we made is that if we are interested in Jupiter kind of giant planet at the distance of the order of that Jupiter has around the Sun, then Gaia is your discovery machine. The selling point was that if we put value of a giant planet protecting inner terrestrial planets, then Gaia is doing very well. Unfortunately Gaia is not able to tell anything of possible small planets inside the orbit of potentially Gaia discovered giant planet. The astrometric signature is simply too small. Final point concerning Gaia is that the exoplanet results are something coming late in the mission. This because the discovery is based on the wobble of the light centroid above the parallax and proper motion which need to be established first.}

\subsubsection{Observational Considerations}

{\em Orbit determination:} Single-epoch imaging observations will only provided a measurement of the {\em projected} semi-major axis. Therefore, imaging observations that do not cover significant fraction of the targeted planet's orbit will not provide a good assessment of the orbit and equilibrium temperature of the planet, unless auxiliary observations (stellar astrometry and/or radial velocity) are available.
Lacking auxiliary observations multi-epoch imaging observations will be required to derive the orbital elements. The number of the observations required for the orbit determination depends on the orbital period of the planet, the uncertainty on the measured position of the imaged planets at each epoch, and the orbital phases covered by the observations. In a study \citet[][]{2017AAS...22914602B} explored how the uncertainties of the orbital parameters depend on the observing rates (number of observation / year, see, also \citealt[][]{2017AJ....153..229B}). Figure~\ref{Fig:OrbitDetermination} shows how the constraints on the orbital parameters of a hypothetical planet are improving for simulated WFIRST observations. A study within the Terrestrial Planet Finder Coronagraph Report (TPF-C) found that, lacking auxiliary information, imaging in at least 5 to 6 distinct orbital phases is required to constrain the orbital semi-major axis of a planet to {$\sim$10\%) or better accuracy.

\begin{figure}[H]
\begin{center}
\includegraphics[width=0.75 \textwidth]{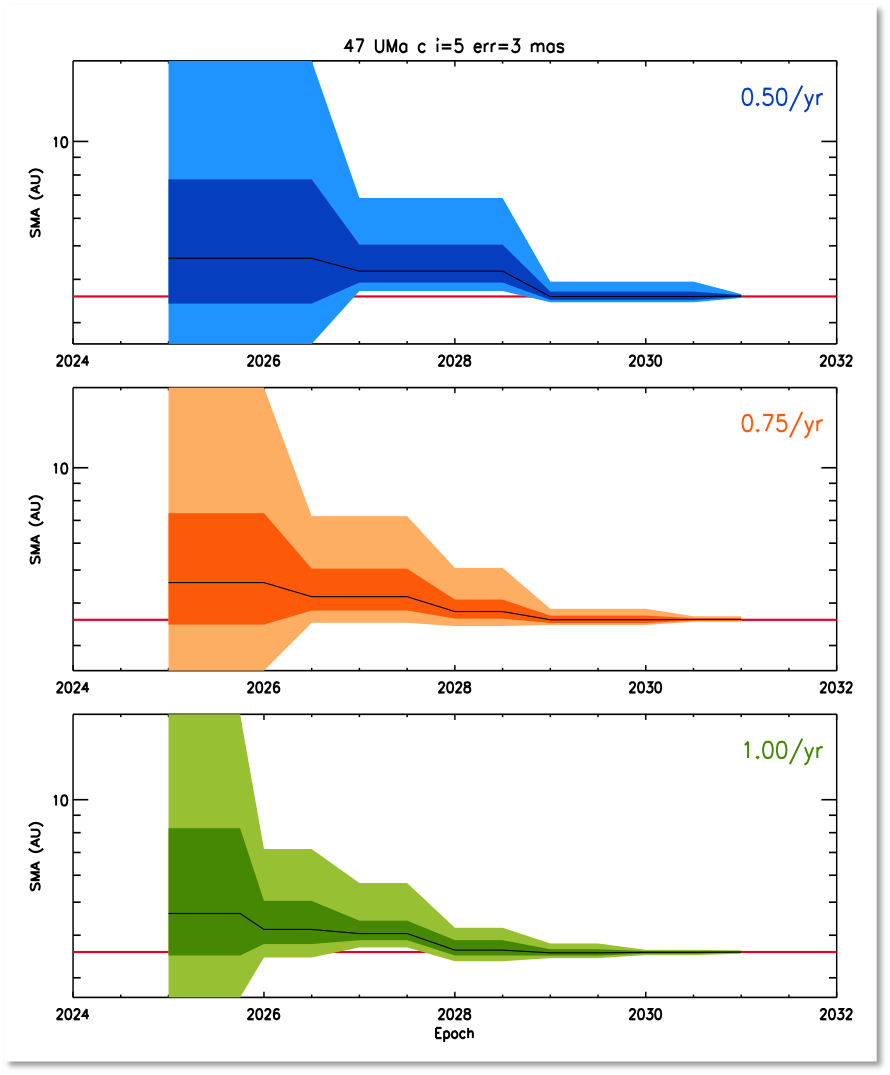}
\caption{Multi-epoch imaging observations will be required to derive the orbital elements.  This figure shows an example for  how the semi-major axis of a hypothetical planet analog to 47 Uma c would improve as new observations are added at different rates of observations (top: 0.5 observations/yr to 1 observation/yr in the bottom, assuming the performance of WFIRST). The red line indicates the actual semi-major axis of the planet, light-- and dark--shaded regions show 68\% and 95\% confidence areas, respectively. With four distinct observations spread across the orbital period the semi-major axis is determined to 5\% precision. From \citet[][]{2017AAS...22914602B}.  \label{Fig:OrbitDetermination}}
\end{center}
\end{figure}

%"Left: reduction of semimajor axis uncertainty with time for a hypothetical analog of 47 Uma c discovered with WFIRST. Panels show different observing rates. Light and dark colors show 95\% and 68\% confidence areas, respectively. We conclude that 1) taking 4 observations spread across the orbital period recovers orbital parameters to 5\% precision, and 2) continuous orbital monitoring is not required."

\begin{mdframed}[backgroundcolor=blue!20] 
{\bf Observational Requirements:} \\
{\em Planetary Orbital Elements:} Visual or infrared imaging to identify the presence and locations of planets in each system. Imaging in at least five to six epochs {\em or} complementary radial velocity or astrometry measurements are required to constrain well the orbital parameters. \\
{\em Planet Type:} Multi-color photometry or spectroscopy are required to establish the nature of each planet (approximate mass and composition). Additional stellar radial velocity and/or astrometry can significantly enhance the capability to determine the planet mass and type. \\
{\em Wavelength:} The ideal choice of wavelength is driven by a trade-off in achievable contrast and sensitivity (to detect the planets and to accurately measure their positions) and by the required spectral coverage to determine their fundamental properties or otherwise classify them. Therefore, if additional radial velocity or astrometry is available a narrower wavelength range is sufficient.\\
{\em Complementary Observations:} a) Stellar radial velocity and astrometry are important as they will identify or place constraints on the existence of additional, unseen planets. b) Stellar radial velocity and astrometry can also provide important constraints enabling the classification of the imaged planets (mass and type). c) Radial velocity and astrometry will also help to improve the orbital solutions for the imaged planets and/or reduce the number of re-imaging visits necessary to establish the orbits of the planets in the system (which are likely to have very different orbital periods, and fitting the multiple orbits may require a larger number of imaging visits).
\end{mdframed}

%\begin{mdframed}[backgroundcolor=yellow!20] 
%\small
%{\bf Pending Questions / Comments: } \\
%1) What statistical constraints will Gaia and future RV surveys provide?\\
%2) Add more info on GMT/ELT/TMT capabilites.\\
%\end{mdframed}

\newpage

\subsection{A2. What are the distributions and properties of planetesimal belts and exo-zodiacal disks in exoplanetary systems and what can these tell about the formation and dynamical evolution of the planetary systems?}

{\bf Contributors:} Daniel Apai, Markus Kasper

Direct imaging missions will provide spatially resolved images of exo-zodiacal disks, possibly composed of narrow and/or extended dust belts. In these belts dust is produced by minor body collisions and the dust belts are dynamically sculpted by the gravitational influence of the star and the planets, grain-grain collisions, as well as radiation pressure (for reviews see, e.g., \citealt[][]{2008ARA&A..46..339W}). In some, apparently very rare, systems gas is also present and may influence the dust distribution. 

The distribution and properties of exo-zodiacal dust belts (or debris disks) are important as they provide information on:\\
{\em i)} The presence, orbits, and masses of unseen planets orbiting in the disks.\\
{\em ii)} The orbits and masses of planets seen in the direct images, but for which orbits are not known.\\
{\em iii)}  The inclination of the disk/planet system.\\
{\em iv)} The formation and evolution of the system, including the past migration and orbital rearrangements of the planets.\\
{\em v)} Compositional constraints on the availability of volatiles/organics in the planetesimal belts and, by inference, in the planets.\\

\begin{figure}[H]
\begin{center}
\includegraphics[width=\textwidth]{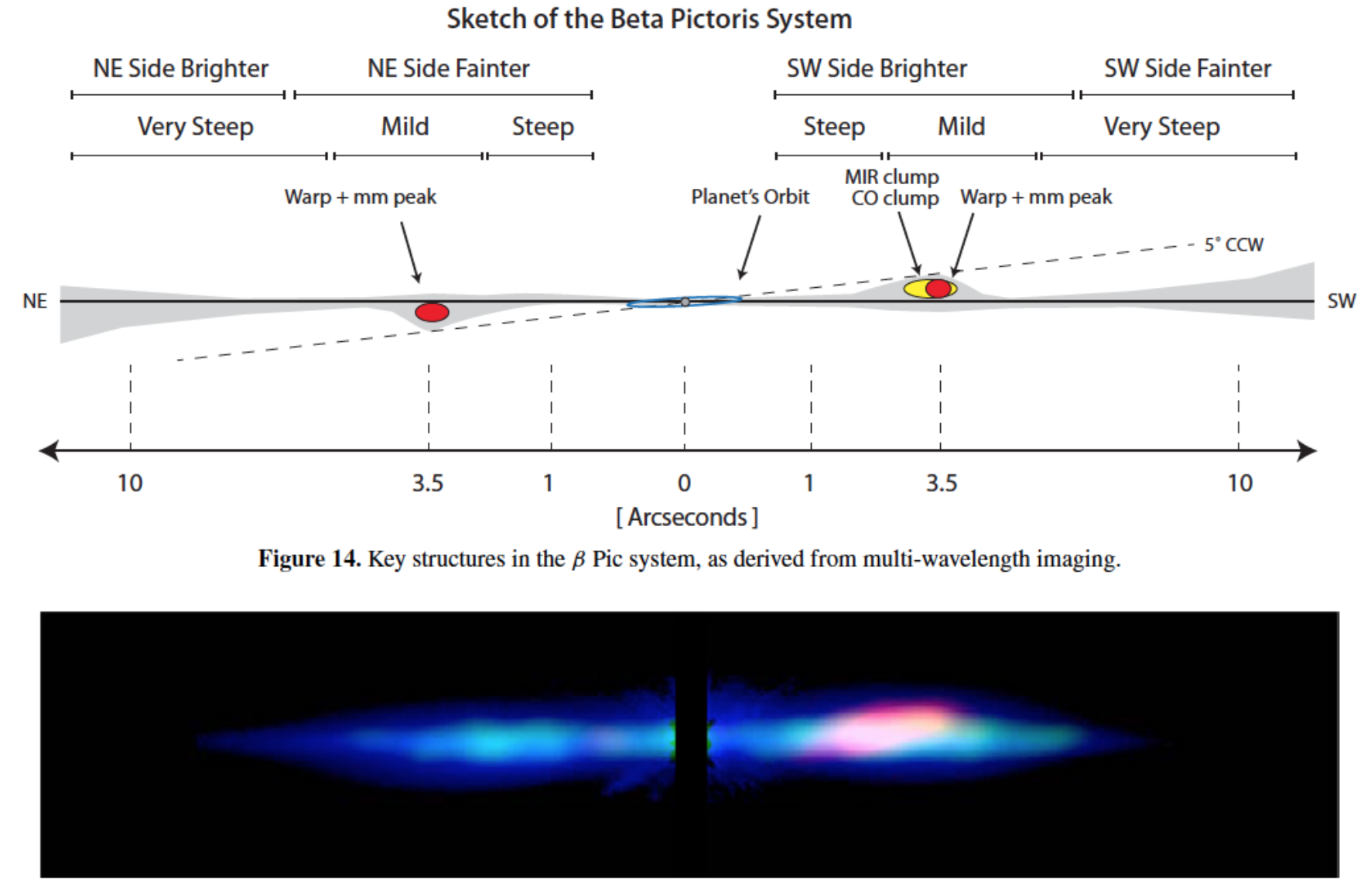}
\caption{Simulations of the structure of the edge-on debris disk around $\beta$~Pictoris correctly predicted the location and mass of the perturber super-Jupiter $\beta$ Pictoris b \citep[][]{1997MNRAS.292..896M}. This system is one of the the best-studied examples of disk-planet interactions. {\em Lower panel:} HST/STIS coronagraphic image (blue), ALMA dust continuum (green), and ALMA CO gas emission (red) illustrate the complex structure of the disk (from \citealt[][]{2015ApJ...800..136A}).
\label{Fig:Apai2015}}
\end{center}
\end{figure}

\begin{figure}[h]
\begin{center}
\includegraphics[width=0.5 \textwidth]{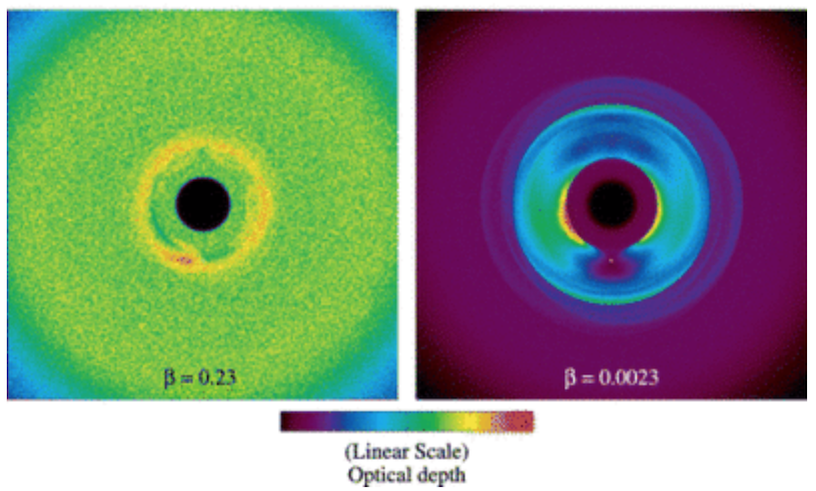}
\caption{Comparison of the optical depths predicted by disk-planet interactions models for a composite cloud formed for a 2 earth-mass planet at 6 au (from \citealt[][]{2008ApJ...686..637S}). The planet, marked with a white dot, orbits counterclockwise in these images. {\em Left:} Optical depth of the smallest particles included in the composite clouds. {\em Right:} Optical depth of the largest particles included in the composite clouds. The largest particles dominate the optical depth in a cloud of particles released with a Dohnanyi crushing law, because they are longer-lived and more likely to be trapped in mean motion resonances than smaller particles.
\label{Fig:Stark2008}}
\end{center}
\end{figure}

\subsubsection{Current Knowledge} 
Currently, large databases of bright debris disks are available for which spatially unresolved thermal infrared observations (spectral energy distributions or SEDs) are available. In addition, for a subset of disks spatially resolved scattered light or thermal emission images are available (see, e.g., Fig.~\ref{Fig:Apai2015}). Mid-infrared spectroscopy of solid state dust features \citep[e.g.,][]{2005Natur.433..133T} and polarimetric imaging provide additional constraints on dust composition and disk structure (e.g., \citealt[][]{2015ApJ...799..182P}). The different wavelengths, the types of emission (continuum, spectral features), and the polarization properties of the light allow us to disentangle the different dust components and study their origins. 

%{\bf  Add a few sentences on general debris disk evolution, collisional events, self-stirring and planetary stirring, orbital rearrangements and bombardments. }

\subsubsection{Sub-questions}
{\em The presence, orbits, and masses of unseen planets:} Detailed simulations of debris disk structures and disk-planet interactions provide predictions for the expected disk structures (see Fig.~\ref{Fig:Stark2008},  e.g., \citealt[][]{1999ApJ...527..918W,2003ApJ...598.1321W,1997MNRAS.292..896M,2008ApJ...686..637S}). In a large set of disks complex structures have been observed which can possible be explained by the influence of yet unseen planets \citep[e.g.,][]{2014AJ....148...59S}; in a very small number of systems disks and planets have been observed together, providing an opportunity to study disk-planet interactions and to validate models (see, e.g., \citealt[][]{2015ApJ...800..136A}  and Fig.~\ref{Fig:Apai2015}).

{\em The orbits and masses of planets seen in the direct images:} With certain direct imaging architectures (e.g., starshades) opportunities for multi-epoch observations may be limited, making it more difficult to verify that point sources are planets and not background sources; and to estimate masses/orbits for the planets from short integrations. Most directly imaged systems are expected to host dust disks, whose structures may be used to verify that the planet candidates imaged are indeed in the system and then to constrain their mass and orbit.

{\em The inclination of the disk/planet system:} For any planet an important  but particularly challenging parameter pair to determine is the inclination/eccentricity pair. These quantities are partially degenerate and can be difficult to disentangle from observations limited to a handful of visits. Resolved debris disks structures can complement measurements of the planet's relative motion to break the degeneracy of inclination/eccentricity. For example, nearly-edge on disks can be recognized even in single-epoch images, which then greatly constrain the available parameter space for the planet's orbit.

{\em The formation and dynamical evolution of the systems:} The mass and position of planetesimal belts can provide powerful constraints on the formation and evolution of planetary systems, including planet migration and/or major orbital rearrangements. For example, the asteroid belt and the Kuiper belt in the Solar System have revealed such orbital rearrangement and potential past instabilities \citep[e.g.,][]{1993Natur.365..819M,2005Natur.435..459T}. In addition, sensitive time-resolved observations in debris disks also have the potential to identify multiple other mechanisms that act on short timescales, such as the aftermath of recent major impacts \citep[e.g.,][]{2014Sci...345.1032M}, dust clumps moving under the influence of radiation pressure, or dust created by planetesimals trapped in resonant structures (e.g., \citealt[][]{2003ApJ...598.1321W,2015ApJ...800..136A,2015Natur.526..230B}).

{\em Compositional constraints on the availability of volatiles/organics in the planetesimal belts:} In each system planetesimal belts are leftovers of reservoirs that likely contributed mass to the planets. Therefore, the planetesimals' compositions may place some constraints on the composition of the planets themselves. It is important to note, however, that the composition of planetesimals need not necessarily match the compositions of rocky planets: for example, in the Solar System none of the known meteorite groups matches the abundance patterns of the bulk Earth or the bulk silicate Earth, although the compositions of Mars and Vesta are consistent with mixtures of known meteorite types \citep[e.g.,][]{2006mess.book..803R}. On the other hand, the fact that terrestrial depletions of many siderophile elements in Earth's primitive upper mantle are matched by predictions of high-temperature, high-pressure partition consistent with the pattern expected from metal-silicate equilibrium and homogeneous accretion of CI-chondritic material \citep[][]{1997E&PSL.146..541R}.

In the observations of extrasolar planetesimal belts of particular interest is the availability of volatiles and organics in the planetesimals, as these components are essential for life as we know yet are likely to be difficult to identify in rocky planets. Volatiles and organics are thought to be heavily depleted in the warm, inner disk regions where habitable planets accrete. Organics and volatile content (interior or as a surface layer) change the optical properties of the dust grains, producing signatures that are detectable at visual and infrared wavelengths \citep[e.g.,][]{2008ApJ...673L.191D,2014ApJ...783...21R,2016ApJ...823..108B}. Recently discovered debris disks with gas content that may be recent or primordial provide an additional opportunity to explore volatile reservoirs in planetesimal belts \citep[][]{2014Sci...343.1490D,2013ApJ...776...77K,2013ApJ...777L..25M}

\subsubsection{Complementary Data} 
Exo-zodiacal disk studies will benefit from:\\
1) WFIRST imaging of debris disks: these studies will be capable of imaging disk structures with an intermediate inner working angle. \\
2) ALMA observations of cold debris disks: will provide mass estimates and combined density-temperature constraints, allowing to constrain disk-planet interactions \citep[][]{2012A&A...544A..61E,2015ApJ...798..124R,2012ApJ...750L..21B,2011MNRAS.413..554M,2016MNRAS.460L..10B} and even identifying CO gas in some debris disks \citep[][]{2014Sci...343.1490D,2013ApJ...776...77K}.  \\
3) LBTI and ELT observations of the warm debris: The three EELT instruments (METIS, MICADO, and PCS)  will contribute to characterizing the exo-zodiacal disks. While METIS will provide detailed images of warm debris disk belts inside the terrestrial planet region at a couple of au, MICADO and ultimately PCS will provide detailed images of debris disks in scattered light.\\
4) JWST observations of warm debris disks: may provide spatially resolved or unresolved near- and mid-infrared spectroscopic information on the dust grains. The spectral information could be used to constrain the grain size distribution and the composition of the warm inner dust disk. \\

\begin{mdframed}[backgroundcolor=blue!20] 
{\bf Observational Requirements} \\
%{\bf Sample size:} Individual or small samples are useful for the characterization of individual disk-planets systems, but medium to larger samples (30-60) are required for studying disk evolution or the distributions of disk properties. \\
{\em Disk Structure:} Multi-wavelength (preferably visual to mid-infrared) high-resolution, high-contrast imaging, spectroscopy, and visual or near-infrared polarimetry. Spectropolarimetry over a broad wavelength range would be ideal for characterizing the dust properties and for disentangling scattering phase functions, and albedo/grain size variations from disk structure (density and vertical scaleheight). Large field of view ($\sim$5--10" or larger) will be necessary to study Kuiper-belt-like disk morphologies in nearby systems.\\
{\em Complementary Observations:} This topic will benefit from a broad range of supplementary observations. Indirect detection methods (stellar astrometry and radial velocity) can reveal or constrain the presence of short- and medium-separation planets. Ground-based adaptive optics observations can provide information on the disk structure and close-in exoplanets. ALMA sub-millimeter continuum and gas line observations can sample disk density variations and gas/dust mass ratio. 
\\
\end{mdframed}

%\begin{mdframed}[backgroundcolor=yellow!20] 
%\small
%{\bf Questions to SAG15: } \\
%Input from WFIRST PS team on what debris disk science do they foresee.
%\end{mdframed}

\newpage

\newpage
\section{Exoplanet Characterization}

\subsection{B1. How do rotational periods and obliquity vary with orbital elements and planet mass/type?}

{\bf Contributors:} Daniel Apai, Nicolas Cowan, Renyu Hu, Anthony del Genio
%{\bf Suggested referees:} Hajime Kawahara

\medskip

	A planet's rotational state refers to both its obliquity and frequency, or equivalently period. Planetary rotation constrains the formation and angular momentum evolution of a planet, especially when comparing statistical samples of diverse planets. Moreover, the rotation of a given planet impacts its climate through diurnal forcing and through the Coriolis forces, and contributes to magnetic field generation.
	
	For example, \citet[][]{2013ApJ...771L..45Y,2014ApJ...787L...2Y} showed that the rotation periods of temperate terrestrial planets changes the inner boundary of the habitable zone by a factor of two in insolation (also see \citealt[][]{2016ApJ...819...84K}). Furthermore, planetary magnetic fields may be important shields against atmospheric loss. As these examples illustrate the rotational state of temperate terrestrial planets directly impacts their habitability.  
	
	We note, that depending on the nature and atmospheric composition of a planet its true rotational period (that of its bulk mass) may or may not be possible to determine observationally (e.g., Venus). For example, while a rocky planet's rotational period may be observed via the observations of surface features, for gaseous planets or rocky planets with optically thick atmospheres the rotational period of the interior may remain hidden and only an "apparent rotational period" may be observed: one that is a combination of the rotational rate and dominant atmospheric motions (winds, circulation). 
	
\subsubsection{State of the Art to Measure Rotational Periods}
As of now little is known about the obliquity and rotational periods of non-synchronously rotating exoplanets. Rotational periods for planets and exoplanets have been determined through four different methods: 
 
{\em a) Phase Curve and Eclipse Mapping for Irradiated Planets:} For some close-in synchronously rotating giant exoplanets the orbital/rotational phase modulation is detectable in the combined light of the star and planet system. 
In particular, modulations during eclipse (planet passing behind the star) allows coarse two-dimensional mapping of the planets: For example, the dayside map of HD 189733b suggests that this hot Jupiter has zero obliquity \citep[][]{2012ApJ...747L..20M,2012A&A...548A.128D}. The eastward offset of the hotspot observed on most hot Jupiters \citep[][]{2007Natur.447..183K,2009ApJ...690..822K,2012ApJ...754...22K, 2010ApJ...723.1436C,2012ApJ...747...82C} is consistent with equatorial super-rotation on a synchronously-rotating planet \citep[][]{2002A&A...385..166S}, but also with slower winds on a non-synchronous planet \citep[][]{2014ApJ...790...79R}. In fact, there is a complete observational degeneracy between the rotation of a gaseous exoplanet and its winds \citep[][]{2011ApJ...729...54C}. \citet[][]{2013ApJ...771L..45Y} showed that tidally locked temperate planets  (with dayside insolation of 220 W/m$^2$) will have a stable cloud pattern resulting from a stabilizing feedback, while non-synchronously rotating but otherwise similar planets will not. The stable water vapor cloud pattern may be detectable in the disk-integrated light curve.

{\em b) Period of the magnetic field's rotation:}  The magnetic field is tracing the rotational periods of the planets' {\em interiors}, which may be different from the latitude-averaged rotational periods measured in their upper atmospheres. In the Solar System, Jupiter's  rotational period is defined by the rotation of its inclined (w.r.t. spin axis) magnetic dipole,  while Saturn's magnetic field exhibits a very small tilt and its rotation period thus remains somewhat uncertain. %For exoplanets, in exceptional cases, the manifestation of inclined magnetic dipoles may be detectable through time-varying auroral emission at UV/optical wavelengths or via modulated synchrotron emission in the radio. 
Recent detections of modulated radio emission from nearby brown dwarfs \citep[e.g.,][]{2016ApJ...818...24K} suggest that very sensitive radio-wavelength observations of extrasolar giant planets may also be used in the future to establish their rotational periods.

{\em c) Absorption line width measurements: } Recently,  CO absorption line width measurements have been used to measure the rotational velocity ($v \sin{i}$) for the directly imaged exoplanet Beta Pictoris b (\citealt[][]{2014Natur.509...63S}) and in the combined star and planet light for hot jupiters (e.g., \citealt[][]{2012ApJ...753L..25R}).
Similar studies for rotational line broadening have been carried out successfully for brown dwarfs \citep[e.g.,][]{2008ApJ...684.1390R}. In order to convert the observed {\em v sin i} into a rotational period, one must know the planet's radius and obliquity. This method is therefore well-suited for brown dwarfs and giant planets (which are all approximately the size of Jupiter), but could prove problematic for lower-mass directly-imaged planets of unknown radius. Furthermore, it is more applicable for systems where constraints exist on the planets' obliquities (primarily derived from rotational modulations observed over multiple orbital phase angles).

{\em d) Rotational photometric/spectroscopic modulations} in hemisphere-integrated light for directly imaged exoplanets (Fig.~\ref{Fig:Zhou2016}, \citealt[][]{ 2016ApJ...818..176Z}) and planetary-mass brown dwarfs (e.g., \citealt[][]{2015ApJ...813L..23B}, ~\citealt[][]{2016ApJ...830..141L}). This method is conceptually identical to method {\em a}, but requires a different observational approach. An excellent recent review by Biller (2017 Astronomical Review) provides an overview on the state of the field and next steps. Brown dwarfs (planetary mass and more massive), are good analogs for directly imaged exoplanets. These observations showed that low-level ($\sim1\%$) rotational modulations in thermal emission are {\em very} common \citep[][]{2014ApJ...782...77B,2015ApJ...799..154M}, and can be used to measure or constrain rotational periods and to study cloud properties \citep[e.g.,][]{2009ApJ...701.1534A,2012ApJ...750..105R,2013ApJ...768..121A}. Similarly, reflected-light observations of Solar System giant planets have also been used to demonstrated that rotational periods and their cloud covers can be characterized (e.g., Jupiter: \citealt[][]{2015ApJ...814...65K}; Neptune: \citealt[][]{2016ApJ...817..162S}).

\begin{figure}[ht]
\begin{center}
\includegraphics[width= 0.5 \textwidth]{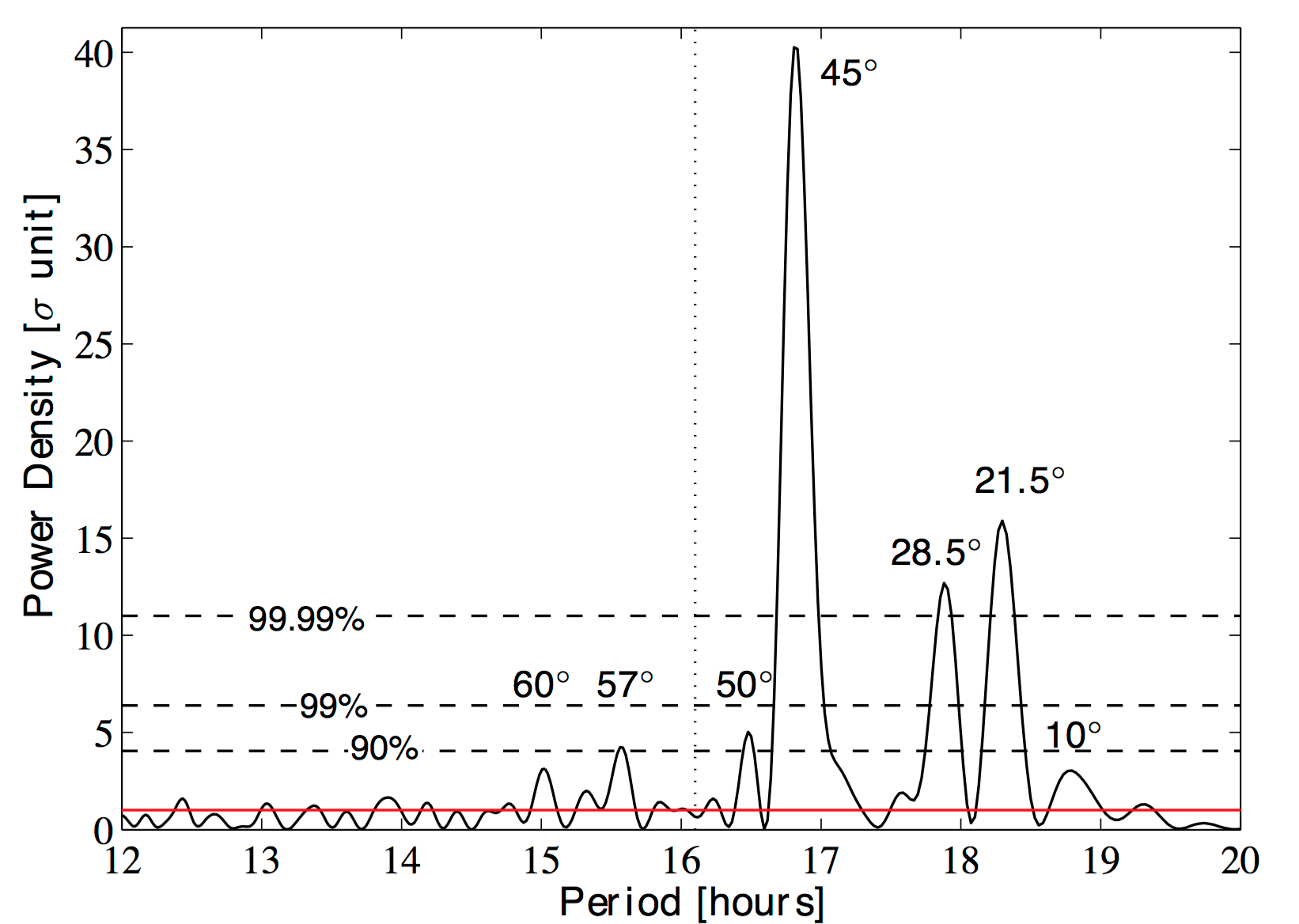}
\caption{Whitened power spectrum from 50-day-long Kepler monitoring of hemisphere-integrated reflected light Neptune, with the most significant peak corresponding to the rotation period. Numbers above some peaks indicate the latitudes on Neptune corresponding to that rotation period based on the zonal velocities. From Simon et al. (2016).\label{Fig:Simon2016}}
\end{center}
\end{figure}

\begin{figure}[ht]
\begin{center}
\includegraphics[width=0.5 \textwidth]{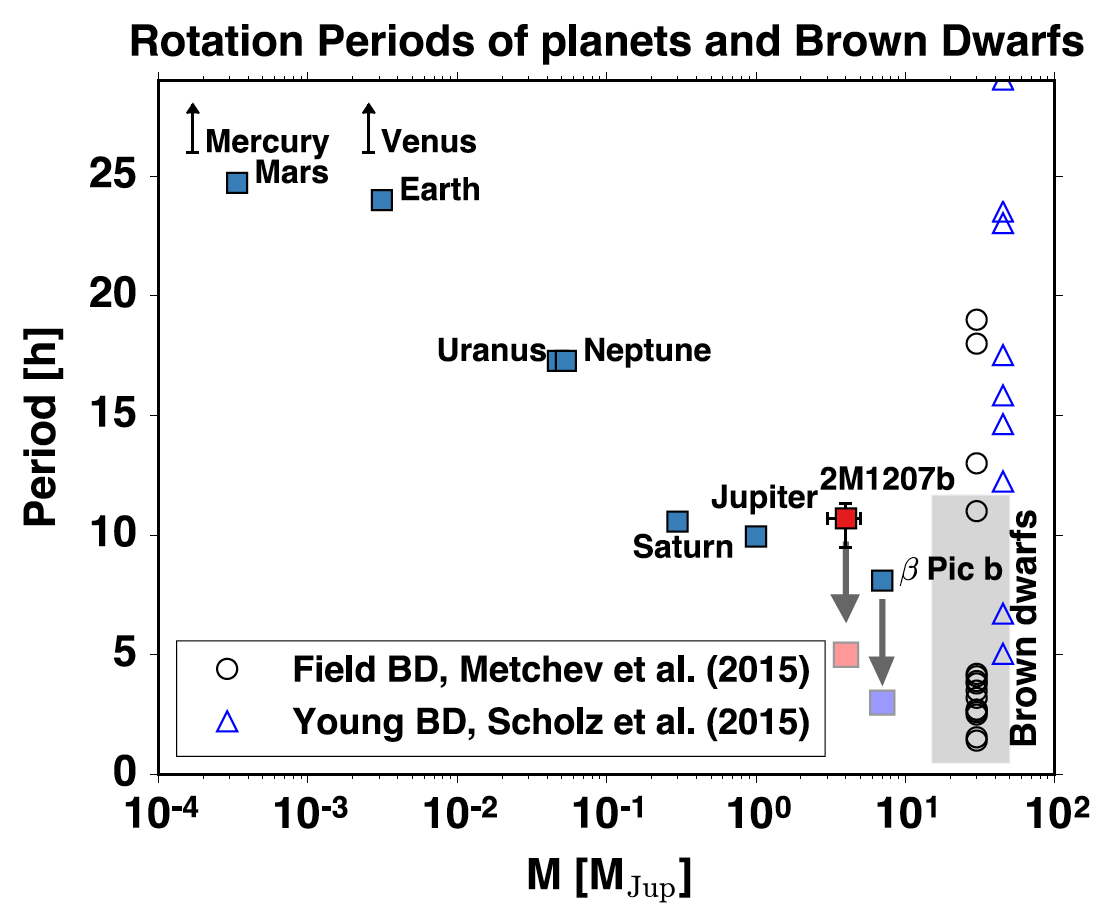}
%\plotone{Figures/Zhou_2016.png}
\caption{Rotation periods provide insights into the properties and formation of planets. A comparison of Solar System planets, directly imaged exoplanets, and brown dwarfs  reveals a characteristic mass-dependent rotation rate for massive planets. The ages of the Solar System planets is 4.56~Gyr; the ages of the directly imaged planets is $<$30~Myr; the ages of the brown dwarfs are few Myr (triangles) and a broad age range for the field objects (triangles). The arrows shows the expected spin-up due to gravitational contraction. From Zhou et al. (2016).   \label{Fig:Zhou2016}}
\end{center}
\end{figure}

Both techniques {\em a} and {\em d} may be applicable for exoplanets directly imaged with next-generation space telescopes. While method {\em b} requires high spectral resolution and provides Doppler information, method {\em d} requires only high signal-to-noise time-resolved photometry and not strongly wavelength-dependent. 
	
\subsubsection{Science Cases}
	
{\bf Habitable Planets (Earth-sized and Super-Earths):} Rotation rates are an important parameter for climate and atmospheric circulation models of habitable planets: they constrain diurnal temperature modulations, determine the strength of the Coriolis force, the nature of the circulation, and thus the location of clouds, influence current and past magnetic field strengths and geometry, and indirectly constrain the atmospheric loss that may have occurred on these planets.  Comparative studies of dynamo-generated magnetic energy densities in Solar System planets, the Sun, and rapidly-rotating low-mass stars show a correlation between the magnetic field strengths and the density and bolometric flux of the objects (see Fig.~\ref{Fig:Christensen2009}, e.g., \citealt[][]{ 2009Natur.457..167C}). These studies argue for a scaling relation, based on Ohmic dissipation, where the field strength is only weakly sensitive to rotation rate, but the rotational rate fundamentally impacts the magnetic field geometry (bipolar vs. multi-polar, \citealt[][]{2010SSRv..152..565C}).
Furthermore, rotational rates may be influenced by the presence of a moon or, if no massive satellite is present, may preserve some information about the accretion history of the planets \citep[e.g.,][]{2007ApJ...658..593S}. While this information is unlikely to be diagnostic for any single planet, patterns in rotation rates may emerge from larger samples of otherwise similar planets.

\begin{figure}[ht]
\begin{center}
\includegraphics[width=0.5 \textwidth]{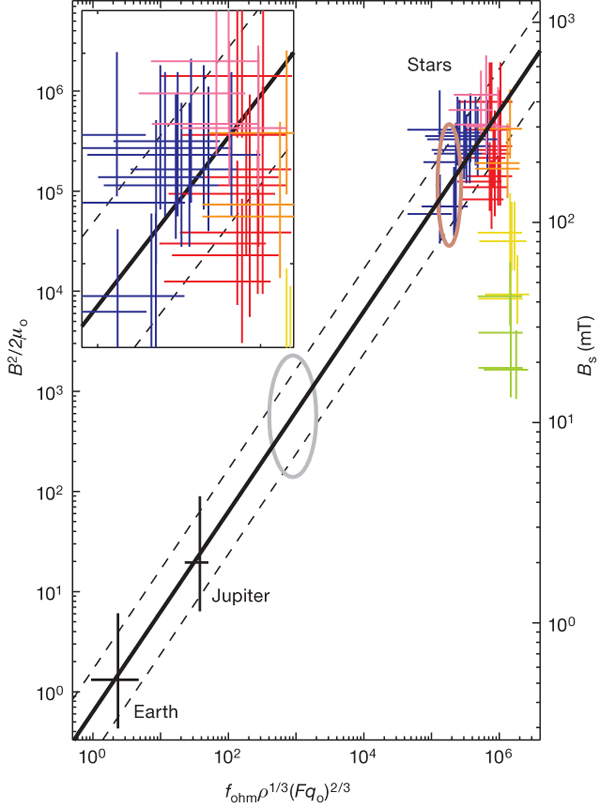}
%\plotone{Figures/Fujii_2015_Figure7.png}
\caption{The comparison between Earth, Jupiter, and stars shows that the magnetic energy density (in the dynamo) strongly correlates with a function of density and bolometric flux (here both in units of J m$^{-3}$). The bar lengths show estimated uncertainty rather than formal error. The stellar field is enlarged in the inset. Brown and grey ellipses indicate predicted locations of a brown dwarf with 1,500 K surface temperature and an extrasolar planet with seven Jupiter masses, respectively. From \citet[][]{2009Natur.457..167C}. \label{Fig:Christensen2009}}
\end{center}
\end{figure}

In addition, the obliquity of habitable planets also has a major impact on the seasonal and diurnal temperature variations and on their climate in general. Obliquity is much more difficult to determine than the rotational rate. However, simulated observations demonstrate that it is possible to determine this quantity from high signal-to-noise reflected light lightcurves obtained at multiple orbital phases.
	
Considerable effort was put into exploring time-resolved observations of Earth, as exoplanet analog. Researchers have used simulated disk-integrated brightness variations of Earth to demonstrate that its rotational period can be estimated, even in the presence of time-varying clouds \citep[][]{2008ApJ...676.1319P,2009ApJ...700.1428O}.  Likewise, such observations spanning multiple orbital phases constrain obliquity (\citealt[][]{ 2010ApJ...720.1333K,2011ApJ...739L..62K,2012ApJ...755..101F,2016MNRAS.457..926S,2016ApJ...822..112K}).  \citet[][]{2016MNRAS.457..926S} showed that although both latitudinal and longitudinal heterogeneities contribute to the obliquity signal, the latter contains more information  (see Figure \ref{Fig:Schwartz2016}).  In principle, the amplitude modulation of rotational variations at only three orbital phases uniquely identifies a planet's obliquity vector (the obliquity and its orientation with respect to the observer's line of sight).  Taking the complementary {\em frequency modulation} approach, \citet[][]{2016ApJ...822..112K} showed that modest signal-to-noise observations spanning most of a planet's orbit could also constrain a planet's obliquity, even if one is agnostic of the planet's albedo map. 
%A comprehensive study by \citet[][]{2016MNRAS.457..926S} demonstrated that planetary obliquity can be constrained from observations at just a few orbital phase angles.

The precision with which the rotational period of an Earth analog can be estimated depends on the wavelengths used and on the temporal baseline over which the data are collected. \citet[][]{2008ApJ...676.1319P} explored this dependence using globally integrated photometric lightcurves for Earth and demonstrates the challenge in establishing accurate rotational periods (see~Figure \ref{Fig:Palle}).

\begin{figure}[ht]
\begin{center}
\includegraphics[width=0.5 \textwidth]{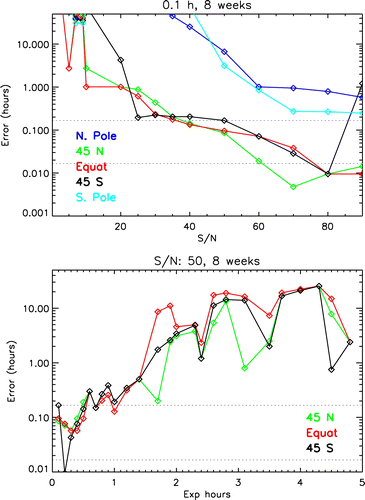}
%\plotone{Figures/Fujii_2015_Figure7.png}
\caption{ Top panel: Error in estimating Earth's rotation rate from the globally integrated photometric light curve. Each point is the error of the averaged rotational period found for 21 yr with different (real) cloud patterns for the same geometries. The five different colors indicate five different viewing angles (i.e., equator means the observer is looking at the Sun-Earth system from the ecliptic plane, the North Pole indicates the observer is looking at the Sun-Earth system from 90$^\circ$ above the ecliptic). All calculations are given for a 90$^\circ$ phase angle in the orbit (i.e., one would see a quarter of the Earth's surface illuminated). In the plot, the top dashed line represents an accuracy in determining the rotational period of 10 minutes and the lower one of 1 minute. Bottom panel: Same as in the top panel, but this time the S/N is fixed and the exposure time is allowed to vary. As in the top panel, an object follow up of 2 months (8 weeks) is considered. From \citet[][]{2008ApJ...676.1319P}. \label{Fig:Palle}}
\end{center}
\end{figure}

\begin{figure}[ht]
\begin{center}
\includegraphics[width=0.9 \textwidth]{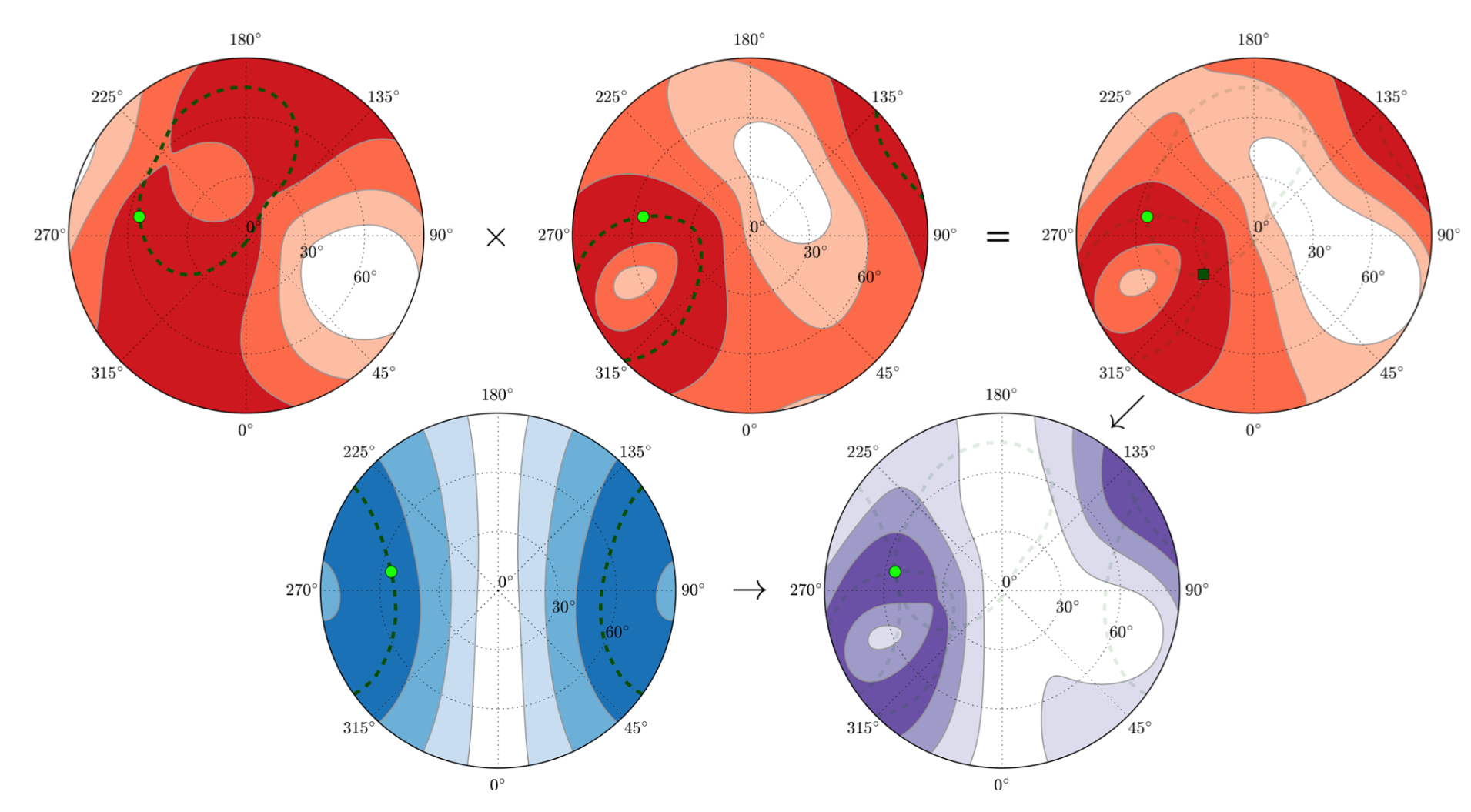}
%\plotone{Figures/Fujii_2015_Figure7.png}
\caption{Predicted confidence regions for planet's spin axis, from hypothetical single-and dual-epoch observations.  Observing a planet at just a few orbital phases can significantly constrain both its obliquity and axial orientation. Obliquity is plotted radially: the centre is   = 0$^\circ$ and the edge is   = 90$^\circ$. The azimuthal angle represents the planetÕs solstice phase. The green circles are the true planet spin axis, while the dark dashed lines and square show idealized constraints assuming perfect knowledge of the orbital geometry and kernel (i.e. no uncertainties). The upper left-hand and centre panels describe planet Q at phase angles 120$^\circ$ and 240$^\circ$, respectively, while the lower left-hand panel incorporates both phases. For the colored regions, 10$^\circ$ uncertainty is assumed on each kernel width, inclination, and orbital phase, while 20$^\circ$ uncertainty is assumed on the change in dominant colatitude. Regions up to 3$\sigma$ are shown, where darker bands are more likely.  From \citet[][]{2016MNRAS.457..926S}. \label{Fig:Schwartz2016}}
\end{center}
\end{figure}
%=================
A special case of rocky planets are those with very thin or no atmosphere (analogous to a "super-Mars" or a "dry Earth", an Earth-like planet that formed essentially dry or lost its atmosphere and water). Such planets may form as a result of extensive atmospheric loss due to evaporation (Hot super-Mars), stellar wind stripping, or impact stripping \citep[e.g.,][]{2015Icar..247...81S}.  Even if the planet is otherwise Earth-sized and it is inside the habitable zone it may be inhabitable if the atmospheric pressure is too low: Water is readily lost if its mixing ratio at the surface exceeds $\sim$20\% by volume, thus a background pressure limit for water stability is about 100 mbar \citep[][]{2014ApJ...785L..20W}.
The ability to measure rotational periods for these planets may provide important insights into the mechanism that led to the complete atmospheric loss. 
Atmosphereless planets are suitable for direct measurements of their rotational periods, because various types of rocky surfaces (i.e., mineral assemblages) have deep and wide albedo features that will introduce photometric rotational modulations in the visible and near-infrared \citep[][]{2012ApJ...752....7H}.

 \citet[][]{2014AsBio..14..753F} used albedo-map generated lightcurves and, where available, observed photometric variations to explore the geological features detectable on diverse Solar System bodies with minor or no atmospheres (Moon, Mercury, the Galilean moons, and Mars).  The study included the evaluation of the light curves and the features that are detectable at wavelengths ranging from UV through visible to near-infrared wavelengths, and also explored the accuracy required to determine the rotational periods of these bodies. Figure~\ref{Fig:Fujii2014} provides an example for the wavelength-dependence of the rotational variability amplitudes in different bodies.

{\bf Gas and Ice Giant Exoplanets:} The rotational periods of gas/ice giants may also be useful for constraining their formation and evolution \citep[][]{1991Icar...89...85T} and important for understanding their atmospheric circulation. Non-axisymmetrically distributed condensate clouds and hazes (photochemical or other origin) will introduce rotational modulations, both in reflected and in thermal emission \citep[e.g.,][]{2016ApJ...817..162S}. In addition, polarimetric modulations introduced by light scattering on heterogeneously distributed dust/haze grains may also be detectable.  Currently, rotational rate estimates exist for close-in exoplanets (assumed to be equal to their orbital periods) and a few measurements exist for directly imaged exoplanets and planetary-mass brown dwarfs. The rotational angular momenta of close-in exoplanets (i.e., synchronously rotating) is reset by tidal interactions and no longer carries information on the intrinsic angular momenta of the objects. In contrast, the angular momenta of non-synchronously rotating exoplanets (such as those probed via direct imaging) carry information about their formation and angular momentum evolution. Photometric modulations have been measured in two near-infrared filters for the $\sim$4--6~M$_{Jup}$ exoplanet 2M1207b \citep[][]{2016ApJ...818..176Z} and led to a rotational period measurement of $10.7_{-0.6}^{+1.2}$ h. CO absorption line rotational broadening measurements for the 10--13 M$_{Jup}$ planet $\beta$~Pictoris b suggests a $v \sin{i}=15$~km/s, which -- assuming an equatorial viewing geometry, age, and mass -- suggests a very similar rotational period. Similarly to these young exoplanets, photometric variations were used to measure the rotational periods of unbound young planetary mass-objects \citep[][]{2015ApJ...813L..23B,2016ApJ...830..141L} and very low-mass brown dwarfs \citep[e.g.,][]{2015ApJ...809L..29S}. The picture emerging -- based on the very limited data -- suggests that super-jupiter exoplanets and low-mass brown dwarfs start with similar angular momenta and during their evolution (cooling and contraction) their rotation rate increases, converging to the extrapolation of the Solar System mass-period relationship (see Figure~\ref{Fig:Zhou2016}).
 
A direct imaging mission capable of obtaining moderately high signal-to-noise ratio photometry of giant exoplanets can study possible trends between planet mass, semi-major axis, and rotational period.
 
{\em Obliquity for gas giants:} For gas giants (with well-constrained radii) combining the rotational period determined from rotational modulations with radial velocity information (line broadening due to rotation) allows constraining or deriving the rotation and inclination of the planet (e.g., \citealt[][]{2016ApJ...819..133A}). Finally, the Fourier spectrum or polarimetry of thermal emission (\citealt[][]{2011ApJ...741...59D,2013MNRAS.434.2465C}) as well as the amplitude and frequency modulation of reflected light rotational variations can provide an obliquity estimate \citep[][]{2016MNRAS.457..926S, 2016ApJ...822..112K}. 

{\bf A Note on Hazy Atmospheres:} Planets with thick haze layers may pose a challenge for rotational signals using methods c and d (line width measurements and temporal photometric/spectroscopic variations) depending on the wavelengths of observations and the origins of molecular absorption or cloud features studied). Because haze particles {\em by definition } are small ($\sim$0.01--1 $\mu$m)  and are not modulated by large-scale condensation-evaporation patterns associated with vertical motions the way clouds are, they sediment more slowly and their residence time in the atmosphere will be much longer than the rotational period ($t_{res} >> P$). This may result in featureless haze layers (e.g., Venus), unless other absorbing constituents that are sensitive to the atmospheric circulation are present.  As haze particles can be generated at higher altitudes than larger particles produced by condensation, the featureless haze layers {\em if optically thick } will mask any heterogenous condensate cloud structure as well as any surface structures. Similarly, optically thick haze layers may cover or weaken the rotationally broadened line profiles in the atmospheres, also limiting the use of Doppler techniques. Therefore, planets enshrouded in thick haze layers may often not be well suited for rotational studies.

The two hazy planets in our solar system are useful cases in point. Venus is shrouded in a $\sim1\mu$m sulfuric acid haze but with dark ultraviolet features due to an unknown absorber that revealed a $\sim$4-day rotational period in ground-based observations \citep[][]{1961AnAp...24..531B, 1975JAtS...32.1045T}.  This was later shown to be due to the atmosphere's superrotation rather than the slow 243 day rotation period of its surface \citep[]{1990JAtS...47.2053R}. Titan is covered by a stratospheric hydrocarbon haze that is featureless except for a seasonally varying hemispheric albedo asymmetry \citep[e.g.,][]{2009DPS....41.0706L}.  The haze obscures the view of tropospheric methane clouds and the surface, but these can be detected in near-infrared imagery \citep[][]{2011GeoRL..38.3203T,2011Icar..216...89R}.
	
\subsubsection{The Science Value of Independently Measured Planet Masses and Radii} 	
{\em Planetary Radius:} For methods that measure rotational velocity rather than period, knowledge of planetary radius and obliquity are required to convert rotational broadening into rotational period.  However, if the goal is to determine the Coriolis forces then rotational broadening is sufficient. For the photometric methods that produce a period estimate, on the other hand, the frequency of diurnal forcing is easily derived, while estimating the Coriolis forces again requires the planetary period.  In general, rotational information is most useful when combined with radius estimates, but some science results can be derived directly from rotational period measurements without complementary observations. Furthermore, observations constraining the planetary orbits may be combined with the obliquity and rotational period to constrain the formation history of low-mass planets. 

{\em Giant Planets:} The radii of mature giant planets is determined by the combination of electron degeneracy pressure (at the highest pressures, $R\propto M^{-1/3}$) and by classic Coulomb forces acting on ions ($R\propto M^{1/3}$) (e.g., \citealt[][]{2010SSRv..152..423F}). Equilibrium models by predict radii variations between $\sim$0.6 R$_{Jup}$(weakly-irradiated giant planets with a mass of 0.15 M$_{Jup}$ and 4.5~Gyr age) to 1.06 R$_{Jup}$ (mass of 11.3 M$_{Jup}$) \citep[][]{2007ApJ...659.1661F}. Masses may be derived from spectral retrieval that includes a fit for surface gravity. 

Radii and masses of smaller planets vary {\em more} than those of giant planets: mass may vary by a factor of $\sim$20 (from Mars to super-Earths): while rotational periods alone will be important and useful for atmospheric circulation models, mass and/or radius measurements would yield important additional science: mass measurements would allow exploring trends between formation mechanisms and angular momentum; and radius estimates (even from mass-radius relationships) would allow calculating Coriolis forces from rotational periods, significantly constraining the atmospheric circulation models.

{\em Planet mass measurements} from radial velocity or astrometry, or gravitational interactions between the planets, can be combined with rotational periods to determine the angular momenta of the giant planets, which may be useful for constraining their accretion history. 

The periodicity in photometric variations is a direct measure of the rotational period, i.e., rotational period measurements do not require mass measurements. However, verifying the predicted trend between angular momentum, orbital period, mass (which potentially constrains the formation history) requires mass and radius measurements.

\begin{figure}[ht]
\begin{center}
\includegraphics[width=0.9 \textwidth]{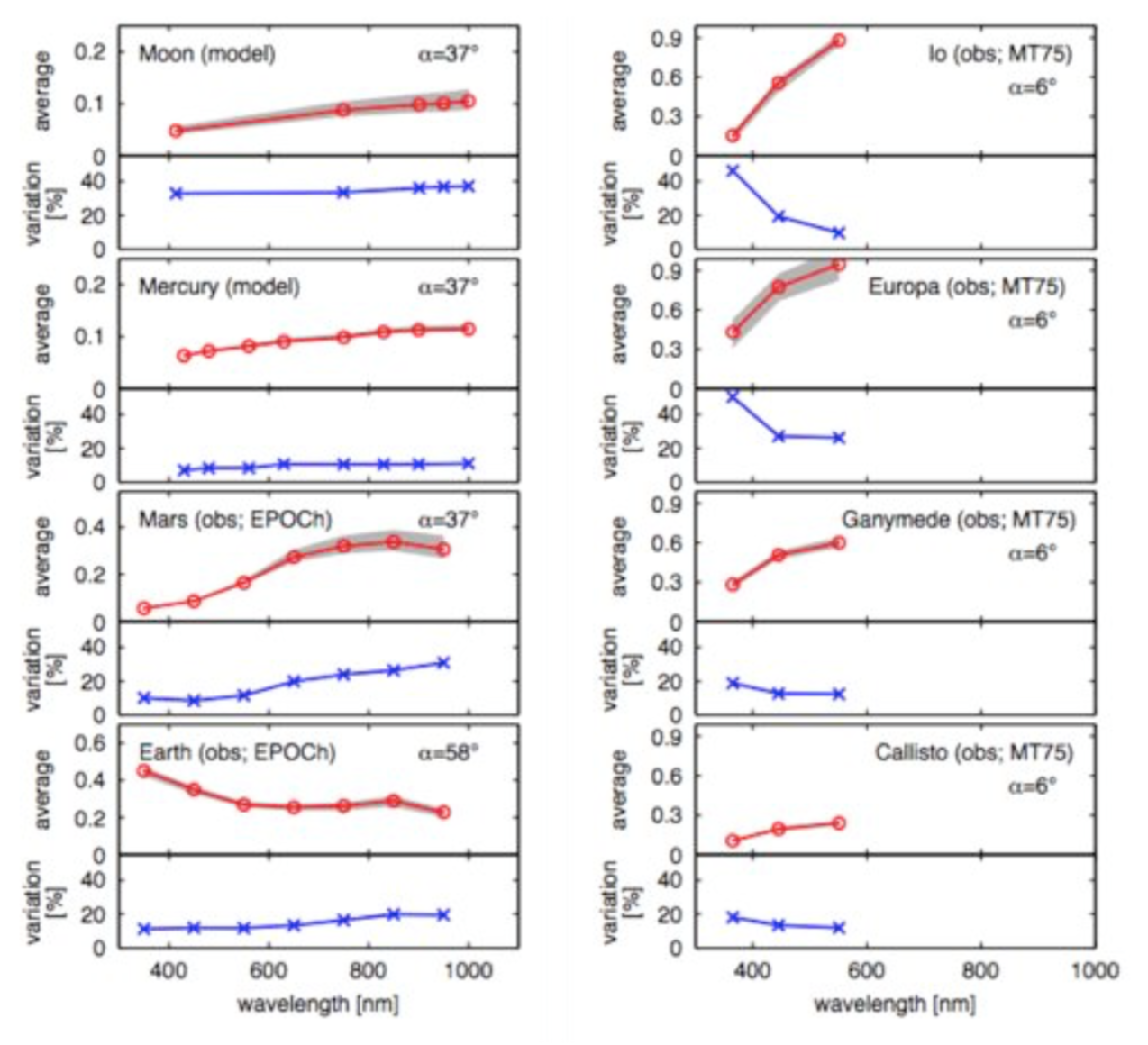}
%\plotone{Figures/Fujii_2015_Figure7.png}
\caption{Albedo and its variations as a function of wavelengths for Solar System bodies with minor or  no atmosphere. From Fujii et al. (2014). \label{Fig:Fujii2014}}
\end{center}
\end{figure}

\smallskip 

\begin{mdframed}[backgroundcolor=blue!20] 
{\bf Observational Requirements} \\
{\bf Observations for Rotational Periods:} Very high spectral resolution for rotational broadening studies {\em or} multi-epoch photometry for disk-integrated lightcurve analysis; observations with a few rotational periods are required to determine rotational period in the presence of rapidly changing cloud cover. If the cloud cover is changing on timescales of the rotational period, longer continuous monitoring will be required to characterize the power spectrum of the objects. \\
{\bf Observations for Obliquity:} photometry over at least one complete rotational phase at {\em multiple} distinct orbital phases (at minimum three phases) required to constrain the obliquity.\\
{\bf Wavelength range:} The rotational modulations and line broadening are present over a broad wavelength range (from optical to thermal infrared). The observations should primarily focus on the wavelengths where the highest signal-to-noise measurements can be reached for a given planet over a given time. Broad-band ("white light") photometric variations can be observed simultaneously with, for example, long-integration time spectroscopic observations of the target planet and can also be collected simultaneously on multiple planets in the same field of view, even without integral field spectroscopic capabilities.
{\bf Supplementary Observations:} Mass measurements (from stellar radial velocity and/or stellar astrometry) will enhance the science value of the rotational measurements, but not required for the determination of the rotational period.  
\end{mdframed}

%\begin{mdframed}[backgroundcolor=yellow!20] 
%{\bf Pending Comments/Questions: } \\
%To what accuracy should rotation periods be determined? This is ill-defined unless one specifies the expended rotation period a priori.  The max dwell time sets an upper limit on the period one can be sensitive to (would like to see $>$1.5 full rotations), while the exposure time sets a lower limit on the rotation period (need $>$5 exposures per rotation).
%Connection between planet formation/evolution and angular momentum (Tremaine 1991; Dones \& Tremaine 1993; Kokubo \& Ida 2007; Miguel \& Brunini 2010; Schlichting \& Sari 2007)?\\
%Kasper: ELT HCI + HRS will allow us to study rotation periods, sample will be small - not sure about statistical relevance
%\end{mdframed}

\begin{table}[htp]
\caption{Expected rotational modulation amplitudes and constraints on rotational period and obliquity for terrestrial and giant exoplanets.}
\begin{center}
\begin{tabular}{p{3.5cm} c c | c c | c }
\cellcolor{blue!25}{\bf Planet Type} & \cellcolor{blue!25}{\bf at Optimal $\lambda$} & \cellcolor{blue!25} Amplitude &  \cellcolor{blue!25} at Acceptable $\lambda$ & \cellcolor{blue!25} Amplitude & \cellcolor{blue!25} Baseline\\
\cellcolor{blue!25} {\em Rotational Period } &\cellcolor{blue!25} & {Opt. $\lambda$ }\cellcolor{blue!25} &\cellcolor{blue!25} &\cellcolor{blue!25} {Accept. $\lambda$ }&\cellcolor{blue!25}\\
\hline
Terrestrial & 0.9 $\mu$m & 25\% & 0.5-10 $\mu$m & 10--35\% & P=3$-$30~h \\
Ice/Gas Giant & 5 $\mu$m & 15\% & 0.3-5.0 $\mu$m & 3\% & P=3$-$20 h \\
\hline
\cellcolor{blue!25}{\em Obliquity } &\cellcolor{blue!25}&\cellcolor{blue!25} &\cellcolor{blue!25} &\cellcolor{blue!25} &\cellcolor{blue!25} \\
Terrestrial &  0.9 $\mu$m& 25\% & 0.5-10 & 10--35\% & 3$\times$P \\
Ice/Gas Giant & 5 $\mu$m& 15\% & 0.3-5.0 $\mu$m & 3\% & 3$\times$P\\ 
\hline
\end{tabular}
\end{center}
\label{T:SciQuestions}
\end{table}

%\tablerefs{
%(7) Fujii 2015
%(8) Gelino \& Marley 2001}

%\clearpage

%% Tables may also be prepared as separate files. See the accompanying
%% sample file table.tex for an example of an external table file.
%% To include an external file in your main document, use the \input
%% command. Uncomment the line below to include table.tex in this
%% sample file. (Note that you will need to comment out the \documentclass,
%% \begin{document}, and \end{document} commands from table.tex if you want
%% to include it in this document.)

%% \input{table}

\clearpage

\subsection{B2: Which rocky planets have liquid water on their surfaces? Which planets have continents and oceans?}
\label{B2LiquidWater}

{\bf Relevance:} Water is not a biosignature itself, but the presence of liquid water is required for life as we know it. Liquid water is not the only factor required for a planet to sustain life, but it is arguably the most important one. Thus, liquid water is a {\em habitability signature}. Establishing which habitable zone planets have liquid water on their surfaces provides an important context for EXOPAG SAG16, which focuses on biosignatures, but will rely on SAG15 for habitability signatures and characterization of habitable planets. 

Our understanding of the distribution of water is surprisingly limited even for the case of Earth, and very incomplete for exo-earths: Currently, water detections (direct and indirect) in extrasolar systems are limited to protoplanetary disks (e.g., \citealt[][]{2008Sci...319.1504C,2008ApJ...676L..49S}), the atmospheres of hot jupiters and hot neptunes (e.g., \citealt[][]{2014Natur.513..526F}), and in disks around white dwarfs fed by tidally disrupted minor bodies (\citealt[][]{2016NewAR..71....9F}); however, no direct or indirect observations exist of water in extrasolar habitable zone Earth-like planets or even in super-Earths.

Simulations of exo-earth observations have been used to demonstrate that rotational phase mapping (time-resolved observations of hemisphere-integrated reflected light from the planet) can reveal the types and distribution of surfaces. Equipped with additional data on the color/spectra of the features and the physical conditions on the planetary surface may be used to identify surface features such as oceans and continents. 

In the following we will discuss two different pathways for identifying liquid water on Earth-like habitable zone planets: 1) via the detection of oceans; and 2) via the detection of water clouds.

\subsubsection{Detecting Oceans}
\label{DetectingOceans}
The traditional habitable zone (HZ) is defined in terms of surface liquid water \citep[][]{1993Icar..101..108K}. Three distinct methods have been proposed to search for large bodies of liquids (oceans) on the surface of a planet:\\

%	{\bf Rotational color variability.} Oceans are darker and have different colors than other surface types on Earth, so the time variations in the color of a spatially unresolved planet can reveal the presence of liquid water oceans. This method relies on the existence of longitudinal heterogeneities in the planet's surface composition (e.g., \citet[][]{2001Natur.412..885F,2009ApJ...700..915C,2010ApJ...715..866F,2011ApJ...739L..62K,2011ApJ...739L..62K}). 
	
	{\bf Polarization.} For planets with low average ocean wind speeds ($\lesssim$ 2m/s) oceans are smoother than other surface types (typically solids) and thus polarize light to a high degree \citep[e.g.][]{2008Icar..195..927W,2010ApJ...723.1168Z,2011ApJ...739...12Z}. For idealized scenarios, the phase variations in polarization are significant, but in practice the effect of oceans is masked by Rayleigh scattering, clouds, and aerosols. For planets with Earth--like average ocean wind speeds ($\approx$ 10 m/s) the ocean surfaces (with the exception of the glint surface) will depolarize the reflected light (due to wind-induced ripples on the oceanic surface). Observations of polarized Earthshine, however, imply that rotational variations in polarized intensity may still be useful in detecting oceans \citep[][]{2012Natur.483...64S}.

	{\bf Specular reflection} The same smoothness that leads to polarization dictates that cceans are also able to specularly reflect light, especially at crescent phases \citep[][]{2008Icar..195..927W}. The signal-to-noise requirements for phase variations are not as stringent as for rotational variations since the integration times can be much longer: weeks instead of hours. However, \citet[][]{2010ApJ...721L..67R} showed that clouds not only mask underlying surfaces, but forward scattering by clouds mimics the glint signal at crescent phases, while atmospheric absorption and Rayleigh scattering mask the glint signature. They proposed using near-infrared opacity windows to search for glint, but this would only be possible if the effects of clouds could be accurately modeled for exoplanets. Moreover, \citet[][]{2012ApJ...752L...3C} showed that crescent phases probe the least-illuminated and hence coldest regions of a planet regardless of obliquity. Insofar as these planets have ice and snow in their coldest latitudes, then this latitude--albedo effect acts as false positive for ocean glint.
	
	{\bf Rotational Color Variability:} Although the faces of extrasolar planets will not be spatially resolved in the foreseeable future, their rotational and orbital motions produce detectable changes in color and brightness.  \citet[][]{2001Natur.412..885F} used simulations of Earth to show that the changing colors of its disk-integrated reflected light encode information about continents, oceans, and clouds.  The inverse problem --- inferring the surface geography of a planet based on time-resolved photometry --- is much more daunting than the forward problem.
	
	Much progress has been made on the {\em exo-cartography} inverse problem since the seminal work of \citet[][]{2001Natur.412..885F}.  The rotational color variations of a planet can be used to infer the number, reflectance spectra, surface area, and longitudinal locations of major surface types (\citealt[][]{2010ApJ...715..866F,2011ApJ...738..184F,2009ApJ...700..915C,2011ApJ...731...76C, 2013ApJ...765L..17C}). Meanwhile, the rotational and orbital color variations of an unresolved planet can be analyzed to create a 2-dimensional multi-color map equivalently a 2D map of known surfaces (\citealt[][]{2010ApJ...715..866F,2011ApJ...739L..62K,2010ApJ...720.1333K,2012ApJ...755..101F}). 
	
\subsubsection{Liquid Water Clouds}
Additional methods may be used to deduce the probable presence of liquid water on the surface of a potentially habitable planet without directly or indirectly detecting an ocean. 
The presence of liquid water on the surface of an exoplanet can be indirectly inferred by the presence of liquid water clouds in the exoplanetary atmosphere. 
With the help of spectroscopy astronomers have detected signs of water vapor on a number of giant exoplanets and brown dwarfs  and even and even water ice clouds on a brown dwarf \citep[e.g., ][]{2016ApJ...826L..17S,2016ApJ...823..109I,2014A&A...565A.124B,2014Natur.513..526F}.

%{\bf Test whether partial pressure of water vapor reaches saturation.}  
%Spectroscopic observations of exoplanets can be used to detect the existence of water vapor on exoplanetary atmospheres \citep[e.g., ][]{2016ApJ...823..109I}. 
%Fitting theoretical models to high resolution spectroscopic observations can help us deduce the temperature--pressure (and atmospheric composition) profile of an observed exoplanetary atmosphere \citep[e.g.,][]{2013cctp.book..367M}. 
%The retrieved temperature--pressure profile can then be used in combination with water condensation curves, to test whether the water vapor pressure can reach saturation in the exoplanetary atmosphere. 
% Additionally, if clouds exist in the exoplanetary atmosphere, the use of spectral and/or spectropolarimetric observations can help us deduce the cloud  top pressure \citep[e.g.,][]{2012ApJ...753..100B, 2014Natur.513..526F}. This 
%pressure can then be compared with the best--fit temperature--pressure profiles to control whether the detected clouds are in the part of the atmosphere where liquid water clouds can exist. 

{\bf Identifying clouds made of liquid water droplets (and not water ice) using polarization.}  On Earth, both liquid water and water ice clouds exist because liquid water is present on the surface. On an exoplanet, though, detection of water ice clouds could only be reliably be interpreted as a signature of surface liquid water if the surface temperature were independently known to be above freezing. Liquid water cloud detection is less likely to be a false positive for surface liquid water, although it could be in the presence of near-surface temperature inversions as may occur near the terminators of synchronously rotating planets.The detection of liquid water clouds can also be achieved with the help of broadband polarimetry. 
The state of polarization of starlight reflected by a planet is highly sensitive to the composition and structure of the planetary atmosphere. Observations of planets of our Solar system show that polarization is a powerful tool in the characterization of the micro- and macro- physical properties of clouds in planetary atmospheres \citep[e.g., ][]{1974SSRv...16..527H, 2010arXiv1010.1171M}. Simulations of the polarization signal of terrestrial and gaseous exoplanets indicate that polarization can also be a powerful tool for the characterization of exoplanet atmospheres \citep[e.g., ][]{2000ApJ...540..504S,  2008A&A...482..989S,2011A&A...530A..69K}. An early example of the power of polarimetry in the characterization of clouds in an exoplanetary 
atmosphere, is the retrieval of the cloud top pressure, and composition and size distribution of cloud droplets in the upper Venusian atmosphere using ground-based, unresolved observations of Venus by \citet[][]{1974JAtS...31.1137H}.

The identification of the state of water clouds on Earth is routinely done with the help of polarization \citep[][]{1995SPIE.2311..171P,2000JGR...10514747G}. The (highly polarized) primary rainbow is a direct indication of the existence of liquid water clouds in a planetary atmosphere. \citet[][]{2007AsBio...7..320B} was the first to suggest the use of the primary rainbow to detect liquid water clouds on exoplanets. \citet[][]{2011A&A...530A..69K} and \citet[][]{2012A&A...548A..90K} presented numerical simulations of  broadband spectra of planets covered by a cloud deck and patchy liquid water clouds respectively, and showed that the rainbow is a robust tool for the detection of liquid water clouds in exoplanetary atmospheres. 

Ice water clouds can interfere with the detection of liquid water clouds in the Earth's atmosphere. Ice clouds can produce highly polarized halos (rainbows), that could mask the primary rainbow of the liquid water clouds and the existence of the liquid water clouds altogether. However, \citet[][]{2012A&A...548A..90K} showed that for a heterogeneous liquid and ice water cloud coverage like the Earth's the primary rainbow of liquid water clouds will still be detectable. Even for extreme cases where optically thick ice clouds cover $\sim$50\% of the water clouds of an exo--Earth the primary rainbow will be detectable. 

For an Earth-like planet orbiting at 1AU around a star at 10 pc the primary rainbow will appear between 30 to 44 milli-arcsec from the parent star (phase angle of $\sim 30^\circ$--$\sim 40^\circ$). To detect the primary rainbow we will need to observe the exoplanet with a spatial resolution of $\sim$2 milli-arcsec. In addition to the very high spatial resolution, very high contrast and sensitivity are also required for these measurements.
%{\bf  More discussion/references required on the feasibility of these proposed observations.}

{\bf Identifying water vapor clouds from time-resolved spectroscopy:}  In the case of Earth, patchy water cloud cover may be identified in the
disk-integrated spectra as a time variation of absorption features by atmospheric molecules. \citet[][]{2013ApJ...765...76F} identified diurnal time
variability of absorption bands of CO$_2$, O$_2$, and H$_2$O which correlates
with cloud cover. It is also found that the variation pattern of H$_2$O
looks different from that of O$_2$ and CO$_2$, and attributed this to the
non-uniform distribution of H$_2$O, which would imply short residence
time of H$_2$O in the atmosphere due to the rapid phase changes in the atmosphere through evaporation from the surface liquid ocean and the
cloud/precipitation processes. 

%These observations may be within the capabilities of the generation of extremely large telescopes.
%\textbf{Question: what is the contrast that ELTs will be able to achieve? Spatial resolution should not be a problem, but will they be able to detect a terrestrial planet at 1AU/10pc?}

\begin{mdframed}[backgroundcolor=blue!20] 
{\bf Observational Requirements:} \\
{\bf Sample size:} Planets can characterized individually, but larger samples of planets will be required to establish clear trends with system parameters.\\
{\bf Observations:} 1) Polarimetry at the appropriate orbital phases to identify specular reflection; {\em or} 2) Time-resolved (rotational) multi-band photometry to identify albedo variations that may indicate water; {\em or} 3) Time-resolved spectroscopy to probe variations in the shape of the water vapor absorption that indicates patchy clouds and, therefore, condensation.\\
b) {\em Presence of Greenhouses gases} and water vapor in the atmosphere: CO$_2$ and H$_2$O have strong features in the near-IR and constraining their abundance is important for correct climate models.
{\bf Supplementary observations for individual planets:} a) {\em Orbital semi-major axis} of a planet is important as it significantly impact the allowed surface temperature range and thus possibility for liquid surface water to be present. The semi-major axis can be constrained either through multiple imaging observations or by combining imaging with supplementary stellar radial velocity or astrometry (see Question A1 for further discussion). 

\end{mdframed}

%\begin{mdframed}[backgroundcolor=yellow!20] 
%{\bf Comments on B2): } \\
%\end{mdframed}

\clearpage

\subsection{B3. What are the origins and composition of condensate clouds and hazes in ice/gas giants and how do these vary with system parameters? }
\label{S:Clouds}
{\bf Contributors:} Daniel Apai, Anthony del Genio, Mark Marley

All Solar System planets with an atmosphere also harbor condensate cloud and/or haze layers. Clouds and hazes influence the pressure-temperature structure of the atmosphere, its emission and transmission spectra, as well as its albedo. Particles or droplets that make up clouds primarily form through condensation and grow via further condensation and/or particle collisions. With grain sizes that may range from a micron to $\sim$millimeter, cloud particles/droplets have short settling time and are typical below the tropopause, where the dynamics of an atmosphere is most likely to saturate volatile constituents in regions of rising motion. Based on different extrapolations of clouds observed on Earth and on other Solar System planets, a range of cloud models have been proposed for giant exoplanets and brown dwarfs (for a review and comparison, see \citealt[][]{2008MNRAS.391.1854H}).
Haze particles (typically $<0.1-1 \mu$m in size) often form via photochemistry-driven (e.g., Venus and Titan) or charged-particles-driven chemical reactions in the upper atmospheres ($<$1 bar); with long residence times these particles often introduce large optical depths to upper atmospheres.
From an observational perspective clouds and hazes may also used as tracers of atmospheric dynamics (circulation, mixing, turbulence).  Presence of haze or cloud layers may also mask the presence of specific atmospheric absorbers even if present at large abundances at pressures higher than the particle layer. 

{\bf Current Knowledge:} Condensate clouds have been observed in brown dwarfs and in hot jupiters, over a very broad range of temperatures and pressures. High-altitude haze layers have been observed for transiting planets ranging from hot jupiters to super-earths and possibly for earth-sized planets, as well as for brown dwarfs. In the following we briefly summarize the key aspects of condensate clouds and haze layers.

{\em Condensate Clouds:} As the atmospheres of exoplanets encompass a very broad temperature range ($\sim$50 to 2,000~K) these atmospheres are expected to harbor a large variety of condensates. For solar compositions the most important condensates include Ca-Ti-oxides, silicates, metallic iron, sulfides, CsCl and KCl, H$_2$O, NH$_4$HS, NH$_3$, CH$_4$ (e.g., \citealt[][]{2002Icar..155..393L}, for a recent review see \citealt[e.g.,][]{2015ARA&A..53..279M}). Most of our current knowledge on cloud properties and compositions come from studies of Solar System planets (most importantly, Earth and Jupiter) and from the abundant samples of brown dwarfs. Water vapor and water ice clouds in Earth can be studied in-situ and via remote sensing; models developed to explain their behavior and properties are often used as a starting point for models of extrasolar clouds \citep[][]{2001ApJ...556..872A}, although it is likely that in some exoplanet and brown dwarf atmospheres cloud formation and properties may be set by different processes (for a review of different cloud models see, e.g., \citealt[][]{2008MNRAS.391.1854H}). 

With over $\sim$3,000 brown dwarfs known these objects provide a easy-to-study analogs of extrasolar giant planet atmospheres. Temperatures of known brown dwarfs range from $\sim$250 K (below freezing point!) to above 2,300~K; an increasing number of known brown dwarfs have very low gravities and masses of only a few M$_{Jup}$, enabling the definition of samples essential for comparative parameter studies. 

Comparative studies of brown dwarfs reveal the presence of silicate cloud layers through prominent infrared color-magnitude changes that occur through the M--L--T--Y spectral type sequence. The sequence itself is primarily set by the presence and absence of prominent gas-phase absorbers and not directly by the presence/absence of clouds (e.g., \citealt[][]{2006ApJ...637.1067B,2006ApJ...648..614C,2012ApJ...753..156K}); however, there is a strong correlation between the spectral type and colors of a given object (e.g., \citealt[][]{2002ApJ...571L.151B,2006ApJ...640.1063B,2008ApJ...689.1327S, 2012ApJS..201...19D}). The general and oversimplified picture that emerged suggests that while the hottest (M-type) brown dwarfs are condensate cloud free, with temperatures below $\sim$1,800~K the atmospheres of L-type brown dwarfs are characterized by thick silicate clouds (resulting in red near-infrared colors between 1--3 $\mu$m); at even lower temperatures (T$<$1,300~K) a transition to silicate cloud-free atmospheres is envisioned. Correspondingly, cool T-type brown dwarfs have blue near-infrared colors (dominated by scattering by gas molecules rather than particles and differential methane opacities in the J, H, and K bands), consistent with the lack of thick clouds in their upper atmospheres (see Figure~\ref{Fig:WagnerCMD}). At even lower temperatures, within the Y spectral type, less refractory and less abundant species, including water ice, are expected to condense out and form clouds \citep[e.g.,][]{2014ApJ...787...78M}. 

\begin{figure}[h]
\includegraphics[width=\textwidth]{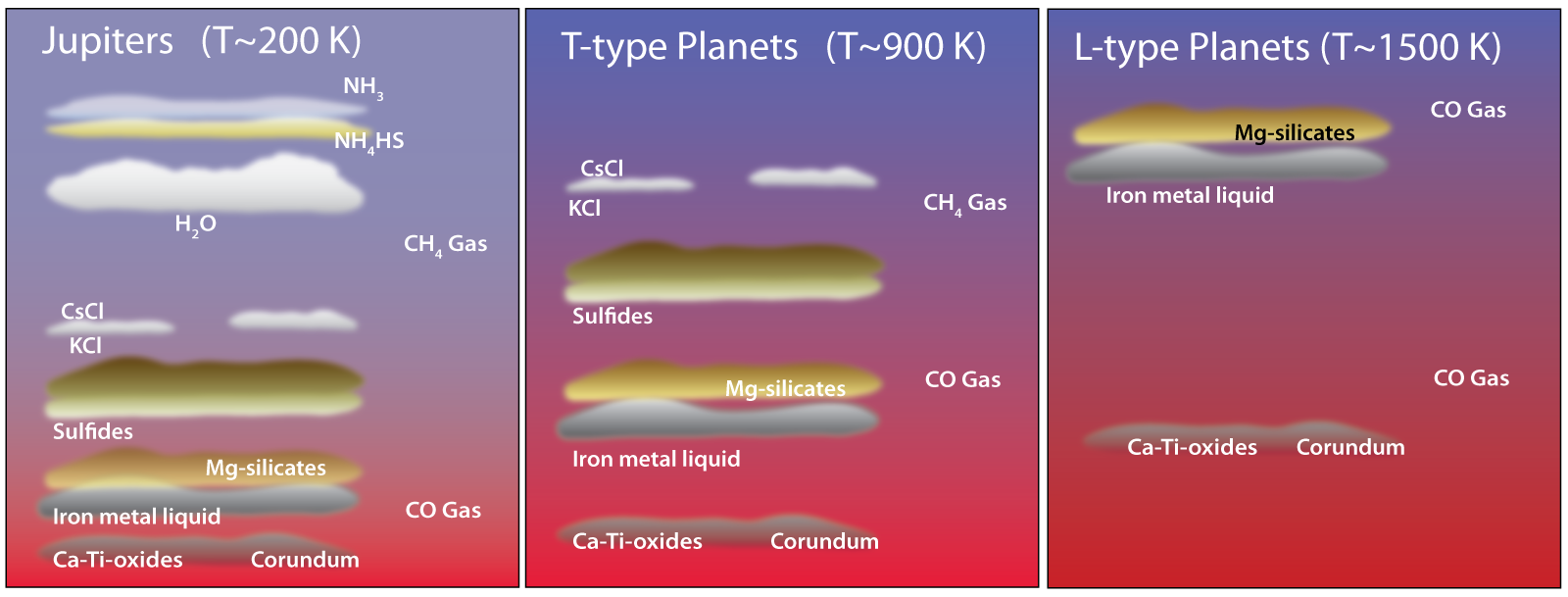}
\caption{Condensate clouds predicted for the upper atmospheres of giant planets of different temperature. By D. Apai, after Lodders (2003). \label{Fig:ApaiClouds}}
\end{figure}

\begin{figure}[h]
\begin{center}
\includegraphics[width=0.8 \textwidth]{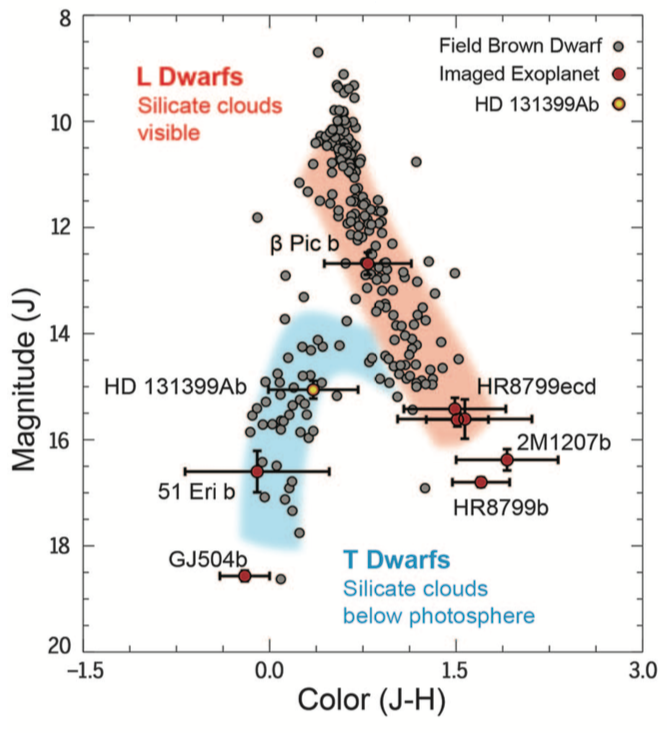}
\caption{Condensate clouds have a fundamental impact on the positions of brown dwarfs and directly imaged exoplanets on the near-infrared color-magnitude diagram. Along the L-type sequence (red) silicate clouds in the upper atmosphere become thicker. The cooler T-dwarfs are bluer because the silicate clouds are below the visible upper atmosphere. Figure from \citet[][]{2016Sci...353..673W}, which is in part based on the parallax database by \citet[][]{2012ApJS..201...19D}. \label{Fig:WagnerCMD}}
\end{center}
\end{figure}

\begin{figure}[ht]
\begin{center}
\includegraphics[width=0.8 \textwidth]{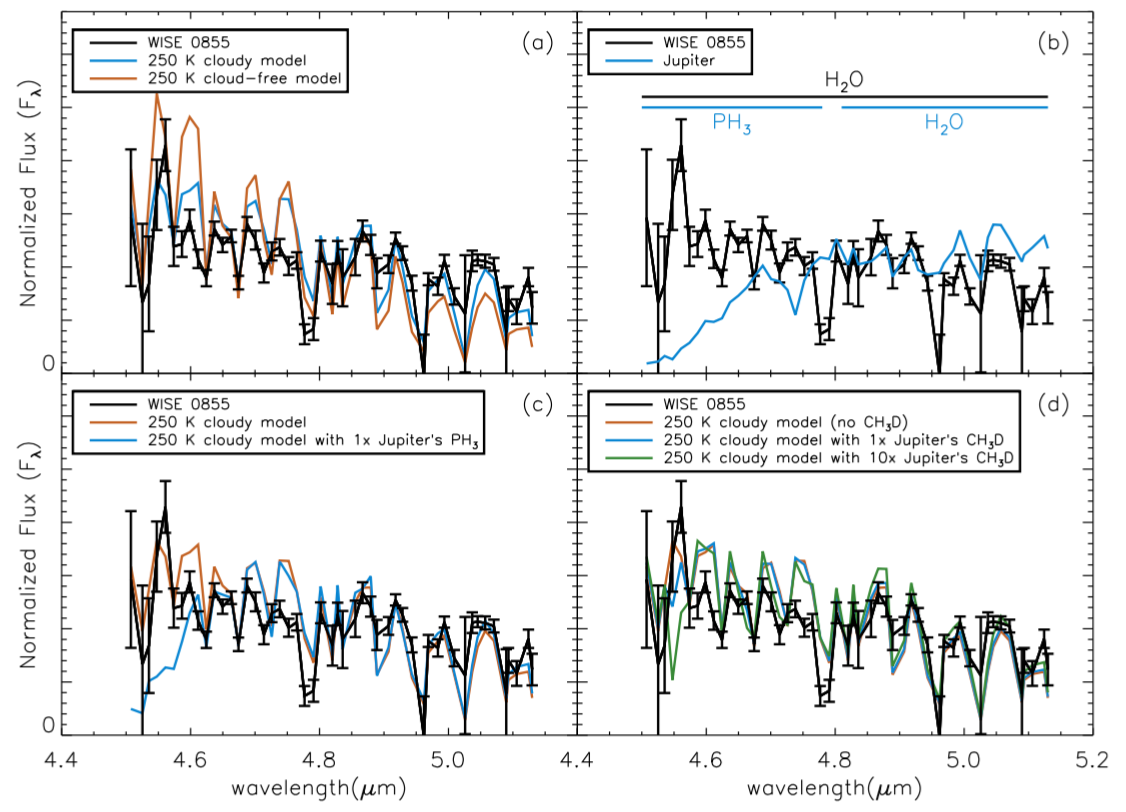}
%\plotone{Figures/Apai_Clouds_LT_Jupiter.png}
\caption{Gemini/GNIRS spectrum of the $\sim250$~K Y-dwarfs WISE0855 shows a series of absorption features attributable to water vapor, muted by clouds (likely water ice). From \citet[][]{2016ApJ...826L..17S}. \label{Fig:SkemerWaterIceClouds}}
\end{center}
\end{figure}

Although there is ample evidence supporting the overall picture described above, it is also clear that the above picture fails to capture the real complexity of cloud properties and atmospheric chemistry in brown dwarfs. Outstanding questions include the large dispersion in color along the L--T sequence (e.g., \citealt[][]{2008ApJ...674..451B}); the unusually red colors of many of the very young brown dwarfs (and a few intermediate-age ones), a likely sign of unusually dusty upper atmospheres (e.g., \citealt[][]{2013ApJ...772...79A,2013ApJ...777L..20L}. Furthermore,  the first detections of water ice clouds has been reported in a Y-dwarf with an effective temperature of only $\sim250$~K  \citep[][]{2014ApJ...793L..16F,2016ApJ...826L..17S}, enlarging the temperature range over which cloud models can be tested.

Recently, time-resolved high-precision observations (photometric and spectroscopic light curves) enabled the comparative studies of different cloud layers within the same objects, breaking the degeneracy between the effects of the multiple atmospheric parameters that may vary between any two brown dwarf (age, composition, temperature, surface gravity, vertical mixing, cloud structure). Space-based (HST and Spitzer) studies with sub-percent photometric precision found that most, if not all, brown dwarfs have heterogeneous (patchy) cloud cover \citep[][]{2014ApJ...782...77B,2015ApJ...799..154M}; ground-based surveys found that the highest amplitude brown dwarfs are at the L/T transition \citep[][]{2014ApJ...793...75R}. Time-resolved spectroscopy of L/T transition dwarfs showed that the spectroscopic variations emerge from the atmospheres characterized by a mixture of warm thin cloud / cooler thick cloud patches, and {\em not} by clouds and cloud holes \citep[][]{2013ApJ...768..121A,2015ApJ...798..127B}. Simultaneous HST (1.1-1.7 $\mu$m) and HST--Spitzer (1.1-1.7 $\mu$m and [3.6] or [4.5]) observations of clouds in L, L/T, and T-type brown dwarfs revealed pressure-dependent (vertical) structures with characteristic patterns for objects of different spectral types \citep[][]{2012ApJ...760L..31B,2013ApJ...768..121A,2015ApJ...798L..13Y}. Most recently, planetary-mass brown dwarfs and companions have also been accessible to rotational modulation studies, providing an opportunity to explore cloud properties as a function of gravity \citep[][]{2015ApJ...813L..23B,2016ApJ...818..176Z,2016ApJ...830..141L}. 

Directly imaged exoplanets and planetary-mass companions cover a spectral type range from early L to mid-T. These objects differ from old, high-gravity brown dwarfs both in the fine structure of their spectra \citep[][]{2011ApJ...733...65B,2011ApJ...735L..39B,2012ApJ...753...14S} and, often, in their broad-band colors (see, e.g., Fig. \ref{Fig:WagnerCMD}), but show some strong similarities to some young brown dwarfs \citep[e.g.][]{2013ApJ...772...79A,2013AJ....145....2F,2016ApJS..225...10F}. From the small sample of directly imaged exoplanets it appears that early L-type exoplanets have colors similar to brown dwarfs with matching spectral types \citep[][]{}, late L and L/T-type exoplanets are often much redder and fainter than brown dwarfs with matching spectral types (e.g., \citealt[][]{2005A&A...438L..25C,2008Sci...322.1348M,2016A&A...587A..58B}), but the coolest T-type exoplanets appear to have colors consistent with those of T-type brown dwarfs \citep[][]{2015Sci...350...64M,2016Sci...353..673W}. This pattern, if verified, would argue for a difference in cloud properties (most significant in late-L and L/T transition objects) between the higher gravity brown dwarfs and the low-gravity exoplanets.

Clouds have also been studied in hot jupiters via transmission and emission spectroscopy, spectral phase mapping, and in reflected light. 
Observations from the Kepler space telescope (dominated by reflected light) argued for a large-amplitude, heterogeneous silicate cloud cover \citep[e.g.,][]{2013ApJ...776L..25D} that avoids the cold trap in the night side of the planet. \citep[][]{2008ApJ...678.1419F} proposed that the presence/absence of silicate clouds in hot jupiters should follow the general sequence observed in brown dwarfs. Although optical-near infrared HST transmission spectra argued for the presence of cloud decks in some hot jupiters (\citealt[e.g.,][]{2013MNRAS.436.2974G,2015MNRAS.446.2428S}), no clear trend (in terms of presence/absence of clouds) emerged from a homogeneous survey of hot jupiters \citep[][]{2016Natur.529...59S}. In contrast, \citet[][]{2016ApJ...817L..16S} suggests that clouds in hot jupiter atmospheres are restricted to regions in the surface gravity/temperature plane.
It is likely that the presence of silicate clouds in the regions probed by transmission (terminator) and emission spectroscopy (dayside) strongly depends on the day-night temperature difference \citep[e.g.,][]{2013ApJ...764..103R}, atmospheric circulation (see Question C1, and e.g., \citealt[][]{2009ApJ...699..564S,2015ApJ...801...95S}), and the importance of potential cold traps \citep[][]{2013A&A...558A..91P}.

Kepler-measured planet phase curves contain contribution of both reflected light and planetary thermal emission. Distinctive phase dependency of the two components may allow them to be separated \citep[][]{2015ApJ...802...51H}, and the reflective component is directly related to the distribution of clouds on the planet. This method has been applied to three hot Jupiters, and they all appear to have heterogeneous silicate clouds (\citealt[][]{2013ApJ...776L..25D}; \citealt[][]{2015AJ....150..112S}). Detailed models involving cloud condensation and general circulation suggest that such heterogeneous clouds are indeed common on hot Jupiters, and the cloud-forming material differs under different temperature regimes \citep[][]{2016ApJ...828...22P}. This knowledge is relevant for direct imaging because (1) it proves that at least some exoplanets have highly reflective clouds and therefore high albedo, and (2) it calls for considering inhomogeneous cloud coverage when interpreting spectra from direct-imaging observations.

{\em Hazes} -- with particles less than 0.1~$\mu$m -- have been argued for in a few objects where the lack of near-infrared absorption features (commonly water) necessitates that the absorption features are muted by high-altitude particles (unless the upper atmospheres of some transiting planets are extremely dry, see \citealt[][]{2014ApJ...791L...9M}). Such strong reduction in water absorption features was seen in the warm sub-neptune GJ1214b \citep[][]{2014Natur.505...69K} and atmospheric models argued for the presence of very small particles at low pressures ($\sim$1~mbar), consistent with photochemical hazes but not with condensate clouds \citep[][]{2015ApJ...815..110M}. The transmission spectra of temperate earth-sized planets in the Trappist-1 system \citep[][]{2016Natur.533..221G,2017Natur.542..456G} (orbiting a very low-mass red dwarf) also appear to be flat, perhaps also influenced by small particles lofted to low pressures \citep[][]{2016Natur.537...69D}. The detection of hazes (based on a very similar water absorption-based evidence) argues for the presence of some haze in L-type brown dwarfs, in spite of the lack of a host star, which argues for non-photochemical haze production \citep[][]{2015ApJ...798L..13Y}, possibly driven by charged particles accelerated by the brown dwarf's magnetic field.

\begin{figure}[ht]
\begin{center}
\includegraphics[width=0.8 \textwidth]{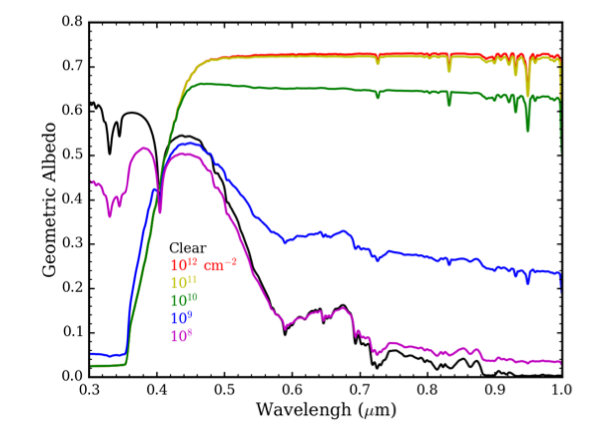}
%\plotone{Figures/Apai_Clouds_LT_Jupiter.png}
\caption{The effect of S$_8$ haze particles on the reflection spectrum of a Jupiter-mass planet as a function of particle size (as labeled). From \citet[][]{2017AJ....153..139G}. \label{Fig:Gao}}
\end{center}
\end{figure}

Figure~\ref{Fig:Gao} shows the impact of S$_8$ hazes on the reflection spectra of Jupiter--mass planets at 2~au separations form a sun-like star. Such a planet would be warm enough to lack substantial ammonia or water clouds, but the sulfur photochemical hazes produced by the destruction of atmospheric H$_2$S \citep[][]{2016ApJ...824..137Z} are in some cases sufficiently abundant to both brighten the spectra at redder wavelengths and to substantially darken the spectra at blue and NUV wavelengths. Labels on the figure point to the column number density of assumed 0.1~$\mu$m radius haze particles. Understanding the effect such photochemical hazes have on giant planet spectra is a prerequisite to ultimately understanding the role hazes play in terrestrial planet spectra.

%Directly imaged exoplanet

%Observational constraints on the origins of condensate clouds include: 
%a) Pressure range where they reside  ; b) Grain size distribution; c) Longitudinal-vertical structure; d) Evolution of cloud cover.

%Understanding the composition of cloud- and haze-forming particles is an important step in developing physical/chemical models for exoplanets.

%1) What data can constrain particle size distribution, pressure levels, composition?

%2) What fundamental parameters (composition, temperature, surface gravity?) are expected to have significant impact on cloud/haze formation and properties?

{\em	Solar System Gas Giants as exoplanet analog observations:} Overlapping Kepler  photometry and Hubble Space Telescope images of Neptune have shown complex time-varying signal whose frequency analysis revealed not only the fundamental rotation rate, but also the level of differential rotation of major mid-latitude cloud features \citep[][]{2016ApJ...817..162S}. Quasi-continuous 20-hour-long two-band optical imaging of Jupiter with the Hubble Space Telescope provided simultaneous high-precision photometry and high-fidelity and high-resolution images \citep[][]{2015ApJ...814...65K}. These authors showed that MCMC-based lightcurve modeling can correctly retrieve the position, size, and surface brightness of the dominant features in the lightcurve, such as the Great Red Spot, even from a single rotation. 

{\bf Example Focused Science Questions:} The study of extrasolar cloud layers is novel and we may not be in the position yet to identify the right set of key questions to ask. Nevertheless, the following list attempts to capture the most important uncertainties of our current models of clouds:\\
{\em i)} What are vertical structures of single and multi-layer clouds formed from different condensates? \\
{\em ii)} What are the grain size distributions and compositions (single-species or compound grains) in the clouds?\\
{\em iii)} Under which conditions do photochemically-- and charged particle-driven haze layers form? How complex can chemistry get in haze layers?\\
{\em iv)} How do condensate clouds form and evolve as a function of fundamental atmospheric parameters?

\begin{mdframed}[backgroundcolor=blue!20] 
{\bf Observational Requirements} \\
%{\bf Sample size:} Medium- to large (20-40 planets total) is likely required to form small groups (4--6) of similar planets in which cloud properties can be compared. \\
{\bf Observations:} Broad wavelength range (visual to mid--infrared) spectro-photometry or spectroscopy is ideal for probing a broad range of pressures. Such data can constrain the number and vertical structures of cloud layers. Time-resolved photometry/spectroscopy (sampling the rotational modulation) is important to assess the properties of non-homogeneous clouds. \\
\end{mdframed}

%\begin{mdframed}[backgroundcolor=yellow!20] 
%{\bf Questions to SAG15:} \\
%1) How challenging are the very different methods described here relative to each other?
%2) How are they best carried out?\\
%Kasper: ELT: A Jupiter analogue (1Rj, 5AU) will just be observable (c$\sim$1e-9). Neptune / Uranus analogues cannot be seen with an ELT.
%\end{mdframed}

%\subsubsection{Science Value of Independently Measured Planet Masses and Radii} 

\clearpage

\subsection{B4. How do photochemistry, transport chemistry, surface chemistry, and mantle outgassing affect the composition and chemical processes in terrestrial planet atmospheres (both habitable and non-habitable)?}

{\bf Contributors:} Caroline Morley, Mark Marley, Daniel Apai

The composition of exoplanetary atmospheres is one of the central questions of exoplanet characterization. Atmospheric composition is influenced by a multitude of factors (both initial and boundary conditions and processes) and, therefore, can provide valuable insights into the formation and evolution of each planet as well as on its present-day status. For example, characterizing the atmospheric composition of habitable zone planets is essential for determining whether they are, in fact, habitable planets --- in other words, that surface conditions allow the presence of liquid water (see also \S,\ref{B2LiquidWater}). 

Direct imaging missions are expected to image a diverse range of planets both in mass (from sub-earths to super-jupiter), in temperature ($<$100 K to $>$500~K), and in composition (H-rich, CO$_2$-dominated, atmosphereless, etc.). In the initial reflected light images, identifying the type of planets in a system may be very difficult (e.g., a small but high-albedo planet may look identical to a large but low-albedo planet; or --- even if the albedos are similar --- a partially illuminated (crescent-phase) giant planet may be very similar to a full disk of a slightly hazy super-earth). It is therefore imperative for direct imaging missions to include some level of planetary characterization as part of a discovery survey.

A multitude of excellent and up-to-date reviews are available on exoplanet atmospheric composition, based on observational evidence (Solar System planets, brown dwarfs, and exoplanets), and on theoretical predictions for the range of possible and expected compositions; therefore, we focus on the questions most salient for direct imaging missions.

{\bf Example Focused Science Questions}\\
{\em i)} a) What are the major and minor constituents of the atmospheres  of rocky planets? \\
{\em ii)} How do the compositions of rocky planet atmospheres vary as a function of mass, bulk composition, and irradiation?\\
{\em iii)} How strongly does mantle outgassing affect rocky planet atmospheres?\\
{\em iv)} Which planets show evidence for primordial atmospheres? \\
{\em v)} How are planetary atmospheres impacted by stellar high-energy radiation and stellar wind?\\

{\bf Example Science Cases and Observations}

Thorough exploration of the possible compositional classes for warm super-earth/neptune atmospheres argues for at least six classes (see Fig.~\ref{Fig:HuAtmosphereTypes}): i) Water-rich atmospheres; ii) co-existing water and hydrocarbon-dominated atmospheres; iii) hydrocarbon-rich atmospheres; iv) oxygen-rich atmospheres; v) CO/CO$_2$-dominated atmospheres; and, vi) H$_2$/He-rich atmospheres. The photochemistry in these atmosphere types has been explored, for example, in \citet[][]{2014ApJ...784...63H}. Many smaller terrestrial planets are likely to have CO$_2$-- or N$_2$--dominated atmospheres, based on solar system experience, with significant amounts of sulfur-bearing gases if volcanic activity is present \citep[][]{2012ApJ...761..166H,2013ApJ...769....6H}.

{\bf Retrieving the Compositions of Diverse Planets}
A key goal of exoplanetary research in the coming decades is to determine the compositions of planets from sub-Earth to super-Jupiter in mass. The environment in which a planet forms and evolves will shape the makeup of its atmosphere, so by studying planet compositions for diverse planets in a range of environments, we study the physics and chemistry of their formation and evolution. 

Direct imaging missions will be capable of detecting reflected light spectra for a variety of planets and thermal emission spectra for self-luminous planets. From these spectra, the atmospheric composition must be retrieved. Retrieval models have been used for decades in the solar system \citep{1977NASCP...4..117R,2000SAOPP...2.....R,2008JQSRT.109.1136I}
 and recently to study exoplanets \citep{
 2009ApJ...707...24M,2011Natur.469...64M,2013MNRAS.434.2616B,2013MNRAS.430.1188B,2012ApJ...749...93L,2013ApJ...775..137L,2014ApJ...793...33L,2012ApJ...753..100B,2013ApJ...778..153B}. Each model combines a radiative transfer scheme, which generates a synthetic spectrum for a given set of input parameters, with a fitting algorithm, often an MCMC algorithm, to fully explore the range of parameters. For an atmospheric retrieval, the parameters of most interest are the abundance of molecules, the temperature structure of the atmosphere, and the extent and composition of clouds and hazes. Most exoplanet retrieval models have been developed for thermal emission or transmission spectroscopy. \citet{2016AJ....152..217L} and \citet{2017PASP..129c4401N} have recently published results demonstrating retrievals using simplified reflected light spectra, and this is an area that will need further additional work before a major direct-imaging mission. 

{\bf Wavelengths of Major Absorption Features}

In order to retrieve the abundance of a particular molecule, the spectrum must probe a wavelength region where that molecule absorbs strongly. We have provided a table of wavelength regions of interest for molecules that are likely to be found in terrestrial atmospheres, reproduced from \citet{2002AsBio...2..153D}. 

\begin{table}[htp]
\caption{Overview of relevant atmospheric bands. \label{T:SpectralFeatures}}
\begin{center}
%\begin{tabular}{|l|c|}
%\begin{tabular}{p{13cm} l p{4cm}|}
\begin{tabular}{l c c c c}
Species & Min. $\lambda$ & Max. $\lambda$ & Ave. $\lambda$ & $\lambda / \Delta\lambda$  \\
              & $\mu$m   &   $\mu$m                     & $\mu$m            &   \\
\hline
 H$_2$O & 33.33		&	50.00	& 	40.00	&	2 	  \\
 H$_2$O &  25		&	33.33	&	28.57	&	4		  \\
 H$_2$O &  17.36 	&  	25 		&	20.49	&	3  \\
 H$_2$O &  6.67 		&	7.37 	&	7.00	  	&	10	\\
 CO$_2$ &  13.33		&	17.04 	&	14.96  	&	4	\\
 CO$_2$	 &  10.10		& 	10.75	&	10.42	&	16	\\
 CO$_2$	 &  9.07		&	9.56		&	9.31		&	19	\\
 O$_3$	 &  9.37		&	9.95		&	9.65		&	17	\\
 CH$_4$	 &  7.37		&	7.96		&	7.65		&	13	\\
 CH$_4$	 &  7.37		&	8.70		&	7.98		&	6	\\
 H$_2$O &  1.79		&	1.97		&	1.88		&	11	\\
 H$_2$O &  1.34		&	1.48		&	1.41		&	10	\\
 H$_2$O &  1.10		&	1.17		&	1.13		&	19	\\
 H$_2$O &  0.91		&	0.97		&	0.94		&	17	\\
 H$_2$O &  0.81		&	0.83		&	0.82		&	35	\\
 H$_2$O &  0.71		&	0.73		&	0.72		&	37	\\
 CO$_2$ &  1.97		&	2.09		&	2.03		&	16	\\
 CO$_2$ &  1.52		&	1.66		&	1.59		&	11	\\
 CO$_2$ &  1.20		&	1.23		&	1.21		&	34	\\
 CO$_2$ &  1.04		&	1.06		&	1.05		&	40	\\
 O$_2$    &  1.26		&	1.28		&	1.27		&	72	\\
 O$_2$    &  0.76		&	0.77		&	0.76		&	69	\\
 O$_2$    &  0.68		&	0.70		&	0.69		&	54	\\
 O$_3$    &  0.53		&	0.66		&	0.58		&	5	\\
 O$_3$    &  0.31		&	0.33		&	0.32		&	16	\\
 CH$_4$  &  2.19		&	2.48		&	2.32		&	8	\\
 CH$_4$  &  1.62		&	1.78		&	1.69		&	10	\\
 CH$_4$  &  0.97		&	1.02		&	1.00		&	20	\\
 CH$_4$  &  0.88		&	0.91		&	0.89		&	32	\\
 CH$_4$  &  0.78		&	0.81		&	0.79		&	29	\\
CH$_4$  &  0.72		&	0.73		&	0.73		&	57	\\
\end{tabular}
\end{center}
\end{table}%

\begin{table}[htp]
\caption{Overview of relevant atmospheric bands. Same as Table~\ref{T:SpectralFeatures}, but ordered by Column 2. \label{T:SpectralFeatures2}}
\begin{center}
%\begin{tabular}{|l|c|}
%\begin{tabular}{p{13cm} l p{4cm}|}
\begin{tabular}{l c c c c}
Species & Min. $\lambda$ & Max. $\lambda$ & Ave. $\lambda$ & $\lambda / \Delta\lambda$  \\
              & $\mu$m   &   $\mu$m                     & $\mu$m            &   \\
\hline
 H$_2$O & 33.33		&	50.00	& 	40.00	&	2 	  \\
 H$_2$O &  25		&	33.33	&	28.57	&	4		  \\
 H$_2$O &  17.36 	&  	25 		&	20.49	&	3  \\
 CO$_2$ &  13.33		&	17.04 	&	14.96  	&	4	\\
 CO$_2$	 &  10.10		& 	10.75	&	10.42	&	16	\\
 O$_3$	 &  9.37		&	9.95		&	9.65		&	17	\\
 CO$_2$	 &  9.07		&	9.56		&	9.31		&	19	\\
 CH$_4$	 &  7.37		&	7.96		&	7.65		&	13	\\
 CH$_4$	 &  7.37		&	8.70		&	7.98		&	6	\\
 H$_2$O &  6.67 		&	7.37 	&	7.00	  	&	10	\\
 CH$_4$  &  2.19		&	2.48		&	2.32		&	8	\\
 CO$_2$ &  1.97		&	2.09		&	2.03		&	16	\\
 H$_2$O &  1.79		&	1.97		&	1.88		&	11	\\
 CH$_4$  &  1.62		&	1.78		&	1.69		&	10	\\
 CO$_2$ &  1.52		&	1.66		&	1.59		&	11	\\
 H$_2$O &  1.34		&	1.48		&	1.41		&	10	\\
 O$_2$    &  1.26		&	1.28		&	1.27		&	72	\\
 CO$_2$ &  1.20		&	1.23		&	1.21		&	34	\\
 CO$_2$ &  1.04		&	1.06		&	1.05		&	40	\\
 H$_2$O &  1.10		&	1.17		&	1.13		&	19	\\
 CH$_4$  &  0.97		&	1.02		&	1.00		&	20	\\
 H$_2$O &  0.91		&	0.97		&	0.94		&	17	\\
 H$_2$O &  0.81		&	0.83		&	0.82		&	35	\\
 O$_2$    &  0.76		&	0.77		&	0.76		&	69	\\
 H$_2$O &  0.71		&	0.73		&	0.72		&	37	\\
 CH$_4$  &  0.88		&	0.91		&	0.89		&	32	\\
 CH$_4$  &  0.78		&	0.81		&	0.79		&	29	\\
CH$_4$  &  0.72		&	0.73		&	0.73		&	57	\\
 O$_2$    &  0.68		&	0.70		&	0.69		&	54	\\
 O$_3$    &  0.53		&	0.66		&	0.58		&	5	\\
 O$_3$    &  0.31		&	0.33		&	0.32		&	16	\\
\end{tabular}
\end{center}
\end{table}%

{\bf The Problem of Clouds and Hazes}

The presence of clouds and hazes is very important for measuring a planet's reflected light, because clouds can strongly scatter light, increasing the albedo and revealing the presence of absorption features, particularly at redder wavelengths. Cloud-free models predict that without clouds, gas- and ice-giant planets would be dark and therefore faint \citep[e.g.][]{1999ApJ...513..879M, 2000ApJ...538..885S, 2015ApJ...815..110M}. 

However, the effect of clouds and hazes (see Section~\ref{S:Clouds} for details) poses perhaps the greatest astrophysical challenge for retrieving robust and precise abundances for molecules in a planet's atmosphere. The location and scattering properties of a cloud are not known ahead of time, and therefore must be retrieved alongside parameters of interest such as the atmospheric composition. However, particularly with limited SNR and limited wavelength range, cloud properties can be degenerate with other properties \citep{2017ApJ...835..198K}. The choice of parameterization becomes very important (e.g., whether the cloud is a single layer or multiple layers, the vertical extent, the optical depth and scattering properties). 
%Detailed modeling work can inform us on these challenges before the launch of a major direct-imaging mission in order to determine how to overcome these challenges. 

\begin{figure}[ht]
\begin{center}
\includegraphics[width=0.5 \textwidth]{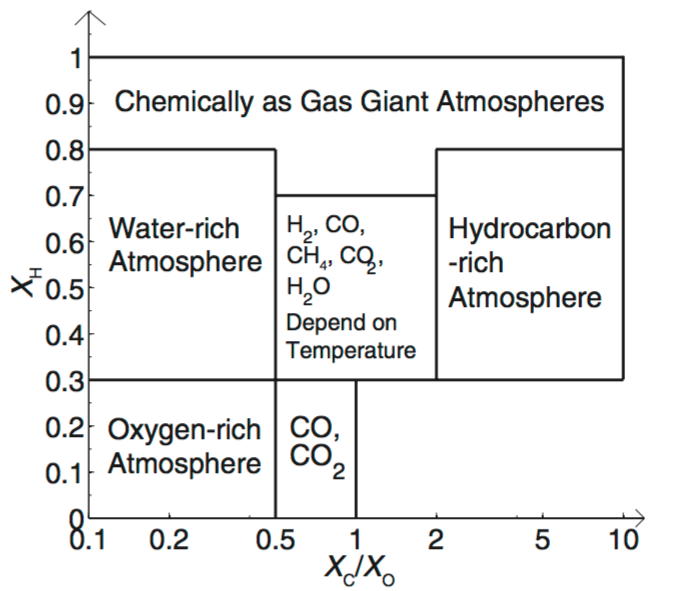}
%\plotone{Figures/Apai_Clouds_LT_Jupiter.png}
\caption{The types of thick atmospheres possible on Super-Earths and mini Neptunes, based on the extensive exploration of chemical reaction networks. For atmospheres {\em not} dominated by H$_2$, different atmosphere classes emerge as a function of the relative abundances of C, O, and H. From \citet[][]{2014ApJ...784...63H}. \label{Fig:HuAtmosphereTypes}}
\end{center}
\end{figure}

\begin{figure}[ht]
\begin{center}
\includegraphics[width=0.5 \textwidth]{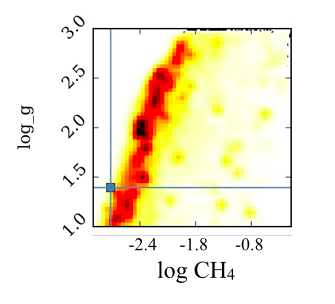}
%\plotone{Figures/Apai_Clouds_LT_Jupiter.png}
\caption{Posterior probability distributions for the gravity of a Jupiter-like planet and the atmospheric methane abundance as found in a retrieval on simulated optical reflectance data (Marley et al. 2014). Note that the retrieved abundance is highly correlated with the gravity since the column number density of absorbers above a reflecting cloud layer is proportional to g$^{-1}$. The ÒtrueÓ CH$_4$ abundance and gravity used in the forward model is shown by the blue lines. Without useful constraints on gravity the range of acceptable CH$_4$ mixing ratios is very large. \label{Fig:Marley2014}}
\end{center}
\end{figure}

\subsubsection{Science Value of Independently Measured Planet Masses and Radii} 
Characterizing planets with reflected light spectroscopy will be greatly aided by knowing their masses from independent observations. For example, \citet{2014arXiv1412.8440M} showed that the methane abundance derived from an albedo spectrum is strongly degenerate with the retrieved gravity (see Figure~\ref{Fig:Marley2014}). In the absence of any constraints on gravity, the methane abundance can be inferred to about a factor of 10 from their nominal SNR$\sim$10 optical light spectrum. However, if the gravity is known to a factor of two, the methane abundance can be measured more precisely, to within a factor of 3 of the true value. Constraining the gravity independently will require both mass and radius measurements. The mass can be measured accurately using both radial velocity (to measure $M$ sin $i$) and imaging (to constrain the inclination), or alternately by using astrometric techniques. Once the mass is known, the radius can be calculated for gas giant exoplanets using an empirical or model mass-radius relationship. The radius can also be measured from the spectrum, particularly by making the measurement at known phase angles \citep{2017ApJ...835..198K}. 

\begin{mdframed}[backgroundcolor=blue!20] 
{\bf Observational Requirements} \\
%{\bf Sample size:} Medium- to large (20-40 planets total) is likely required to form small groups (4--6) of similar planets in which cloud properties can be compared. \\
{\bf Observations:} The planets in or close to the habitable zones will have low temperatures. Therefore, detecting their thermal emission would require sensitive 10$\mu$m observations. Lacking that capability reflected light spectroscopy will be optimal for characterizing the atmospheres of these planets.\\
{\bf Wavelength Coverage:}  The key, possibly detectable atmospheric features expected in these planets will be a subset of H$_2$O, CO/CO$_2$, O$_2$, O$_3$, while contributions of H$_2$ and N$_2$ are expected but will remain difficult to detect due to the lack of the permanent dipole moments of these molecules. The characterization of Rayleigh-scattering will require observations at short wavelengths ($<$0.5 $\mu$m), the detection of O$_2$ will requires wavelength coverage in the visual (0.5--0.7 $\mu$m), H$_2$O is detectable over a broader wavelength range (0.7$-2\mu$m), but the broader and more prominent CH$_4$ bands require longer-wavelength observations (1.78, 2.48, or 7.98 $\mu$m). Almost all atmospheric constituents as well as Rayleigh scattering will require only low to moderate spectral resolution (R$\sim$50), with the notable exception of O$_2$ which requires R$\sim$150. The characterization of atmospheres with optically thick (and large covering factor) haze and cloud layers will be challenging and it may be beneficial to exclude such targets from spectroscopic observations that will likely require very long integrations. Pre-selecting such planets may be possible on the basis of appropriately-chosen broad-band photometry.
 \\
\end{mdframed}

\clearpage

\section{Exoplanetary Processes}

\subsection{C1. What processes/properties set the modes of atmospheric circulation and heat transport in exoplanets and how do these vary with system parameters?}

{\bf Authors:} Daniel Apai, Nick Cowan, Ravi Kopparapu, Anthony del Genio, Thaddeus Komacek

Atmospheric circulation plays a key role in redistribution of the energy in exoplanet atmospheres. Depending on typical wind speeds, rotational velocity, insolation, latent heat released during condensation, and other system parameters different atmospheric circulation regimes are expected on planets that can be studied with direct imaging missions. For potentially habitable exoplanets atmospheric circulation will determine the day-night heat differential and the equator-pole temperature difference. Understanding the presence and size of Hadley cells can also provide important insights into how water vapor (or other condensibles) may be distributed in habitable planets. 

\begin{figure}[h]
\begin{center}
\includegraphics[width=\textwidth]{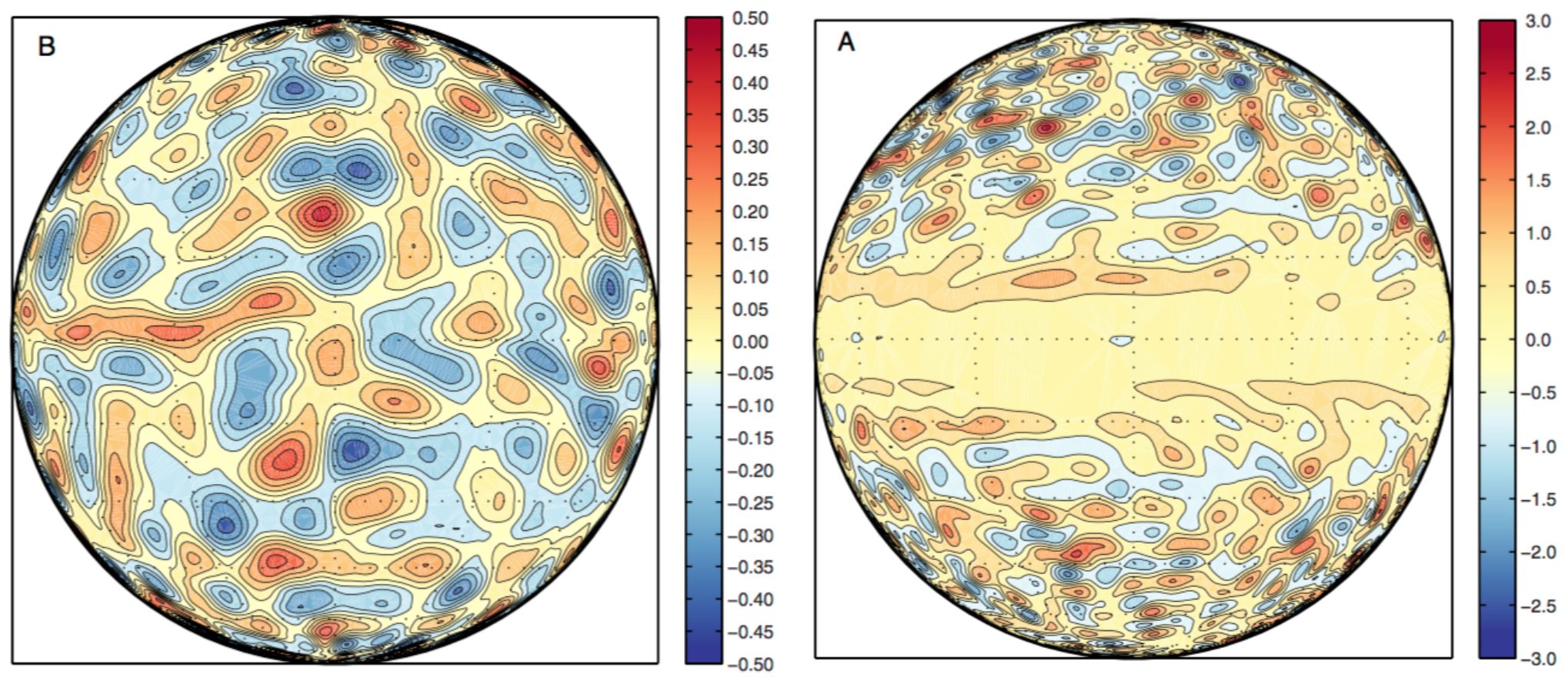}
\caption{Depending on the relative importance of rotational speed, wind speed, and vertical heat transport, simple models predict two different regimes of circulation for giant planets: vortex-dominated (left) and jet-dominated (right). From \citet[][]{2014ApJ...788L...6Z}.
\label{Fig:Zhang2015}}
\end{center}
\end{figure}

\begin{figure}[h]
\begin{center}
\includegraphics[width=\textwidth]{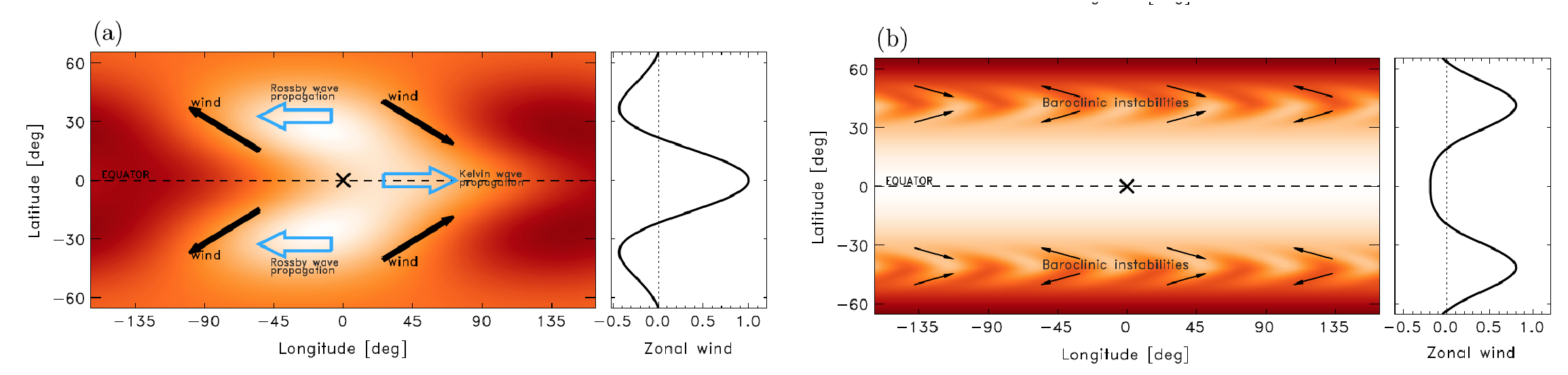}
\caption{Dependence of the atmospheric circulation on rotation rate. Panel (a) shows a slowly-rotating, highly irradiated hot-Jupiter
planet with strong day-night temperature difference and a strong eastward equatorial super-rotating jet. Panel (b) shows rapidly 
rotating warm Jupiters that are weakly irradiated. These planets develop eddy driven zonal jets that peak at mid-latitudes
rather than at the equator.  From \citet[][]{2015ApJ...801...95S}.
\label{Fig:Showman2015}}
\end{center}
\end{figure}

\begin{figure}[h]
\begin{center}
\includegraphics[width=\textwidth]{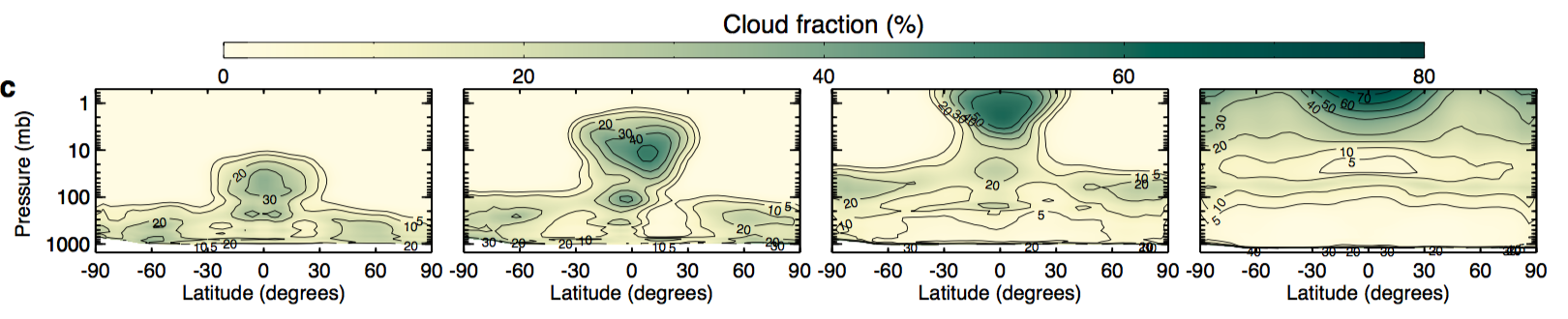}
\caption{Moist/water-rich atmosphere simulations from \citep[][]{2015JGRD..120.5775W}. The four panels indicate the amount of cloud water content  on a planet at different insolation levels (or, alternately, how close to an inner edge of the HZ a planet is located). From left to right, the solar insolation varies:  S$_{0}$ (current Earth insolation), 110\% of S$_{0}$, 112.5\% S$_0$ and 121\% of S$_0$. This is for an Earth-size planet around a Sun-like star.\label{Fig:Wolf2015}}
\end{center}
\end{figure}

\begin{figure}[h]
\begin{center}
\includegraphics[width=0.5 \textwidth]{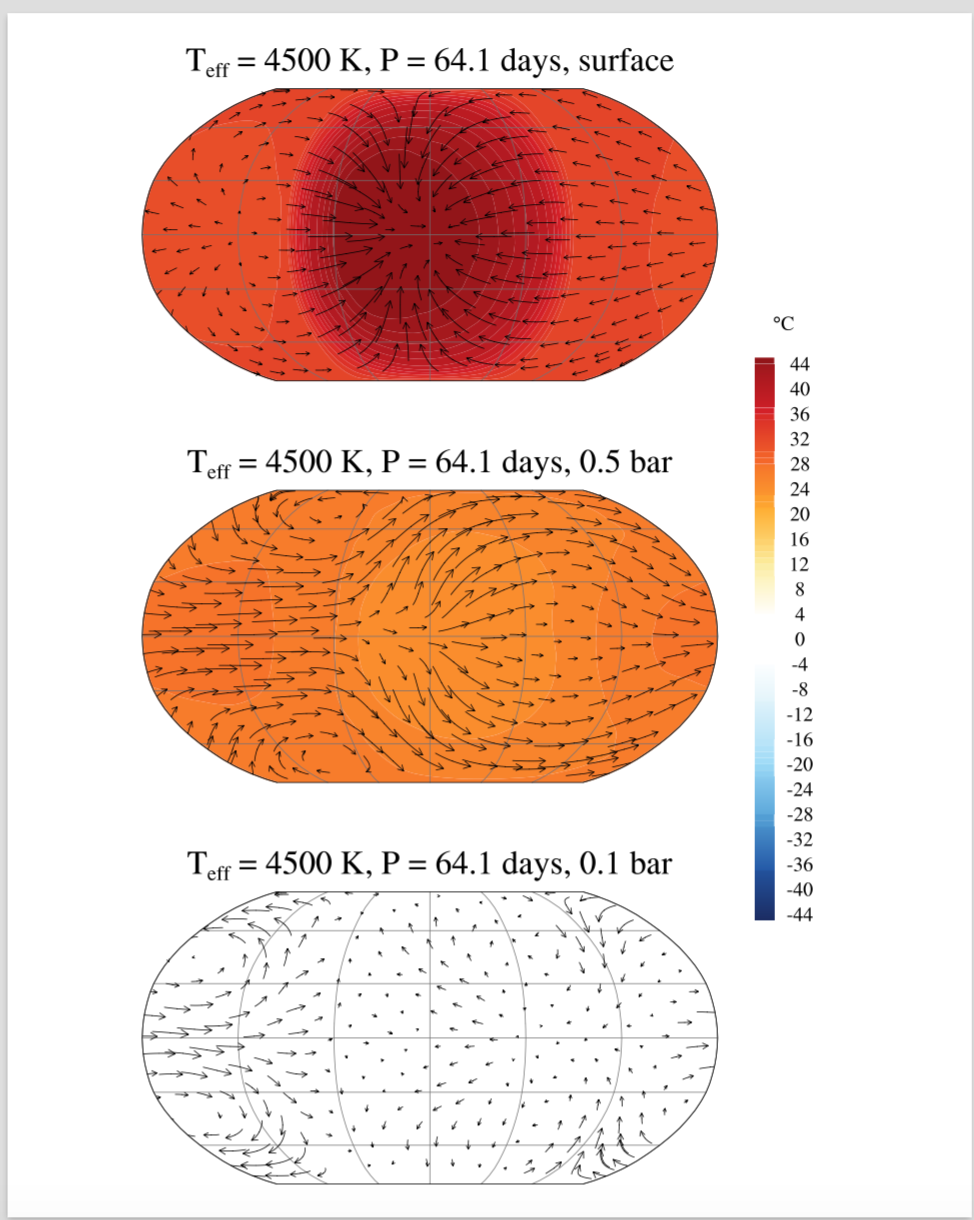}
\caption{Temperature and horizontal wind vectors at the surface, 0.5 bar, and 0.1 bar levels for an Earth-mass planet in a slow-rotating regime near the inner edge of the habitable zone around a K-dwarf. Slowly rotating planets develop sub-stellar clouds that increase the albedo of the planet. Inflow along the equator and from the poles into the substellar point at the center is also shown. From \citet[][]{2016ApJ...819...84K}.
\label{Fig:Kopparapu2016}}
\end{center}
\end{figure}

Understanding atmospheric circulation in habitable exoplanets is an important component in establishing a correct climate model for them.
As of now, atmospheric circulation has modeled in the Solar System planets and a small sample of brown dwarfs, hot jupiters and lower-mass exoplanets (see Figs~\ref{Fig:Wolf2015} and \ref{Fig:Kopparapu2016}, \citealt[][]{2013ApJ...771L..45Y,2013Natur.504..268L,2011AsBio..11..443A,2011ApJ...733L..48W, 2014ApJ...788L...6Z,2016ApJ...819...84K,2014ApJ...785...92K,2014AsBio..14..241W}). The nature of the atmospheric dynamics depends on the thickness of the planet's atmosphere, its rotation rate, the distance of the planet from the star and several other factors. A more comprehensive study of different atmospheric circulation regimes of exoplanets still lacks, but important steps have been taken for rocky exoplanets in a simplified general circulation model by \citet[][]{2015ApJ...804...60K}.

{\bf Gas Giants:} Though there is no comprehensive prediction for how the atmospheric circulation varies with planetary parameters (e.g., incident stellar flux, rotation rate, atmospheric mass and composition), there exist theoretical predictions for how the circulation of tidally-locked planets varies with these parameters. These models have been developed both for rocky \citep{2016ApJ...825...99K} and gaseous \citep{2016ApJ...821...16K,2017ApJ...835..198K} tidally-locked exoplanets, and enable prediction of both the day-to-night temperature contrast and characteristic wind speeds. Notably, the day-to-night temperature contrast can be teased out from the amplitude of an observed infrared phase curve, whether or not the planet is transiting. In the case of terrestrial planets, the inference of the phase curve amplitude can tell us if an atmosphere exists, given the possibility of collapse of the atmosphere on the nightside \citep{2015ApJ...802...21K,2016ApJ...825...99K}. If there is an atmosphere, an observed phase curve amplitude can lead to estimation of the surface pressure. In the case of hot Jupiter atmospheres, there exists a general trend of increasing phase curve amplitude with increasing incident stellar flux \citep{2011ApJ...729...54C,2013ApJ...776..134P,2015MNRAS.449.4192S,2016ApJ...821...16K}, which agrees with the theoretically predicted trend \citep{2017ApJ...835..198K}. However, to date there are only 9 low-eccentricity hot Jupiters with measured infrared phase curves. Future phase curve observations of a large sample ($\sim$25--50) of these planets (possibly with JWST or a dedicate transiting exoplanet telescope) would inform us whether or not the trend of increasing day-to-night temperature contrast with increasing incident stellar flux is general, and if so can test theories for how the atmospheric circulation of hot Jupiters varies with incident stellar flux and rotation rate.

The atmospheric circulation patterns of planetary atmospheres can be characterized broadly from the planetary rotation rate; Earth exhibits three major circulation cells, while planets with a more rapid rotation rate  and/or larger radii (such as gas giants) show five or more circulation bands \citep[]{1982Natur.297..295W}. Knowledge of an exoplanet's rotation rate would provide a strong constraint on the large-scale dynamical features that should occur, given the planet's orbital distance from its host star \citep[][]{2010JAMES...2...13M}.

Different  circulation regimes can exist in the atmospheres of extrasolar planets depending upon the incident flux and rotation rate of planet. For example, \citet[][]{2015ApJ...801...95S} showed that the canonical hot-Jupiter regime ($0.03 - 0.05$ AU), with a large day-night temperature gradient and a fast east ward equatorial jet, transitions at lower stellar fluxes ($\sim$1 AU) and/or faster rotation to a regime with small longitudinal temperature variations and peak wind speeds occurring in zonal jets at mid- to high latitudes.

Furthermore, at a given stellar flux, a greater than factor of two in rotation rate difference between synchronous and 
non-synchronous causes can potentially be discerned in the light curves of hot-Jupiters, providing a way to identify regime transition from highly irradiated to weakly irradiated planets.

Observational characteristics such as variation in thermal emission from orbital phase curves, and net Doppler-shift obtained from high-resolution spectra (as a function of orbital phase), in principle, provide a means to constrain the rotation rate for some hot-Jupiter planets \citep[][]{2014ApJ...790...79R}. Although these techniques may not be individually suited to distinguish the rotation rates, the combination of these two techniques may show observable differences with rotation rate.

{\bf Earth-like planets:} Planets in and around the habitable zone (HZ) of low-mass stars are expected to be in synchronous rotation, though thermal tides can cause asynchronicity on some planets \citep[][]{2015Sci...347..632L}.
Such planets can further be classified as {\em slow-rotators} (where the Rossby deformation radius is equal or greater than planetary 
radius) and {\em fast-rotators} (where the Rossby deformation radius is less than planetary radius). Planets in synchronous orbits that are 
also slow-rotators may develop a shielding cloud presence beneath the substellar point, which can increase the inner habitable limits of the planet \citep[][]{ 2013ApJ...771L..45Y}. However, rapidly-rotating planets tend to smear out this cloud deck, which limits much of this shielding effect \citep[][]{2016ApJ...819...84K}. A comprehensive recent study of stratospheric moisture of synchronously rotation planets with a global circulation model (Fujii et al., under review; \citealt[][]{2017arXiv170405878F}) found a gradual transition to moist climate (as opposed to rapid transition as seen for fast-rotating planets). This effect is qualitatively consistent with the results by \citep[][]{ 2013ApJ...771L..45Y}. Fujii et al. find that stratospheric water increases while substellar temperature is still low and attributing this to the interplay between the vertical transport of water vapor and radiative heating by upper H$_2$O.

On Earth, the mean meridional circulation, or Hadley circulation, is responsible for the poleward transport of energy at low latitudes; however, on 
synchronously rotating planets, the Hadley circulation provides an incomplete diagnosis of energy transport because the Hadley 
circulation itself changes direction between the hemisphere eastward and westward of the substellar point \citep[][]{2015MNRAS.446..428H} and a significant day-night circulation develops when the radiative time scale is shorter than the length of the solar day \citep[][]{2016GeoRL..43.8376W}. Rather than the Hadley circulation, the mean zonal circulation (or Walker circulation) provides a better 
metric for synchronous rotators to examine the efficacy of heat transport between the substellar and antistellar points. For slow 
rotators, the Walker circulation reaches to the night side of the planet, but for rapid rotators, the Walker circulation by itself is 
limited in longitudinal extent. In such cases, a cross-polar circulation also provides energy transport between the day and night side 
to keep the atmosphere from freezing-out or collapsing \citep[][]{1997Icar..129..450J, 2015MNRAS.446..428H}.

Recent three-dimensional climate modeling studies of Earth-like planets predict that rapidly rotating planets undergo a sharp transition 
between temperate and moist greenhouse climate states \citep[][]{2015JGRD..120.5775W,2016NatCo...710627P}.  \citet[][]{2015JGRD..120.5775W} argue that 
this transition is associated with a fundamental change to the radiative-convective state of the atmosphere.   When the mean surface 
temperature approaches $\sim 330$ K, the lower atmosphere becomes opaque to infrared and thermal radiation due to increasing water vapor 
mixing ratios.  The lower atmosphere heats due to solar absorption in the near-IR.  Simultaneously, the lower atmosphere cannot 
efficiently cool to space due to the closing of the 8-13 $\mu$m water vapor window region.  Combined, this results in a net positive 
radiative heating rates in the near surface layers, creating a ubiquitous temperature inversion across the planet.  The inversion 
suppresses boundary layer convection, reducing clouds and the planetary albedo at the climatic transition.  As climate warms further, 
the low atmosphere becomes increasingly hot and dry (i.e., low relative humidity), but upper atmosphere water vapor mixing ratios 
become large and a zonally uniform, albeit patchy, cloud deck develops.

Figure \ref{Fig:Wolf2015} shows the evolution of the zonal mean cloud water content (kg m$^{-3}$) for an Earth-like planet under 
increasing stellar fluxes, varying from the present day Earth insolation up to a $21 \%$ increase.  For the present day Earth climate 
(Fig. \ref{Fig:Wolf2015}, leftmost panel), clouds are confined to pressures greater than $\sim 200$ mb, with the thickest clouds 
located at mid-latitudes.  For moist greenhouse atmospheres, the lower atmosphere becomes cloud free, while the primary cloud deck 
becomes zonally uniform and is pushed higher in the atmosphere.  For an Earth-like planet with a mean surface temperature of 
$ \sim 363$ K, the cloud water peaks near $\sim 50$ mb (Fig. \ref{Fig:Wolf2015}, rightmost panel).  Clouds are well known to 
obscure exoplanetary spectra due to their significant broadband opacity.  Thus we may be able to differentiate habitable Earth-like 
atmospheres from moist greenhouse atmospheres, based on the pressure level of the primary cloud deck.  However, note that an 
Earth-like planet at $\sim 363$ K, would have moist stratosphere ($\sim 6 \times 10^{-2}$ H$_{2}$O mixing ratio at 0.2 mb), and thus 
would be expected to lose an Earth ocean of water to space within several hundred million years.  Moist and runaway greenhouse 
atmospheres are thus transient phenomena.  

Interestingly, \citet[][]{2016ApJ...819...84K} found that the above described radiative-convective transition also occurs on slow and 
synchronously rotating Earth-like planets, which are expected around low mass stars.  While rapidly rotating planets can maintain 
climatological stability beyond this transition due to cloud adjustments in the upper atmosphere, this transition is catastrophic 
for planets located near the inner edge of the habitable zone around low mass stars.  As noted above, synchronously rotating planets are effectively 
shielded from the host star by thick convectively produced clouds located around the substellar point.  These planets can remain habitable despite incident stellar fluxes up to twice that of the present day Earth \citep[][]{ 2014ApJ...787L...2Y,2016ApJ...819...84K}.  However, the 
radiative-convective transition and subsequent onset of the near surface inversion stabilizes the substellar atmosphere, and thus the 
convective cloud deck rapidly dissipates.  Even a small dent in this substellar cloud shield then lets in a tremendous amount of solar 
radiation, destabilizing climate towards an immediate thermal runaway.

%\begin{mdframed}[backgroundcolor=yellow!20] 
%{\bf Questions to SAG15:} \\
%To what level can the atmospheric circulation be constrained for different types of planets?\\
%What hypotheses / toy circulation models should be tested for gas giants? \\
%What hypotheses / toy circulation models should be tested for habitable super-earths / earths?\\
%What data type and cadence is required or best suited for characterizing circulation?\\
%How does the atmospheric circulation in tidally locked planets around M-dwarf stars affect habitability?\\
%Where is the transition region from slow to rapid rotators in tidal-locked planets around low-mass stars?\\
%\end{mdframed}

\begin{mdframed}[backgroundcolor=blue!20] 
{\bf Observational Requirements} \\
%{\bf Sample size:}  Medium-sized sample of rocky Earth-sized planets interior, within, and beyond the habitable zone with different rotational rates to sample circulation patterns as a function of irradiation and rotational period. For example, a sample of $\sim$36 planets divided in nine categories (low/medium/high rotational rates and high/moderate/low irradiation) may be used to evaluate circulation patters as a function of these two parameters. However, planets with thick and homogeneous cloud/haze patterns are not suitable for these observations and will not contribute to the sample. \\
{\bf Observations:} Continuous time-resolved photometric or spectroscopic observations over one or more rotational periods are required to constrain the spatial distribution of cloud structures, which may be used as a proxy for atmospheric circulation. Hot and young planets (gas and ice giants) may be observed in the near-infrared (thermal emission), but cooler planets will require reflected light observations either in the visual -- near-infrared regime. The correct interpretation of the atmospheric circulation will require establishing the rotational periods of the planets. For planets with surfaces composed of constant and time-varying features (such as Earth's continents/oceans and cloud cover) complete rotations covered at multiple orbital phase angles will allow the derivation of rough two-dimensional maps and the determination of the obliquity. 
{\bf Wavelength range:} Atmospheres with different composition will benefit from observations at different wavelengths (see discussion in Question B4). Patchy clouds and hazes, as well as surface compositional variations (e.g., continents and oceans) translate into albedo variations over a broad wavelength range (from the visual to the near-infrared). However, rotational mapping will typically require relatively precise photometry in integrations that are an order of magnitude shorter than the rotational period. Therefore, the choice of wavelength will be driven by the star-planet contrast and the system sensitivity.\\
{\bf Supplementary Observations:} Establishing the planet's mass will be important as a context for interpreting the atmospheric circulation and heat transport constraints gained from the rotational phase mapping observations.
\end{mdframed}

\newpage

\subsection{C2. What are the key evolutionary pathways for rocky planets and what first-order processes dominate these?}

{\bf Contributors:} Nick Cowan, Daniel Apai, Yuka Fujii, Renyu Hu, Peter Plavchan

The two earth-sized rocky planets in the Solar System, Earth and Venus, likely started with very similar initial mass, orbit, and composition, but their evolutionary paths have strongly diverged. 
This divergence could be result of differences in the early solidification phase, during which relatively small differences in insolation levels may lead to major differences in water loss levels  \citep[][]{2013Natur.497..607H}.
Mars, although substantially different in its mass and orbit, has again followed a different evolutionary trajectory, even though it is thought that surface conditions on early Mars, at least temporarily or episodically, may have sustained wide-spread aqueous activity on the the surface, perhaps resembling the early Earth. With the large number of rocky planets that may be observable with a capable future direct imaging mission, the range of evolutionary histories could be explored.

%Hamano et al. (2013) (http://adsabs.harvard.edu/abs/2013Natur.497..607H) and Hamano et al. (2015)
%(arXiv:1505.03552). In the former, she studied how the early evolution of the planet with magma ocean leads to the remarkable
%difference between Venus and Earth. In 2015 paper, they simulated the spectra of magma-ocean planets. They %also predicted the
%possibility of prolonged magma ocean stage depending on the initial amount of water, so I think such planets are interesting targets to
%observe, and can potentially provide useful insights into Venus-Earth problem.

{\em The question naturally emerges: What key evolutionary pathways exist for rocky planets and what factors determine which of these pathways a given planet will follow? }

{\em Attractors and Divergence in the Phase Space of Rocky Planet Evolution: } It is reasonable to describe the momentary state of a given rocky planets with a set of $n$ fundamental parameters and explore the evolution of the planet in this $n$-dimensional phase space. Each planet's history and future evolution is thought of as a trajectory. Fundamental parameters could include, but are not limited to, planet mass, radius, atmospheric pressure scale height, orbital parameters, atmospheric composition, rotation rate, magnetic field strength, etc. Which trajectory a planet follows will depend not only in its momentary location in the phase space, but also by the effect of a set of feedback loops (both positive and negative) as well as on a few environmental variables (e.g., stellar luminosity and incident optical and UV flux).

When describing planet evolution in such a manner, several obvious questions are identified: {\em 1) How sensitive are the trajectories to initial parameters and/or perturbations to the system? 2) What is the importance of a planet's past, e.g., which volumes of the phase space are uni-directional (e.g., irreversible water loss)?
3) Are there preferred evolutionary end-states (attractors) or is the surface defined by coeval planets smooth? 4) What is the importance of quasi-monotonic evolution driven by a small number processes vs. random walk driven by a multitude of competing processes?}

Exploring the past history and current state of rocky planets allows the system-level study of rocky planet evolution and will be essential for understanding the occurrence rate of truly earth-like planets and to place the physical processes that drive planet evolution on Earth to the broader context of exo-earths.

\subsubsection{Science Value of Independently Measured Planet Masses and Radii} 

The science goal is achievable without precise mass measurements: a medium or large sample of rocky exoplanets for which most of the other key parameters are known would likely suffice to establish the topology of the phase space.  

However, precise mass measurements would significantly contribute to the understanding of the planets' properties. In case the phase space is highly complex and its projection to a lower-dimensional (observed) phase space does not allow the identification of the key processes that drive the evolution, expanding the projected phase space by a new dimension (mass) may break the degeneracy between different processes that lead to similar evolutionary outcomes.

A bulk density determination with a precision of 10\% would distinguish among different terrestrial planet composition models (e.g., \citealt[][]{2015ApJ...801...41R}). For an Earth analog in a HZ orbit (reflex velocity of K$\sim$9 cm/s), determining the mass to $\sim$10\% requires $\sim$1 cm/s radial velocity precision on a time-scale of a year.  Such a capability is not currently possible from the ground. Both CODEX and G-CLEF for the ELT and GMT respectively are being designed for an instrument systematic uncertainty of $\sim$2 cm/s (Plavchan et al. SAG8 report).  A current generation radial velocity survey may observe a single target for 5 minutes to reach a photon noise precision of $\sim$1 m/s (e.g., HIRES on Keck), and thus allowing a single facility to observe on the order of 100 stars in a single night.  However, a photon noise of 1 cm/s will require significantly longer integration times.  For example, a $\sim$1 cm/s photon noise is reached for an hour-long integration on a 10-m telescope at V$\sim$4 mag (Plavchan et al. SAG8 report). Thus, mass determination of exoplanets at this precision will require significant amount of telescope time.  This will limit the number of targets that can be observed with competed facilities such as the ELT and GMT. It is also not certain that stellar activity will limit the radial velocity sensitivity at $\sim$1 m/s.  Developing the data analysis tools, cadence and wavelength coverage for stellar activity in radial velocity spectroscopic time-series is an active area of research.

%\begin{figure}[h]
%\begin{center}
%\includegraphics[width=0.5 \textwidth]{Figures/Kopparapu_2016.png}
%\caption{Temperature and horizontal wind vectors at the surface, 0.5 bar, and 0.1 bar levels for an Earth-mass planet in a slow-rotating regime near the inner edge of the habitable zone around a K-dwarf. Slowly rotating planets develop sub-stellar clouds that increase the albedo of the planet. Inflow along the equator and from the poles into the substellar point at the center is also shown. From \citet[][]{2016ApJ...819...84K}.
%\label{Fig:Kopparapu2016}}
%\end{center}
%\end{figure}

\begin{mdframed}[backgroundcolor=blue!20] 
{\bf Observational Requirements:} \\
{\bf Exoplanet characterization:} Atmospheric pressure and composition (see Question B4), orbital parameters (see Question A1), bulk composition (see Supplementary Observations), surface temperature estimate, and stellar parameters. Furthermore, identifying patterns in the present-day distributions of the rocky planets -- essential for identifying major evolutionary pathways -- will require a large sample of planets, commensurate with the expected complexity of the evolutionary pathways.\\
{\bf Supplementary Observations:} Stellar radial velocity and/or astrometric observations will be essential for establishing the mass of the planet to help constrain the bulk composition. Detailed understanding of the stellar spectrum, activity, and evolution will be important to explore the evolution of the atmospheric mass loss of the planet and the evolution of the atmospheric photochemistry. 
\end{mdframed}

%\begin{mdframed}[backgroundcolor=yellow!20] 
%{\bf Comments:} 
%Solicit input from: Robert Boschia, Valerio Lucarinia, Salvatore Pascalea (2013); William Moore, Leslie Rogers, Adrian Lenardic, Dorian Abbot, Dan Fabrycky
%To explore: Toy model for testing hypothesis of smooth distribution vs. attractors
%\end{mdframed}

\newpage

\subsection{C3. What types/which planets have active geological activity, interior processes, and/or continent-forming/resurfacing processes? }

{\bf Contributors:} Stephen Kane, Daniel Apai, Nick Cowan

Planetary interior processes and geological activity play an important role in coupling Earth's atmosphere to its crust and providing a long-term stabilizer for Earth's climate. The source of Earth's atmosphere and volatiles are mostly products of outgassing after the loss of the primary atmosphere.
Developing reliable climate models to determine the habitability of potentially habitable planets will likely require assumptions about the geological activity and the level of coupling between the planet's crust and atmosphere \citep[e.g.,][]{ 2012ApJ...756..178A,
2016GGG....17.1885F}. However, interior processes are obviously very difficult to probe via spatially unresolved remote sensing.

The influence of geological activity on planetary climate is most clearly understood for the case of Earth. On geologic timescales, continental crust production participates in the stabilization of the Earth's climate through its role in carbonate weathering feedback \citep[][]{1993Icar..101..108K}. Chemical weathering of silicate minerals on land in the presence of water causes the slow removal of CO$_2$ from the atmosphere, which is eventually deposited on the ocean floor as carbonate compounds. Without the continual re-injection of new CO$_2$ by volcanoes, the atmospheric stock of CO$_2$ would be slowly depleted. However, the rate of CO$_2$ removal by silicate weathering is temperature dependent, so that in the presence of a steady source of volcanic CO$_2$, weathering interacts with the greenhouse properties of CO$_2$ to produce a negative feedback on planetary temperature. This interaction, whereby warmer conditions lead to increased drawdown of CO$_2$ and a consequent weakening of the greenhouse effect (and vice versa), is believed to play an important role in stabilizing planetary temperatures in the presence of a main-sequence star which is increasing in luminosity over Ga timescales. It is because of this process that it has been argued that volcanism and geological activity are necessary conditions for sustained life on a planet. 

{\bf Current Knowledge:}  Two methods have been proposed to detect geological activity on a rocky exoplanet. First, \citet[][]{2010ApJ...708.1162K} suggested that volcanic emission of SO$_2$ can be detected remotely. However, it has been found that the volcanic sulfur emission would most likely lead to formation of sulfur and/or sulfate aerosols in the atmosphere, leading to muted transmission and thermal emission spectral features \citet[][]{2013ApJ...769....6H}. The sulfur-bearing aerosols may be detected via direct imaging, and indicate volcanic activity on the planet. Second, \citet[][]{2012ApJ...752....7H} suggest that fresh volcanic surfaces and surfaces solidified from a magma ocean have prominent spectral features at 1~$\mu$m and 2 $\mu$m, produced by Si-O bonds in mineral lattices. Surfaces aged by either space or aqueous weathering do not have these features. Therefore, some specific spectral features can imply recent volcanic activities on a rocky exoplanet.

Studies of terrestrial climate and volcanism focus primarily on the effects of volcanism on surface temperature, which we are unlikely to be able to estimate for most exoplanets. However, volcanically forced anomalies in surface temperature are coupled to anomalies in emission temperature, which can be targeted for follow-up observations. Thus, if volcanism can be identified on an exoplanet it may represent the most promising method for estimation of climate sensitivity outside of the Solar System. Furthermore, active volcanism may be able to significantly extend the outer boundary of the habitable zone \citep[][]{2017ApJ...837L...4R}.

The distinctive effect of volcanic eruptions on the transmissivity of atmospheres is related to the force of their explosions. Typically, processes on Earth that produce aerosols in the atmosphere affect only the troposphere. Aerosols are quickly washed out of the troposphere by rain, and thus a sustained impact on atmospheric transmissivity requires a near-continual source of the aerosol or its precursor gas. Many small eruptions do not reach the stratosphere, however the largest explosive volcanic material can, in contrast, inject SO$_2$ directly into the stratosphere, where it reacts to form sulphate aerosols \citep[e.g.,][]{2010ApJ...708.1162K}. 

Because the stratosphere is very dry and the particle sizes are small, these aerosols can persist in the stratosphere for several years, until they are removed by the natural overturning circulation of the stratosphere \citep[][]{2007GeoRL..3423710R}. Stratospheric air rises in the tropics and then migrates towards the pole where it sinks. Because of this, aerosols from tropical eruptions typically persist in the stratosphere for about two years, while aerosols from high-latitude volcanism persist for only one year \citep[][]{2007GeoRL..3423710R,2014GeoRL..41.7838T}. 

Previous work shows a link between exoplanet compositions and stellar compositions \citep[e.g.,][]{2010ApJ...712..974R} such that stellar compositions may be used to approximate the relative abundances of non-volatile/refractory elements in exoplanet interiors. Stars in exoplanetary systems show a wide variation in composition \citep[][]{2014AJ....148...54H}. In particular, some composition parameters with large variability such as Mg:Si ratios, are likely to have a first order effect on the minerals that compose exoplanetary interiors and thus the melting behavior, magma composition generated from these planetary mantles, and their volatile solubility. Certain compositional components, such as alkalis, have also been shown to greatly increase the H$_2$O solubility \citep[e.g.,][]{2001E&PSL.192..363B,2004JVGR..134..109L} in natural melts, and highlight the necessity of measuring volatile solubility behavior across a broad range of melt compositions. Magmatic volatile solubility is highly dependent on temperature, which also varies with mineralogy. 
%For example, we hypothesize that planets with bulk mantle compositions that are dominated by pyroxene and feldspar type minerals (Mg:Si$<$1) will have lower melting temperatures and thus higher H$_2$O solubility than planets dominated by olivine. 

On Earth, in addition to the pressure- and compositional-dependence of volatile solubility in magmas, the explosivity of a given eruption is dependent on the overall volatile concentration (dominated by H$_2$O and CO$_2$), magma supply rate, vent geometry, and source pressure of the magma body \citep[e.g.,][]{1980JVGR....8..297W,1999BVol...60..583P,2004BVol...66..735M}. The most explosive eruptions on Earth tend to be those at convergent plate boundaries where there are abundant volatiles involved in magma genesis sourced from the subducting plate, and some types of intraplate volcanism where interactions with reservoirs of volatiles in the crust produce highly explosive caldera eruptions. In addition, flood basalts and other volumetrically large outpourings of magma common in a planets early history may be a significant source of atmospheric volatiles \citep[][]{2012E&PSL.317..363B}. As such, the lack of tectonics on exoplanets does not preclude extreme volcanism that may produce detectable signatures. 

\subsubsection{Geological Activity and Plate Tectonics on Extrasolar Rocky Planets}

The terrestrial and venutian mantle convection, plate tectonics, and mantle outgassing are influenced by the initial bulk abundance of the planet and are particularly sensitive to the radioisotopic abundances; mantle outgassing and planetary evolution are particularly sensitive to the the modes of the tectonics (e.g., stagnant lid vs. plate tectonics), internal temperature distribution, and lid thickness (e.g., \citealt[][]{2013cctp.book..473O}).
The extrapolation of models of planetary evolution and plate tectonics to extrasolar rocky planets is challenging. A particularly relevant question is how plate tectonics may operate in super-Earths: on one hand, the higher heat flux (due to their intrinsically higher mass-to-surface ratio) should lead to stronger mantle convection \citep[e.g.,][]{ 2007ApJ...670L..45V, 2011E&PSL.310..252V}. On the other hand, based on a visco-elastic models of mantle convection and crust formation, \citet[][]{2007GeoRL..3419204O} find that increasing the planet's radius (and mass) will decrease the ratio of driving-to-resistive forces (see~Fig.~\ref{Fig:ONeill2007}), which reduces the likelihood of mobile plate tectonics in super-Earths and argues for the stagnant lid (or episodic tectonics) in these planets.

Furthermore, for a given planet models also suggest time-dependence and sensitivity to initial conditions: the thermal state of the post-magma ocean mantle is a key parameter that determines the subsequent evolution of the planet (possibly but not necessarily through i) hot stagnant-lid, ii) plate tectonics, then to iii) cold stagnant lid regime). Depending on the planet's transition from the magma ocean stage different evolutionary paths are possible and there may only be a limited time available for Earth-like plate tectonics \citep[][]{2016PEPI..255...80O}.

\begin{figure}[H]
\begin{center}
\includegraphics[width=0.3 \textwidth]{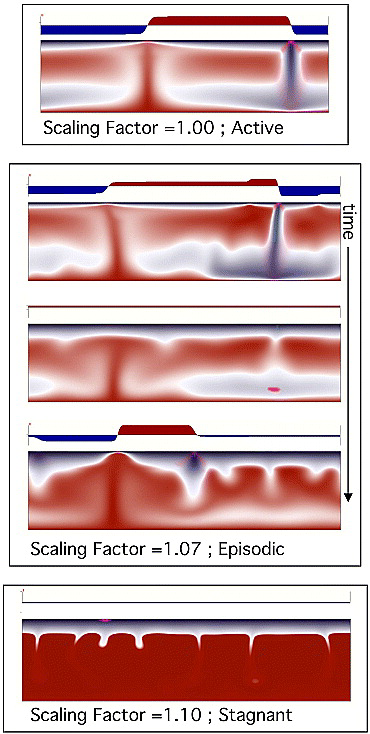}
\includegraphics[width=0.65 \textwidth]{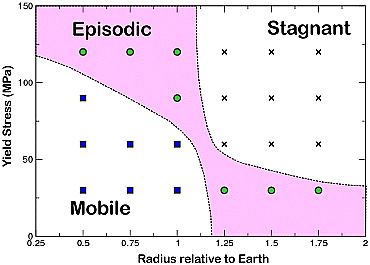}
\caption{Convection as a function of stellar radius and Byerlee-style pressure-dependent yield stress. The models include internal heating, a constant friction coefficient, and gravity matching the planetary mass. Larger radius results in greater buoyancy forces, but also increased fault strength due to increased pressure. Thus planets with larger radii again tend to be in an episodic or stagnant regime, depending on the absolute yield stress. From O'Neill \& Lenardic (2007).
\label{Fig:ONeill2007}}
\end{center}
\end{figure}

\subsubsection{Observational Methods}

While major geological processes usually unfold on timescales not accessible to long-range remote sensing, the {\em  results} of these processes are detectable and, in some cases, may be unambiguously identifiable. For example, in the case of Earth the presence of multiple large land-masses and oceans (detectable via time-resolved observations, e.g., \citealt[][]{2009ApJ...700..915C}) reveals that a continent-forming process acts on timescales shorter than water-driven land erosion and provides a characteristic scale for the continental plates. Another Earth-based example is the accumulation of atmospheric absorbers characteristic of volcanic outgassing (e.g., SO$_2$: \citealt[][]{2010ApJ...708.1162K}). Other, non-Earth-like, planets may offer other detectable signatures of geological activity.

In the following we briefly discuss four representative possibilities:\\
{\em i)} Continents and Oceans from Surface Maps\\
%{\em ii)} Atmospheric Absorbers from Volcanic Outgassing\\
{\em ii)} Water Clouds as Tracer of Topography\\
{\em iii)} Planetary-Scale Surface Geology\\
%{\em iv)} Cloud formation as Tracer of Topography and Erosion\\

{\em Continents and Oceans from Surface Maps:}  Simulated observations of Earth as an exoplanet demonstrate that with appropriate rotational- and orbital phase-resolved precision, multi-band photometric data can be use to identify the presence and one-dimensional and two-dimensional distribution of oceans and landmasses (see also Section~\ref{DetectingOceans}, e.g.,  \citealt[][]{2009ApJ...700..915C,2010ApJ...715..866F,2012ApJ...755..101F}).  
In a planet where large bodies of liquid water (i.e., an ocean) is present, a hydrological cycle is active, and land masses (continents) are detected, land erosion must arguably occur; the timescales for the erosion may be, to the first order, estimated based on terrestrial silicate weathering and erosion rates. The existence of the continents demonstrates that the time-scale of continent-formation is comparable or faster than their erosion. Based on a simplified model of water cycling and continent formation, \citet[][]{2014ApJ...781...27C} argues that continents and oceans may be common even among super-earths with high abundances of water. Such first-principle-based models may be combined with the scales of oceans and continents derived from observations to test whether active continent formation (e.g., plate tectonics) is required for a given planet.

{\em Water Clouds as Tracer of Topography:}
\citet[][]{2008ApJ...676.1319P} combines an Earth reflectance model with observed cloud distributions to calculate local and global (disk-integrated) reflectance photometric observations. They show that the dynamical cloud distributions introduces variable photometric signal that may reduce the value of auto-correlation in determining rotational periods, i.e., planets without strong auto-correlation signal in their time series may imply either near-complete cloud cover (such as Venus) or very chaotic weather. The comparison of the observed terrestrial cloud distribution also reveals a stable cloud component that is highly correlated with continents and topography, offering an indirect probe to topography from disk-integrated reflectance photometry.

 {\em Planetary-scale Surface Geology:} 
 Common mineral assemblages that make rocky planet surfaces have distinctive spectral features in the visible, near-infrared, and thermal infrared wavelengths. Broadband photometry of atmosphere-less rocky exoplanets can therefore tell their surface types \citep[][]{2012ApJ...752....7H}. For example, water-altered silicate surfaces (e.g., clays) will produce narrow absorption bands at 1.8 and 2.3 $\mu$m owing to the OH incorporated in the solids. For another example, the location of the peak in the 7-10 $\mu$m band of a silicate rock tells its silica content, which can be used to distinguish primary versus secondary crust on a rocky planet.
 \citet[][]{2014AsBio..14..753F} used albedo-map generated lightcurves and, where available, observed photometric variations to explore the geologic features detectable on diverse Solar System bodies with minor or no atmospheres (Moon, Mercury, the Galilean moons, and Mars).  The study included the evaluation of the light curves and the features that are detectable at wavelengths ranging from UV through visible to near-infrared wavelengths, and also explored the accuracy required to determine the orbital periods of these bodies. Figure~\ref{Fig:Fujii2014} provides an example for the wavelength-dependence of the rotational variability amplitudes in different bodies.
 
 Amplitude variations at the level of 5--50\% have been reported introduced by features of diverse nature (volcanism, space weathering, planetary weathering, impact excavation, tectonic deformation). In some cases data with the appropriate wavelength coverage can be used to identify some of these features or narrow down the possible origins. 
 
%\citet[][]{2012ApJ...752....7H} calculates theoretical near- and mid-infrared spectra for rocky planet surfaces to explore the potential for spectroscopic detection of surface types. They conclude that the near-infrared spectroscopy has the potential to identify ultramafic surfaces, hydrated surfaces, and water ice; while mid-infrared spectroscopy may detect rocky surfaces.
 
%{\em Volcanic Outgassing}

%{\em  Cloud formation as tracer of topology:} 

\subsubsection{Complementary Datasets} 
We identify three complementary datasets that are critically important for modeling the interior and activity of extrasolar rocky planets:
\begin{itemize}
\item {\em Stellar abundances:} A proxy for the relative refractory elemental abundances that may be present in the planets \citep[e.g.][]{2016ApJ...819...19T}. In particular, stellar abundance patterns may be used to identify outlier systems in terms of elemental abundances. 
\item {\em Stellar/system age:} The age of the system is an important parameter in assessing the evolution of the rocky planets: it can help  in constraining the evolutionary state of the planets (heat flux and time available for volatile loss and resurfacing processes). The stellar ages will likely come from a combination of stellar gyrochronology and astroseismology \citep[e.g.][]{2014MNRAS.441.2361V,2014A&A...572A..34G,2015Natur.517..589M,2016Natur.529..181V}. 
\item {\em Mass and radius of the planet:} These fundamental physical parameters have major impact on the initial energy budget, thermal evolution of the planet, atmospheric/volatile losses, and force balances. Observations of the stellar reflect motion (precision stellar astrometry and/or radial velocity) along with the imaging observations will be important for establishing the planetary orbit and planet mass. Radius measurements will be very difficult for non-transiting planets and the best estimates will likely come from reflected light measurements {\em if} the planetary albedo can be deduced or constrained. Joint constraints on the mass and radius of the planets will provide a constraint on the bulk density, which will be important for assessing the range of interior structures possible for the planets.
\end{itemize}

\subsubsection{Science Value of Independently Measured Planet Masses and Radii: Very High} 
Exploring the planetary-scale geophysics of rocky planets will likely be among the most challenging aspects of characterizing extrasolar rocky planets. Yet, understanding the geophysics and interior activity of these planets may well turn out to be essential for correctly and robustly interpreting atmospheric biosignatures.
The rocky planet's mass is one of the most fundamental parameter that influences heat flux, pressure, and horizontal forces acting on the lithosphere. Given the sensitivity of plate tectonics models to planet mass, it is likely that determining the planet mass with a precision of $\sim$10\% is required for establishing a robust geophysical model. 
\newpage

\begin{mdframed}[backgroundcolor=blue!20] 
{\bf Observational Requirements:} \\
{\bf Exoplanet characterization:} Establishing detailed rotational maps, e.g., identifying the presence and distribution of oceans and continents are essential for this science questions (see Question B3, B4, C1). Detailed atmospheric composition is required for assessing for evidence of large-scale volcanism. Very high signal-to-noise time-resolved mid-infrared spectroscopy is required to attempt to identify mineralogical features and to spatially resolve surface compositional variations. \\
{\bf Supplementary Observations:} Stellar reflex motion (astrometry/radial velocity) will be essential for determining the planet mass, one of the most fundamental properties determining the interior structure. Detailed characterization of the system, including the stellar age and abundance patterns, will provide important context for the interpretation of possible evolutionary paths and the range of interior activity expected.
\end{mdframed}

\appendix

\begin{center}
\Huge Appendix
\end{center}

\section{SAG15 Charter}
Future direct imaging missions may allow observations of flux density as a function of wavelength, polarization, time (orbital and rotational phases) for a broad variety of exoplanets ranging from rocky sub-earths through super-earths and neptunes to giant planets. With the daunting challenges to directly imaging exoplanets, most of the community's attention is currently focused on how to reach the goal of exploring habitable planets or, more specifically, how to search for biosignatures. 

Arguably, however, most of the exoplanet science from direct imaging missions will not come from biosignature searches in habitable earth-like planets, but from the studies of a much larger number of planets outside the habitable zone or from planets within the habitable zone that do not display biosignatures. These two groups of planets will provide an essential context for interpreting detections of possible biosignatures in habitable zone earth-sized planets. 

However, while many of the broader science goals of exoplanet characterization are recognized, there has been no systematic assessment of the following two questions: \\1) What are the most important science questions in exoplanet characterization apart from biosignature searches? \\
2) What type of data (spectra, polarization, photometry) with what quality (resolution, signal-to- noise, cadence) is required to answer these science questions? \\

We propose to form SAG15 to identify the key questions in exoplanet characterization and determine what observational data obtainable from direct imaging missions is necessary and sufficient to answer these. 

The report developed by this SAG will explore high-level science questions on exoplanets ranging from gas giant planets through ice giants to rocky and sub-earth planets, and -- in temperatures -- from cold ($\sim$200 K) to hot ($\sim$2,000 K). For each question we will study and describe the type and quality of the data required to answer it. 

For example, the SAG15 could evaluate what observational data (minimum sample size, spectral resolution, wavelength coverage, and signal-to-noise) is required to test that different formation pathways in giant planets lead to different abundances (e.g. C/O ratios). Or the SAG15 could evaluate what photometric accuracy, bands, and cadence is required to identify continents and oceans in a habitable zone Earth-sized or a super-earths planet. As another example, the SAG15 could evaluate what reflected light data is required to constrain the fundamental parameters of planets, e.g. size (distinguishing earth-sized planets from super- earths), temperature (cold/warm/hot), composition (rocky, icy, gaseous), etc. 

SAG15 will not attempt to evaluate exoplanet detectability or specific instrument or mission capabilities; instead, it will focus on evaluating the diagnostic power of different measurements on key exoplanet science questions, simply adopting resolution, signal-to-noise, cadence, wavelength coverage as parameters along which the diagnostic power of the data will be studied. Decoupling instrumental capabilities from science goals allows this community-based effort to explore the science goals for exoplanet characterization in an unbiased manner and in a depth beyond what is possible in a typical STDT. 

We envision the SAG report to be important for multiple exoplanet sub-communities and specifically foresee the following uses:
1) Future STD teams will be able to easily connect observational requirements to missions to fundamental science goals; \\
2) By providing an overview of the key science questions on exoplanets and how they could be answered, it may motivate new, dedicated mission proposals;\\
3) By providing a single, unified source of requirements on exoplanet data in advance of the Decadal Survey, the science yield of various missions designs can be evaluated realistically, with the same set of assumptions. 

Our goal is to carry out this SAG study by building on both the EXOPAG and NExSS communities. 

We aim to complete a report by Spring 2017 and submit it to a refereed journal, although this timeline can be adjusted to maximize the impact of the SAG15 study for the ongoing and near- future STDTs and other mission planning processes. 

Synergy with a potential future SAG proposed by Shawn Domagal-Goldman: While the SAG proposed here will include studies of habitable zone rocky planets, it will focus on planets without significant biological processes. A future SAG may be proposed by Shawn Domagal- Goldman to explore biosignatures; if such a SAG is proposed, we envision a close collaboration on these complementary, but distinct problems.

\newpage

\section{Methods of Collecting and Organizing Input}

{\bf Updates:} Throughout the project the SAG15 team has provided up-to-date information on the report's status and next steps to different constituents (EXOPAG, EXOPAG EC, NExSS, exoplanet community, STDTs) via the following channels:

\begin{itemize}
\item{The SAG15 website always containing the up-to-date report draft and links to all relevant documents}
\item{Monthly telecons open to anyone in the exoplanet community}
\item{Minutes of most telecons were circulated on the SAG15 mailing list to keep all members abreast of the progess}
\item{Emails sent to the NExSS group and EXOPAG groups}
\item{Status updates provided to the EXOPAG community at every AAS meeting during the project}
\item{Presentation/hackathlon session during the NExSS Face-to-Face meeting in May 2016}
\item{Representatives of the LUVOIR and HabEx STDTs on the SAG15 team and attended telecons}
\item{The up-to-date version of the SAG15 report was shared with the LUVOIR STDT}
\item{A brief presentation by Marley at the LUVOIR STDT meeting in Aug 2016 reviewed the progress of SAG15}
\end{itemize}

{\bf Soliciting Input:}
SAG15 has solicited and collected input from the different constituents (EXOPAG, EXOPAG EC, NExSS, exoplanet community, STDTs) through the following channels:
\begin{itemize}
\item{Presentations at the EXOPAG/AAS meetings}
\item{Presentations to the NExSS community}
\item{Emails sent to the NExSS group and EXOPAG groups}
\item{Targeted emails soliciting input from scientists with required expertise}
\item{Input collected from the NExSS Biosignatures and SAG16 workshop}
\item{Input collected from hackathlon session at NExSS Face-to-Face meeting (25 participants)}
\item{Representatives of the LUVOIR and HabEx STDTs on the SAG15 team, attended telecons, and provided updates on progress}
\item{The advanced draft of the report circulated in Oct 2016 in the EXOPAG, NExSS communities and sent to topical experts}
\end{itemize}

{\bf SAG15 Website:} The SAG15 website (http://eos-nexus.org/sag15/) was established right after the approval of SAG15 by the Astrophysics Subcommittee. The website contains links to the SAG15 report draft, providing step-by-step overview on the evolution of the report as well as a copy of the up-to-date report. 

%% To help institutions obtain information on the effectiveness of their
%% telescopes, the AAS Journals has created a group of keywords for telescope
%% facilities. A common set of keywords will make these types of searches
%% significantly easier and more accurate. In addition, they will also be
%% useful in linking papers together which utilize the same telescopes
%% within the framework of the National Virtual Observatory.
%% See the AASTeX Web site at http://aastex.aas.org/
%% for information on obtaining the facility keywords.

%% After the acknowledgments section, use the following syntax and the
%% \facility{} macro to list the keywords of facilities used in the research
%% for the paper.  Each keyword will be checked against the master list during
%% copy editing.  Individual instruments or configurations can be provided 
%% in parentheses, after the keyword, but they will not be verified.

%{\it Facilities:} \facility{Nickel}, \facility{HST (STIS)}, \facility{CXO (ASIS)}.

%% Appendix material should be preceded with a single \appendix command.
%% There should be a \section command for each appendix. Mark appendix
%% subsections with the same markup you use in the main body of the paper.

%% Each Appendix (indicated with \section) will be lettered A, B, C, etc.
%% The equation counter will reset when it encounters the \appendix
%% command and will number appendix equations (A1), (A2), etc.

\newpage
\section{Contributing to the SAG15 Report}

The SAG15 Report (Science Questions for Direct Imaging Missions) is a community-based effort and it is open to anyone interested in contributing to the report or to the discussions that shape the report. Input is welcome from any members of the exoplanet community, regardless of academic degree, position, level of experience, nationality, or affiliation. Everyone who has participated in discussions leading to the report will be identified as a SAG15 Team member and those who contributed significantly to the report will be identified as authors. Comments are welcome at any time, but are most useful if they follow our report development plan; therefore, if you are interested in contributing, please, join our mailing list, participate in the telecons, and follow the guidelines below on how to format your input.

{\bf Joining the SAG15 Team:} If you would like to join the SAG15 team, please, email to SAG15 Chair Daniel Apai (apai@arizona.edu). We will add you to the SAG15 mailing lists and you will receive invitations to the monthly telecons and will be kept up-to-date on the SAG15 progress.

{\bf Input for the SAG15 Draft Report:} Any level of input is helpful, but the most useful is if you provide a balanced, quantitative, and fully referenced assessment of an aspect that is missing or not thoroughly covered in the current draft. Note, that by this point we have converged on the broad science questions so, if at all possible, plan your contribution to fit within the existing categories.

{\em The latest version of the draft:} The SAG15 website will always contain the latest version: http://eos-nexus.org/sag15/

{\em How to Format your input?}
The SAG15 report is typeset in Latex compiled with PdfLatex.

1) Please send fully referenced paragraphs that can be inserted into the latex source text. 

2) Figures: Please send figures as PDF or PNG files, along with fully referenced captions and source.

3) References: Please, send reference info as bibcodes,  i.e., \verb| 1905LowOB...1..134L| and in the latex text refer them
by bibcode:  \verb| \citep[][]{1905LowOB...1..134L},| \\ \verb|\citet[][]{1905LowOB...1..134L}|, or 
\verb|\citealt[][]{1905LowOB...1..134L}.|
  
4) Original text: We will submit the report to a refereed journal; our manuscript must be original. Therefore, please, do not re-use text from your or other's publications.

5) Please, be specific: identify what should be changed and exactly how. 

{\em Example input}

The following is an example for the input that is most useful:

\begin{verbatim}
Insert the following to Section 3.2.1 after the second paragraph:

"Additional observations by \citep[][]{1925ApJ....62..409H} provided 
supporting evidence, as shown in Figure~\ref{Fig:Label}."

Add the following references to the SAG15 library:
1925ApJ....62..409H

And use the attached .pdf figure for {Fig:Label}."

\end{verbatim}
\clearpage
%=============================================

\section{Relevant Past Reports and Resources}

\noindent \href{https://exoplanets.nasa.gov/exep/exopag/overview/}{Exoplanet Exploration Program} 

\vskip 0.5cm

\noindent {\bf Astrophysics Strategy Documents}

\noindent \href{https://smd-prod.s3.amazonaws.com/science-blue/s3fs-public/atoms/files/secure-Astrophysics_Roadmap_2013.pdf}{Astrophysics Roadmap: Enduring Quests, Daring Visions}

\noindent \href{http://sites.nationalacademies.org/bpa/BPA_049810}{The 2010 Astrophysics Decadal Survey}

\vskip 0.5cm

\noindent {\bf Upcoming Missions}

\noindent \href{https://wfirst.gsfc.nasa.gov}{WFIRST}

\noindent \href{http://www.jwst.nasa.gov}{JWST}

\noindent {\bf STDT and SWG Reports}

\noindent \href{https://exoplanets.nasa.gov/files/exep/tpfI414.pd}{Technology Plan For Terrestrial Planet Finder Interferometer}

\noindent  \href{https://exoplanets.nasa.gov/files/exep/TPFIswgReport2007.pdf}{Terrestrial Planet Finder Interferometer Science Working Group Report}

\noindent \href{https://exoplanets.nasa.gov/files/exep/STDT_Report_Final_Ex2FF86A.pdf}{Terrestrial Planet Finder Coronagraph Science and Technology Design Team Report}

\noindent \href{https://exoplanets.nasa.gov/exep/stdt/Exo-S_Starshade_Probe_Class_Final_Report_150312_URS250118.pdf}{Exo-S Final Report}

\noindent \href{https://exoplanets.nasa.gov/exep/stdt/Exo-C_Final_Report_for_Unlimited_Release_150323.pdf}{Exo-C Final Report}

\noindent \href{http://static1.squarespace.com/static/558adc44e4b002a448a04c1a/t/55a411dee4b0543aa4ede4f2/1436815838795/hdst_report2_071315.pdf}{From Cosmic Birth to Living Earths (AURA Report on Future of UVOIR Astronomy)}

\noindent \href{http://cor.gsfc.nasa.gov/copag/rfi/149_newworlds_Cash_EOS.pdf}{The New Worlds Observer}

\vskip 0.5cm

\noindent {\bf Study Analysis Group Reports}

\noindent \href{https://exoplanets.nasa.gov/exep/exopag/sag/}{EXOPAG Study Analysis Groups Website}

\noindent \href{http://adsabs.harvard.edu/abs/2012PASP..124..799R}{Debris Disks \& Exozodiacal Dust (Aki Roberge and the SAG1 Team}

\noindent \href{http://arxiv.org/abs/1303.6707}{Exoplanet Flagship Requirements and Characteristics (Noecker, Greene and the SAG5 Team}

\noindent \href{http://adsabs.harvard.edu/abs/2015arXiv150301770P}{Requirements and Limits of Future Precision Radial Velocity Measurements (Latham,  Plavchan, and SAG8 Team)}

\noindent \href{https://exep.jpl.nasa.gov/files/exep/ExoPAG-SAG9-Final.pdf}{Exoplanet Probe to Medium Scale Direct-Imaging Mission Requirements and Characteristics (Soummer and SAG9 Team)}

\noindent \href{https://arxiv.org/abs/1409.2759}{Preparing for the WFIRST Microlensing Survey (Yee and the SAG 11 Team)}

\clearpage

%% If you are not including electonic art with your submission, you may
%% mark up your captions using the \figcaption command. See the
%% User Guide for details.
%%
%% No more than seven \figcaption commands are allowed per page,
%% so if you have more than seven captions, insert a \clearpage
%% after every seventh one.

%% Tables should be submitted one per page, so put a \clearpage before
%% each one.

%% Two options are available to the author for producing tables:  the
%% deluxetable environment provided by the AASTeX package or the LaTeX
%% table environment.  Use of deluxetable is preferred.
%%

%% Three table samples follow, two marked up in the deluxetable environment,
%% one marked up as a LaTeX table.

%% In this first example, note that the \tabletypesize{}
%% command has been used to reduce the font size of the table.
%% We also use the \rotate command to rotate the table to
%% landscape orientation since it is very wide even at the
%% reduced font size.
%%
%% Note also that the \label command needs to be placed
%% inside the \tablecaption.

%% This table also includes a table comment indicating that the full
%% version will be available in machine-readable format in the electronic
%% edition.
\bibliographystyle{aa}       % APS-like style for physics
\bibliography{SAG15refsnew}   % name your BibTeX data base

\end{document}